\documentclass[fleqn,usenatbib]{mnras}

\usepackage{newtxtext,newtxmath}

\usepackage[T1]{fontenc}

\DeclareRobustCommand{\VAN}[3]{#2}
\let\VANthebibliography\thebibliography
\def\thebibliography{\DeclareRobustCommand{\VAN}[3]{##3}\VANthebibliography}

\usepackage{color, colortbl}
\definecolor{Gray}{gray}{0.9}

\usepackage{graphicx}	
\usepackage{amsmath}	
\usepackage{subcaption}
\usepackage{array}

\title[AGN radiation effects on CGM]{AGN radiation imprints on the circumgalactic medium of massive galaxies} 
\author[A. Obreja et al.]{
Aura Obreja$^{1}$\thanks{E-mail: obreja@usm.lmu.de},
Fabrizio Arrigoni Battaia$^{2}$,
Andrea V. Macci\`{o}$^{3,4,5}$
and Tobias Buck$^{6,7}$
\\
$^{1}$Universit\"ats-Sternwarte M\"unchen, Scheinerstraße 1, D-81679 M\"unchen, Germany\\
$^{2}$Max-Planck-Institut f\"ur Astrophysik, Karl-Schwarzschild-Straße 1, D-85748 Garching bei M\"unchen, Germany\\
$^{3}$New York University Abu Dhabi, PO Box 129188, Abu Dhabi, United Arab Emirates\\
$^{4}$Center for Astro, Particle and Planetary Physics (CAP3), New York University Abu Dhabi\\
$^{5}$Max-Planck-Institut f\"ur Astronomie, K\"onigstuhl 17, D-69117 Heidelberg, Germany\\
$^{6}$Universit\"at Heidelberg, Interdisziplin\"ares Zentrum f\"ur Wissenschaftliches Rechnen, Im Neuenheimer Feld 205,  D-69120 Heidelberg, Germany\\
$^{7}$Universit\"at Heidelberg, Zentrum f\"ur Astronomie, Institut f\"ur Theoretische Astrophysik, Albert-Ueberle-Straße 2,  D-69120 Heidelberg, Germany
}

\date{Accepted 2023 October 27. Received 2023 October 27; in original form 2023 August 21}

\pubyear{2023}

\begin{document}
\label{firstpage}
\pagerange{\pageref{firstpage}--\pageref{lastpage}}
\maketitle

\begin{abstract}
Active Galactic Nuclei (AGN) in cosmological simulations generate explosive feedback that regulates star formation in massive galaxies, modifying the gas phase structure out to large distances. Here, we explore the direct effects that AGN radiation has on gas heating and cooling within one high-resolution $z=3$ dark matter halo as massive as a quasar host ($M_{\rm h}=$10$^{\rm 12.5}$M$_{\rm\odot}$), run without AGN feedback. We assume AGN radiation to impact the circumgalactic medium (CGM) anisotropically, within a bi-cone of angle $\alpha$. We find that even a relatively weak AGN (black hole mass $M_{\rm\bullet}=10^{\rm 8}$M$_{\rm\odot}$ with an Eddington ratio $\lambda=0.1$) can significantly lower the fraction of halo gas that is catastrophically cooling compared to the case of gas photoionized only by the ultraviolet background (UVB). Varying $M_{\rm\bullet}$, $\lambda$ and $\alpha$, we study their effects on observables. A 10$^{\rm 9}$M$_{\rm\odot}$ AGN with $\lambda=0.1$ and $\alpha\approxeq60^{^{\rm o}}$ reproduces the average surface brightness (SB) profiles of Ly$\alpha$, \ion{He}{ii} and \ion{C}{iv}, and results in a covering fraction of optically thick absorbers within observational estimates. The simulated SB$_{\rm\ion{C}{iv}}$ profile is steeper than observed, indicating that not enough metals are pushed beyond the very inner CGM. For this combination of parameters, the CGM mass catastrophically cooling is reduced by half with respect to the UVB-only case, with roughly same mass out of hydrostatic equilibrium heating up and cooling down, hinting to the importance of self-regulation around AGNs. This study showcases how CGM observations can constrain not only the properties of the CGM itself, but also those of the AGN engine.
\end{abstract}




\section{Introduction}
\label{sec:intro}

Super massive black holes (SMBHs, $\ge$10$^{\rm 5}$M$_{\rm\odot}$) can inhabit all types of galaxies, from dwarfs \citep[e.g.][]{Reines:2013} to bright cluster galaxies \citep[e.g.][]{Gear:1985}. Their strong gravitational field, very spatially localized, has only been directly measured in our own galaxy \citep[e.g.][]{Genzel:1997}, or in the very nearby universe \citep[e.g.][]{EHT:2019}. At large distances (redshifts) their presence is typically inferred from the strong radiation field emitted from their vicinity during periods of accretion, hence the name AGN \citep{Lynden-Bell:1969}. 

AGNs are distinguished from other astrophysical radiation sources by: extremely high bolometric luminosities (up to $L_{\rm bol}\lesssim$10$^{\rm 49}$erg~s$^{\rm -1}$), a very strong evolution of their luminosity function \citep[e.g.][]{Shen:2020}, and SEDs that are detectable across all wavelengths, from radio to gamma rays \citep[e.g.][]{Bianchi:2022}. Their extreme luminosity makes them valuable cosmic beacons to find and explore the properties of diffuse gas even in the highest redshift universe, both in emission and in absorption against their light \citep[e.g.][]{Farina:2019,D'Odorico:2023}. The combination between the AGN's high power and the compactness of their emitting regions implies very high energy densities, bound to have an effect on their host galaxies \citep[e.g.][]{Bower:2008}. Current observational signposts of AGN effects on their host galaxies include: highly collimated jets on scales up to hundreds of kpc \citep[e.g.][]{Waggett:1977,Biretta:1983}, outflows/(ultra)fast winds on scales of the order of kpc to tens of kpc \citep[e.g.][]{Pounds:2003,Ganguly:2007,Ajello:2021}, high energy bubbles and cavities in the circumgalactic medium anchored on the galaxy's center \citep[e.g.][]{Su:2010,Predehl:2020}. On the other hand, scaling relation between SMBH masses and galaxy properties like bulge mass \citep{Magorrian:1998}, central stellar velocity dispersion \citep{Ferrarese:2000,Gebhardt:2000}, and hot circumgalactic medium temperature and X-ray luminosity \citep{gaspari2019x,Lakhchaura:2019} provide further contextual evidence for such effects. All these observations taken together suggest that the energy and momentum generated by the accretion heats, ionizes and pushes outwards the gas surrounding the AGN, thus regulating the SMBH growth and star formation in the galaxy \citep[e.g.][]{Begelman:2005,Fabian:2012}.

One of the strongest indirect evidences in favor of an active role of AGNs in galaxy evolution comes from semi-analytical models, which have shown that the dependency of stellar mass on dark matter halo mass is best described by a double power-law \citep[e.g.][]{Moster:2018,Behroozi:2019}. This translates into a particular halo mass ($\approx$10$^{\rm12}$M$_{\rm\odot}$) for which the efficiency of star formation is maximum. In both semi-analytical studies and cosmological simulations of galaxy formation, this power-law dependency of galaxy stellar mass on dark matter halo mass can only be reproduced if the hydrodynamics equations take into account not only the gas cooling and star formation, but also the back-reaction of gas under energetic phenomena like supernovae at low halo masses \citep[SNe, e.g.][]{Stinson:2006,DallaVecchia:2008,Oppenheimer:2008,Keller:2014} and AGN at high halo masses \citep[e.g.][]{Springel:2005,Sijacki:2007,Booth:2009,Debuhr:2011,Costa:2014,Steinborn:2015,Weinberger:2017}. The impact of these energetic phenomena on the formation and evolution of galaxies are generally referred to as feedback processes. In this framework, simulations, and especially cosmological simulations that also trace the formation of galaxies within the cosmic web, are a powerful tool to follow and understand the complex astrophysics of structure build-up through cosmic time. 

In cosmological simulations, AGN feedback is generally divided in a quasar (radiative) and a radio (jet) mode, none of which can be modeled from first principles due to resolution limitations. Quasar mode feedback (regime of high accretion rates) has been  implemented in two distinct ways: as a \textit{thermal} energy transfer from the accreting SMBH to the nearby gas \citep[e.g.][]{Springel:2005,DiMatteo:2005,Booth:2009,Steinborn:2015}, and as a momentum (\textit{kinetic}) transfer from the AGN photons to the nearby gas \citep[e.g. radiation driven winds][]{Costa:2014,Costa:2018}. In these radiative feedback schemes, the energy and/or momentum imparted to the neighboring gas is proportional to the SMBH's bolometric luminosity, which is assumed to scale linearly with the accretion rate. The radio mode feedback is observationally based on the relativistic jets associated with low accreting AGN \citep[e.g.][]{Blandford:2019,Kondapally:2023}, and it is thought to be responsible for the large high energy bubbles observed around galaxies \citep[e.g.][]{Guo:2012,Yang:2022}. The implementations of radio mode feedback in cosmological simulations is severely limited by resolution, as the modeling of the jets requires general relativistic magneto-hydrodynamical simulations with very high resolutions \citep[e.g.][]{Ressler:2017}. For this reason, this feedback mode is implemented also as a momentum transfer in a very approximate manner, and often referred to as kinetic feedback \citep[e.g.][]{Sijacki:2007,Weinberger:2017,Dave:2019}.  A third, much less explored feedback mode, dubbed \textit{electromagnetic} (EM) feedback by \citet{Vogelsberger:2013} aims to quantify the direct photoionization/photoheating of gas by AGN radiation in simple toy models or isolated galaxy simulations \citep[e.g.][]{Ciotti:1997,Ciotti:2007,Sazonov:2005,Choi:2012,Gnedin:2012,Xie:2017}. In cosmological simulations, \citet{Kim:2011} implemented EM feedback by solving the radiation transfer on-the-fly assuming a 2~keV monochromatic spectrum that scales with the AGN's bolometric luminosity, while \citet{Vogelsberger:2013} used a universal, time independent AGN spectrum, to compute the photoionization and photoheating rates of hydrogen and helium, under the assumption that gas is optically thin to the AGN radiation.   

While the various flavours of AGN feedback models reproduce fairly well galaxy properties, they produce diverging results for the amount of baryons locked in virialized structures. In particular, current AGN feedback models reduce the baryon fractions in massive halos \citep[e.g.][]{Wright:2020,Sorini:2022} in comparison with the cosmic mean \citep[e.g. 0.157,][]{Planck:2014}. This prediction might be in tension with X-ray observations at low redshifts (e.g. \citet{Chadayammuri:2022}, but see also \citet{Comparat:2022} for a different conclusion). Mapping the hot CGM in galaxies is currently becoming a reality \citep[e.g.][]{Nicastro:2023}, and the development of new instruments will, hopefully, allow measuring the baryon fraction across galaxy populations \citep[e.g.][]{Schellenberger:2023}. Another observational estimate in disagreement with current simulations is the non-thermal to thermal pressure ratio in galaxy clusters, whose value is significantly smaller than what such simulations predict \citep[e.g.][]{Sayers:2021}. This indicates that current AGN implementations introduce too much turbulence in the interstellar/circumgalactic medium. All in all, CGM observations across the full electromagnetic spectrum are needed to constrain AGN feedback models, and galaxy formation models in general.  

A powerful method to estimate the properties of the multi-phase CGM, is the study of absorption features in the spectra of bright background sources, i.e. quasars \citep[e.g. see review by ][]{Tumlinson:2017}. In this method, one can statistically probe various impact parameters within the CGM regions of foreground galaxies. The subsequent modeling of the absorption features can put constraints on the nature of the absorber, and therefore on the gas properties in the CGM of the foreground system like, e.g., the ionization state and metallicity \citep[e.g.][]{Prochaska:2013,Werk:2014,Lau:2016}. Studies of quasars' CGM in absorption at $z=2-3$ have found evidence for large reservoirs ($>$10$^{\rm 10}$~M$_{\rm\odot}$) of cool gas with relatively high metallicities ($>0.1~Z_{\rm\odot}$) \citep[e.g.][]{Prochaska:2013,Lau:2016}, and with kinematics mostly in agreement with those of the hosting dark matter halos \citep[e.g.][]{Prochaska:2014,Lau:2018}. Importantly, the distribution of optically thick \ion{H}{i} systems around quasars is found to be highly anisotropic, with the transverse direction preferred over the line of sight (\citealt{HennawiProchaska:2007}). This fact suggests that quasars illuminate their surrounding anisotropically. 

Besides absorption studies, a complementary and more direct method to constrain the effects of AGN radiation on the CGM is provided by line emission maps. Fortunately, the high redshift CGM is now accessible with sensitive integral field unit (IFU) spectrographs like the Multi-Unit Spectroscopic Explorer on the Very Large Telescope \citep[MUSE,][]{Bacon:2010}, or the Keck Cosmic Web Imager on Keck \citep[KCWI,][]{Morrissey:2018}, which can directly probe rest-frame UV emission lines from the high-z cool CGM. The usually strongest emission line is the hydrogen Lyman~$\alpha$ (Ly$\alpha$) transition at 1215.67\AA~restframe. In particular, MUSE and KCWI have facilitated an explosion in the number of surveyed objects for CGM emission at high-$z$, resulting in the detection of extended Ly$\alpha$ glows around a few hundred quasars \citep[e.g.][]{Borisova:2016,ArrigoniBattaia:2019,Cai:2019,Farina:2019,Fossati:2021}. These Ly$\alpha$ nebulae around quasars have been long ago predicted \citep[e.g.][]{Rees:1988}, but the level of emission is so faint that only in the last decade they started to be routinely observed. The largest homogeneous sample of Ly$\alpha$ glows around quasars is the QSOMUSEUM survey \citep{ArrigoniBattaia:2019} which targeted 61 quasar fields at $z\sim3$. This survey found a wide range of nebula morphologies and sizes, from an asymmetric 'enormous Lyman alpha nebula' (ELAN) spanning $\sim$300~kpc to small symmetric ones that barely reach the CGM region \citep{ArrigoniBattaia:2023}. These observations are a unique probe into the CGM of quasars, as Ly$\alpha$ traces cool gas ($T\sim10^4$K), which can subsequently fuel star formation and AGN activity on galaxy scales. A detailed analysis of the ELAN in the QSOMUSEUM survey showed that the cool gas is likely inspiraling towards the central galaxy, and has velocity dispersions in agreement with pure gravitational motions within a $\sim$10$^{\rm 12.5}$M$_{\rm\odot}$ dark matter halo \citep{ArrigoniBattaia:2018}. How this cool and likely dense \citep[e.g.][]{Hennawi:2015,ArrigoniBattaia:2015} gas manages to survive in the hot (virial temperatures of the order of 10$^{\rm 6.5}$K) and harsh environment of quasars is an active topic of research \citep[e.g.][]{Bennett:2020,Costa:2022,Gronke:2022}.

In this work, we quantify, in post-processing the intrinsic and observable effects induced by photoionization and photoheating from an accreating SMBH on the CGM of a high resolution cosmological $z=3$ quasar host halo. The paper is structured as follows. In Sections~\ref{sec:sim} and \ref{sec:model_spectra} we describe the simulation and the spectra of the radiation sources considered. Section~\ref{sec:cloudy} explains how we construct the photoionization models, while Section~\ref{sec:geometry} quantifies the intrinsic effects of an anisotropic AGN radiation field on CGM heating and cooling rates. Our comparisons with CGM observations in absorption and emission are given in Section~\ref{sec:absorbtion} and \ref{sec:emission}, while in Section~\ref{sec:discussion} we discuss the implications for on-going and future observational efforts to map the CGM of high redshift quasars. Finally, Section~\ref{sec:summary} summarizes our results. This study is a first step in the development of an updated AGN EM feedback model for cosmological simulations.

\begin{figure}
    \centering
    \includegraphics[width=0.235\textwidth]{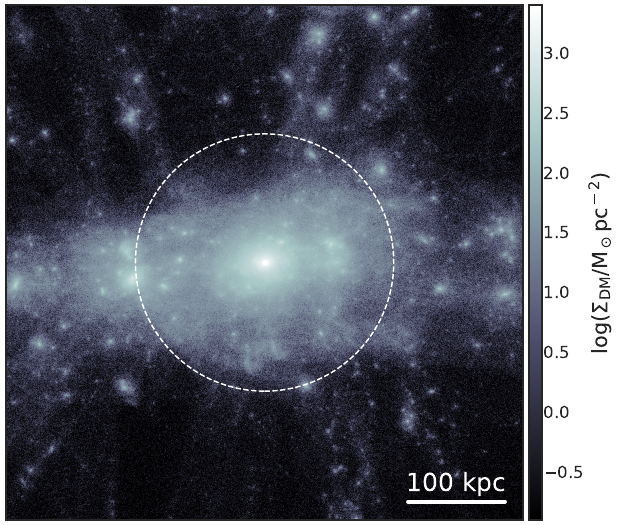}
    \includegraphics[width=0.235\textwidth]{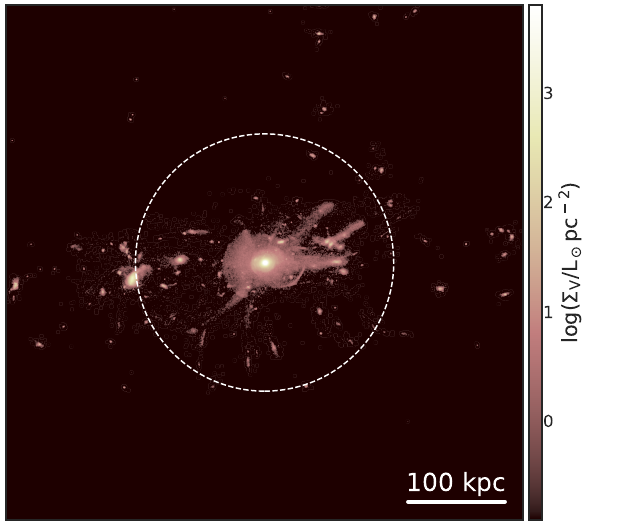}\\
    \includegraphics[width=0.235\textwidth]{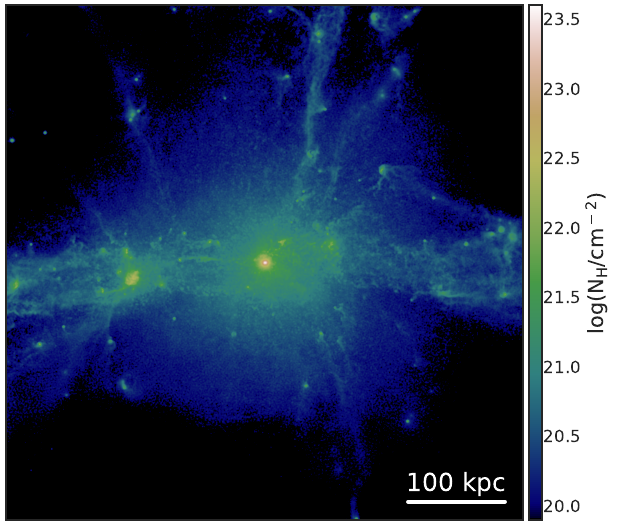}
    \includegraphics[width=0.235\textwidth]{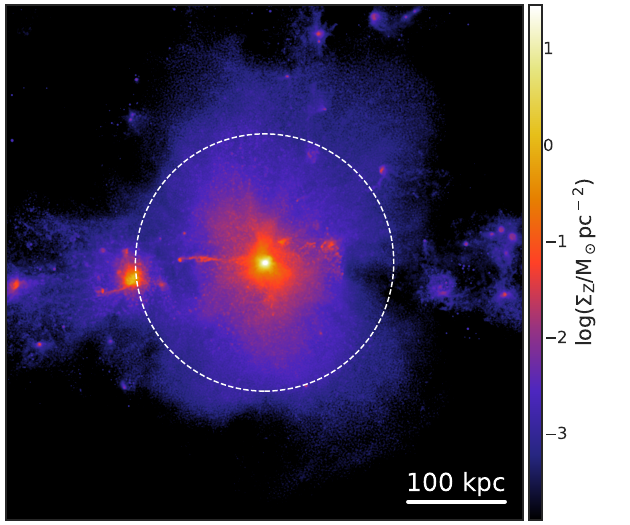}\\
    \caption{Dark matter and baryonic component maps for the simulated galaxy g3.16e12 at $z\sim3$. In clockwise direction from top-left 
    we show: dark matter surface mass density, stellar surface brightness in rest-frame V-band, metal surface mass density, and hydrogen column density. The dashed white circles show the position of the 134~kpc virial radius.}
    \label{fig:g316e12maps}
\end{figure}

\begin{figure}
    \centering
    \includegraphics[width=0.49\textwidth]{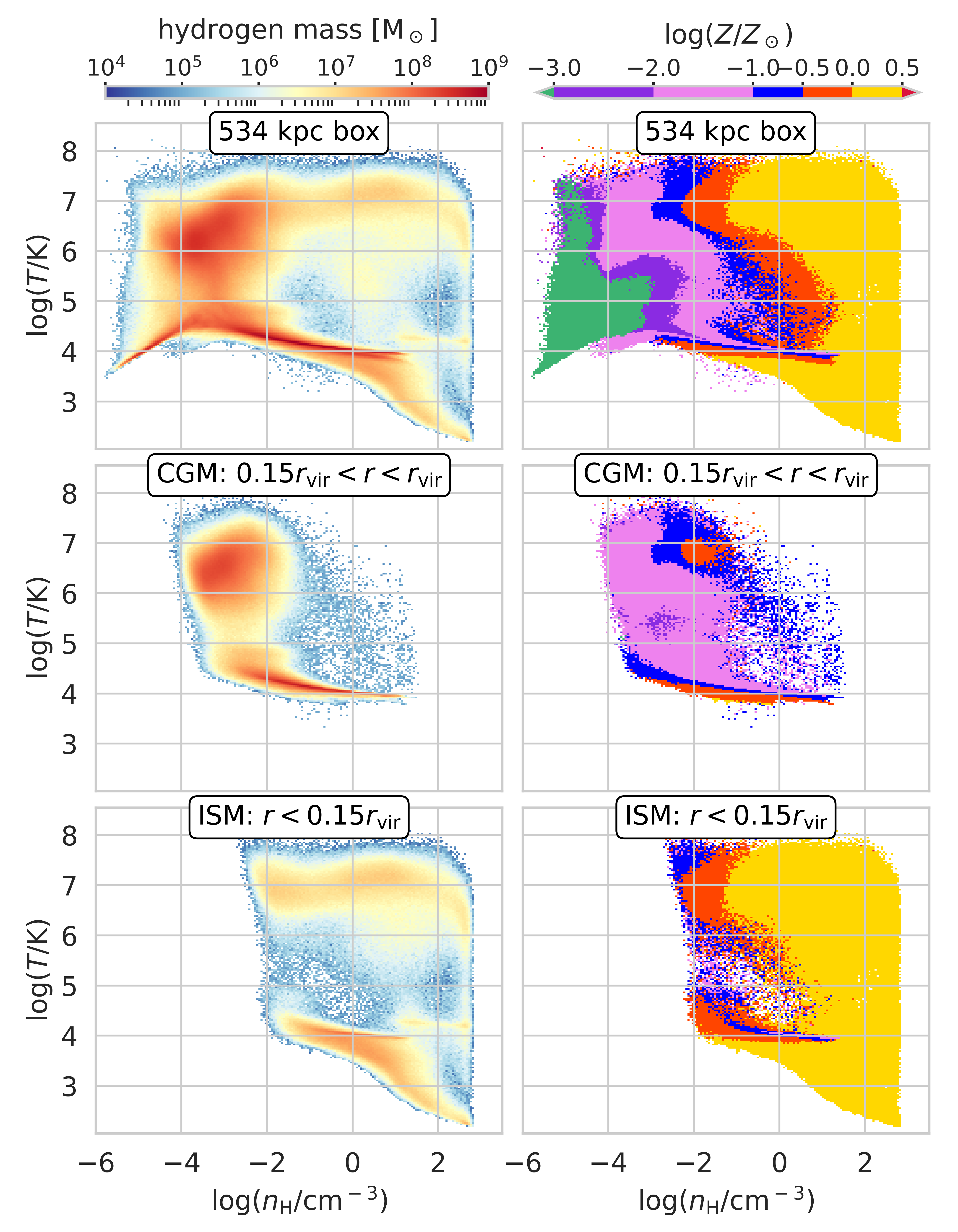}
\caption{Density--temperature phase space of simulation g3.16e12 within a 4$\times r_{\rm 200}$ box (top), and within the CGM of the main halo, $0.15r_{\rm 200}<r<r_{\rm 200}$ or $20~{\rm kpc}<r<134~{\rm kpc}$ (center), and the ISM of the main halo $r<0.15r_{\rm 200}$ (bottom). The color bar in the left panels quantify the mass in hydrogen, while the one for the right panels the median metallicity in 2D density-temperature bins.}
    \label{fig:g316e12phasespace}
\end{figure}

\begin{figure}
    \centering
    \includegraphics[width=0.39\textwidth]{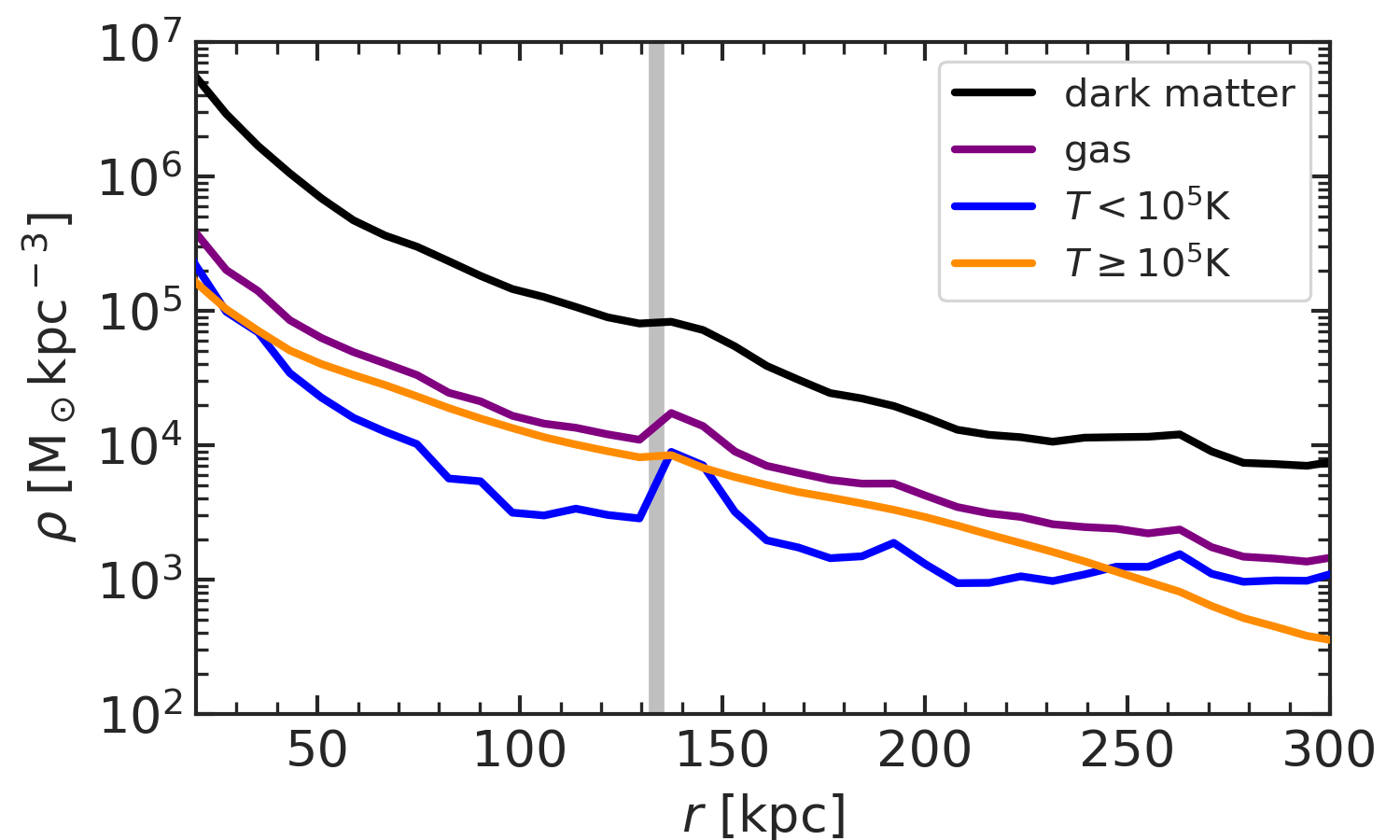}\\
    \includegraphics[width=0.49\textwidth]{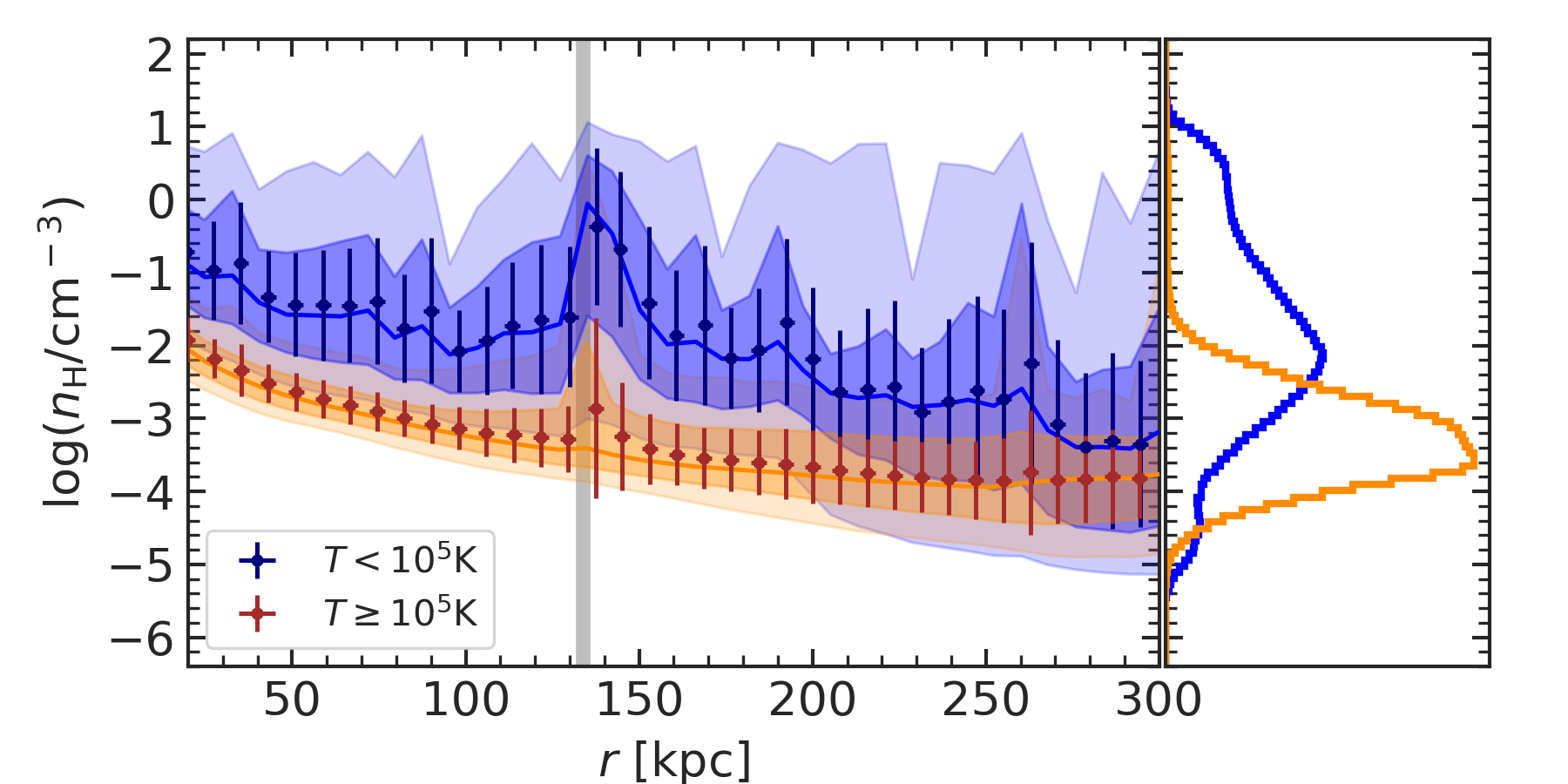}\\
    \includegraphics[width=0.49\textwidth]{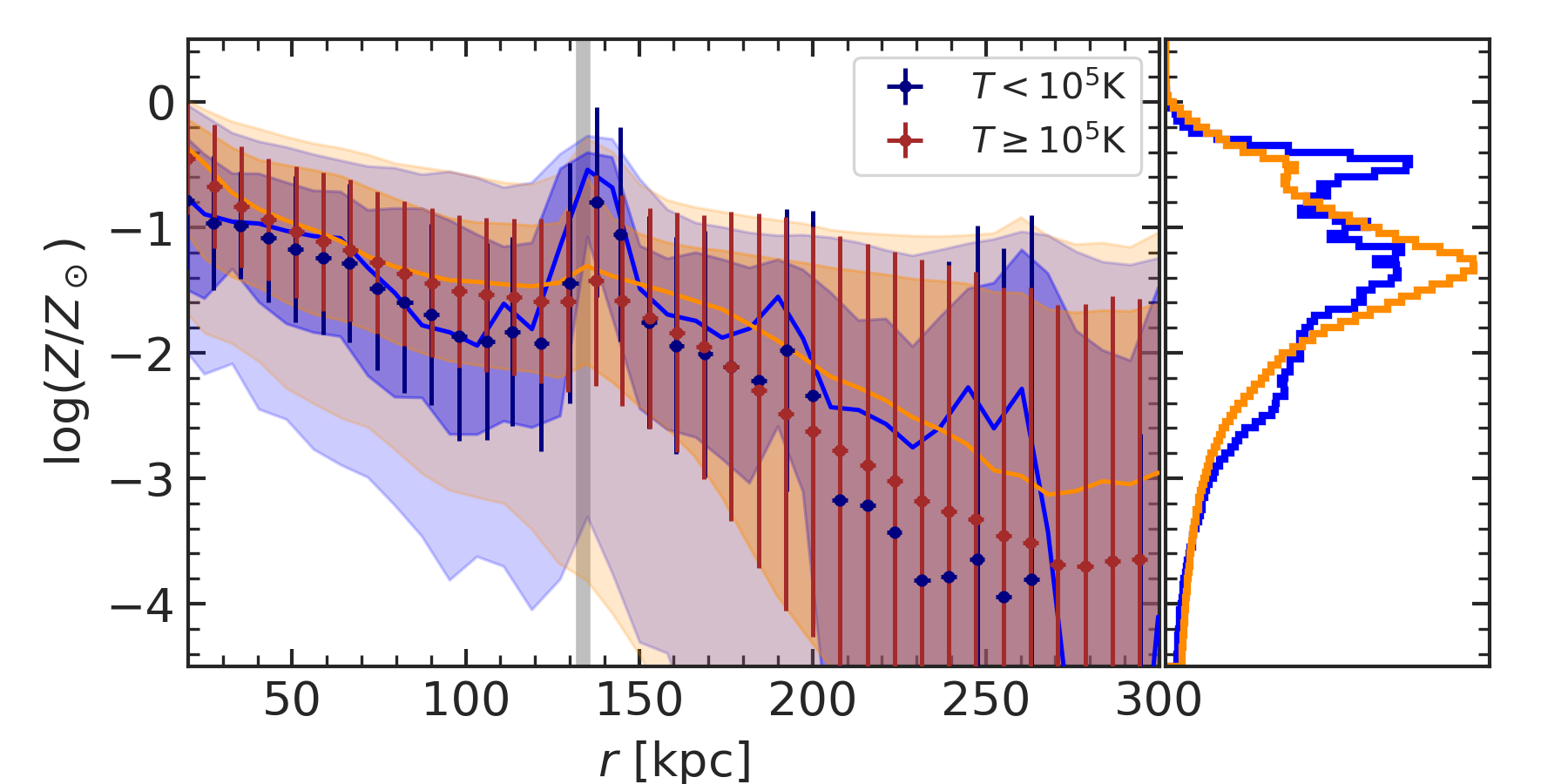}\\
    \caption{\textbf{Top}: Density profiles of dark matter (black), cool ($T<$10$^{\rm 5}$K, blue) and hot ($T\geq$10$^{\rm 5}$K, orange) gas, and all gas (purple). \textbf{Center}: number density of hydrogen as a function of radius for the cool and hot gas phases. \textbf{Bottom}: metallicity as a function of radius for the cool and hot gas phases. In the middle and bottom panels, the points and their errors give the mean and standard deviation, the solid lines give the medians, and the shaded regions enclose from the 16$^{\rm th}$ to the 84$^{\rm th}$ percentiles (darker shade), and from the 2$^{\rm nd}$ to the 98$^{\rm th}$ percentiles (lighter shade). We also show the collapsed density distribution functions of hydrogen number density and metallicity in the right panels. The vertical grey line marks the virial radius $r_{\rm 200}$. In all panels, we exclude the ISM region $r<$~20~kpc, where $r$ is the 3D radius.}
    \label{fig:radial_profiles}
\end{figure}

\section{The simulated halo}
\label{sec:sim}

We use a zoom-in cosmological simulation of a quasar host halo at $z\sim3$ to study in post-processing the radiation effects on the CGM of a central supermassive black hole. This simulation has been run without AGN feedback, using the exact same galaxy model as NIHAO \citep[all free parameters are set to the values in ][]{Wang:2015}, including the same cold dark matter cosmology \citep{Planck:2014}. The code used to simulate this galaxy, named g3.16e12, is the updated version \citep[\texttt{Gasoline2},][]{Wadsley:2017} of the N-body smoothed particle hydrodynamics code \texttt{Gasoline} \citep{Wadsley:2004}. Stellar particles, representing single stellar populations, are formed stochastically from dense ($n_{\rm H}>10.3$~cm$^{\rm -3}$) and cold ($T<$15000~K) gas such that a Kennicutt-Schmidt type of relation is recovered. Gas is metal enriched by supernovae type I and type II \citep{Raiteri:1996}, computed assuming the initial stellar mass function of \citet{Chabrier:2003}, and the metal diffusion and mixing follows the prescriptions of \citet{Shen:2010} and \citet{Wadsley:2008}. Non-equilibrium cooling of hydrogen and helium is computed on-the-fly \citep{Shen:2010}, while for the metal line cooling the gas is assumed in photoionization equilibrium with the UVB of \citet{Haardt:2012}. Stellar particles further impact their surrounding gas through two types of feedback: i) heating prior to SNe II phase also known as 'early stellar feedback' \citep{Stinson:2013}, and ii) blast-waves from the SNe II events \citep{Stinson:2006}. To prevent the rapid dissipation of the SNe II thermal energy transferred to the gas \citep{Stinson:2006}, the cooling of gas particles affected by this feedback is delayed by 40~Myr. The zoom-in region of this simulation has been run with mass resolutions of 6.5$\rm\times$10$^{\rm 5}$M$_{\rm\odot}$ and 3.6$\rm\times$10$^{\rm 4}$M$_{\rm\odot}$, and with gravitational softening lengths of 168~pc and 72~pc for the dark matter and gas particles, respectively. This resolution allows us to follow the gas distribution up to densities of $\sim$10$^{\rm 3}$cm$^{\rm -3}$.     

Figure~\ref{fig:g316e12maps} shows the dark matter (top left), stellar (top right), gas mass (bottom left), and gas-phase metal mass (bottom right) distributions in a $4\times r_{\rm 200}$-box centered on the quasar host halo at $z=3$. The dashed white circle marks $r_{\rm 200}=134$kpc, defined using the \texttt{Amiga Halo Finder} \citep{Knollmann:2009}. The gas column density panel (bottom left) shows nicely the cosmic-web filaments feeding this halo, and the in-falling galaxy sitting roughly at $r_{\rm 200}$ from the center of the main halo. The stellar light surface brightness (top right) more clearly displays the on-going formation process, while many of the dark matter substructures visible just outside of the virial radius are clearly devoid of stars (top left). At this redshift ($z=3$), the dark matter halo mass is $M_{\rm h}$=2.84$\times$10$^{\rm 12}$M$_{\rm\odot}$, while the virial star and gas masses are $M_{\rm \star}$=1.09$\times$10$^{\rm 11}$M$_{\rm\odot}$ and $M_{\rm gas}$=3.25$\times$10$^{\rm 11}$M$_{\rm\odot}$, respectively. This implies an integrated baryon conversion efficiency $\varepsilon\equiv M_{\rm\star}/(M_{\rm\star}+M_{\rm gas})$=0.25, which is within the 1$\sigma$ scatter of the  abundance matching predictions for this galaxy's dark matter halo mass and redshift: $\varepsilon(M_{\rm h},z=3)=0.19^{\rm +0.08}_{\rm -0.06}$ \citep{Moster:2018}.   

Figure~\ref{fig:g316e12phasespace} shows the mass weighted phase space for the selected box (top), as well as for what is typically referred to as the CGM (center), and ISM (bottom) regions\footnote{The choice of $r=0.15r_{\rm 200}=20$kpc to separate the ISM from the CGM is common in the literature, and for our particular simulation it encloses most (96\%) of the stellar mass within $r_{\rm 200}$.}. The left and right panels show the phase space weighted by the hydrogen and median metallicity, respectively. Both the CGM and the ISM regions show two distinct blobs. In the bottom left panel, the lower blob is the cool/cold ISM, while the upper one is made by the ISM gas heated by SNe. The ISM gas heated by SNe also carries a significant amount of the metals, as shown in the bottom right panel. The two blobs in the central left panel represent the hot corona (upper blob) and the "cooling flows" (lower blob) in the CGM. The other important feature in phase-space is visible as a straight line ($T\propto n_H^{\gamma}$ with $\gamma>0$) in the region of low temperature and low density of the full box upper panels, and represents gas of the IGM.     

Looking at the phase diagram of CGM gas, we see that $T\sim$10$^{\rm 5}$K can separate relatively cleanly the cool from the hot phase. Therefore, in the two lower panels of Figure~\ref{fig:radial_profiles} we show how gas density and metallicity varies with the radius for the two phases. The hydrogen number density for the hot gas drops smoothly with $r$ (orange curve and points in the central panel), with an almost log normal distribution in $n_{\rm H}$ when considering all the region with 20~kpc~$\approx$~0.15$r_{\rm vir}<r<$~300~kpc. Contrary, the radial variation of $n_{\rm H}$ for the cool gas also drops with $r$, but it is significantly more wiggly as it also incorporates the ISM of nearby smaller galaxies. At fixed $r$ the distribution of $n_{\rm H}$ for the cool phase is significantly more wide than for the hot one, and has the median at larger $n_{\rm H}$ values. On the other hand, the radial distributions of metallicities (bottom panel) for the two phases are very similar. In the upper panel of Figure~\ref{fig:radial_profiles} we also show how the mass of dark matter, hot and cool gas, and total gas is distributed with $r$. 

To summarize, g3.16e12 at $z=3$ is a high resolution galaxy (resolved densities in the ISM of $\sim$10$^{\rm 3}$cm$^{\rm -3}$), with a halo mass within the observed range for quasar hosts \citep[e.g.][]{Lau:2016,Petter:2023}, and with a stellar mass compatible with the constraints from abundance matching \citep[e.g.][]{Moster:2018}. The simulation has been run without on-the-fly AGN feedback. In the next sections, we post-process the simulation to study the effects of AGN radiation (from a central SMBH with mass compatible with the halo mass) on its CGM's intrinsic properties, as well as on observational constraints like column densities of various ions and surface brightness in various emission lines.

\section{Theoretical model spectra}
\label{sec:model_spectra}

In a galaxy there are a few different types of local ionizing sources (e.g. young massive stars, AGN, X-ray binaries, hot gas emitting Bremsstrahlung), and few cosmological galaxy simulation studies have tried to incorporate at least the heating/ionizing effect of some of them \citep[e.g.][]{Kannan:2014,Kannan:2016,Obreja:2019,Hopkins:2020,Costa:2022}. While most works have focused on the ionizing continuum from stellar sources, few studies have looked at the effects of ionizing radiation from AGN \citep[e.g.][]{Kim:2011,Vogelsberger:2013}, but none studied the EM feedback alone, making it impossible to disentangle its effects from those induced by the widely used thermal and kinetic AGN feedback models.      

AGN have been classified in a myriad of sub-classes based on features observed in their spectra and/or their wavelength of first detection. In time, only three properties emerged as fundamental: the radiative efficiency, the presence/absence of a jet, and the galaxy host environment \citep[][]{Antonucci:1985,Antonucci:1993,Bianchi:2022}. The radiative efficiency measures how efficient the accretion is, and it is quantified by the ratio between the observed and Eddington luminosity $\lambda=L/L_{\rm Edd}$, where $L_{\rm Edd}=1.3\times10^{46}(M/10^8M_{\rm\odot})$erg~s$^{\rm -1}$ is the maximum luminosity of a mass $M$ accreting isotropically. The third property -- galaxy host environment -- was the key one for what it is now known as the Unification Model \citep{Antonucci:1983}. The idea behind this class of models is that all types of AGN have the same nuclear engine, and that an obscuring axisymmetric structure (a dusty torus) around the SMBH defines how the AGN spectrum appears to a distant observer. The presence of such an optically thick torus, on parsec scales and aligned with the symmetry axis, obscures the inner broad line region (typical in the spectra of type I AGN) when the observer line-of-sight (LOS) makes a sufficiently large angle with the symmetry axis, while leaving unaffected the outer narrow line region (typical in the spectra of type II AGN). As the multi-wavelength coverage of AGN spectra has grown, it became clear that this class of models is too simplistic \citep[e.g.][]{Combes:2021}, and that different absorbers on different scales are needed to reproduce well high resolution high sensitivity AGN spectra \citep[e.g.][]{Gaspari:2017}. Notwithstanding these complications on small scales, the radiation from an AGN is currently though to illuminate anisotropically the surrounding CGM with its so-called ionization cones.

\subsection{AGN spectra}
\label{subsec:AGN_spectra}

For the spectral energy distribution (SED) of accreting SMBHs, we use the \emph{\sc{optxagnf}}\footnote{The code that can generate this type of spectrum can be found as a local model in the publicly available \emph{\sc{XSPEC}} spectral fitting package \citep{Arnaud:1996}.} model \citep{Done:2012}.
This theoretical SED model is made of three components:
\begin{itemize}
 \item a pseudo-thermal accretion disk \citep{Novikov:1973},
 \item a non-thermal power-law component at high energies produced by Compton up-scattering from an 
 optically thin and hot medium,
 \item a soft X-ray excess component produced by Compton up-scattering from an 
 optically thick, lower temperature medium. 
\end{itemize}
This model is energetically self consistent, in the sense that the primary and only source of energy is the constant mass accretion rate $\dot{M}$ accretion from a Novikov-Thorne disk onto a black hole (BH). The schematic view of the AGN emission
using such a model is shown in Figure 5 of \citet{Done:2012}.
This model has nine parameters, some of which are partially degenerate:
\begin{enumerate}
 \item black hole mass $M_{\rm\bullet}$,
 \item AGN luminosity in units of Eddington luminosity $L/L_{\rm Edd}$,
 \item black hole spin parameter $a$,
 \item the coronal radius $R_{\rm cor}$ in units of the gravitational radius
 $R_{\rm g}=GM_{\rm BH}/c^{\rm 2}$, which gives the innermost edge of the disk that is visible,
 \item the outer radius of the disk $R_{\rm out}$ in units of $R_{\rm g}$, 
 \item the temperature $T$ of the optically thick gas producing the soft X-ray excess,
 \item the optical depth $\tau$ of the optically thick gas,
 \item the fraction of the corona energy emitted in the power-law component, 
 $f_{\rm PL}$ (the fraction of the corona energy emitted as the soft X-ray excess is $1-f_{\rm PL}$)
 \item the high energy power-law index $\Gamma$.
\end{enumerate}

\begin{figure}
    \centering
    \includegraphics[width=0.49\textwidth]{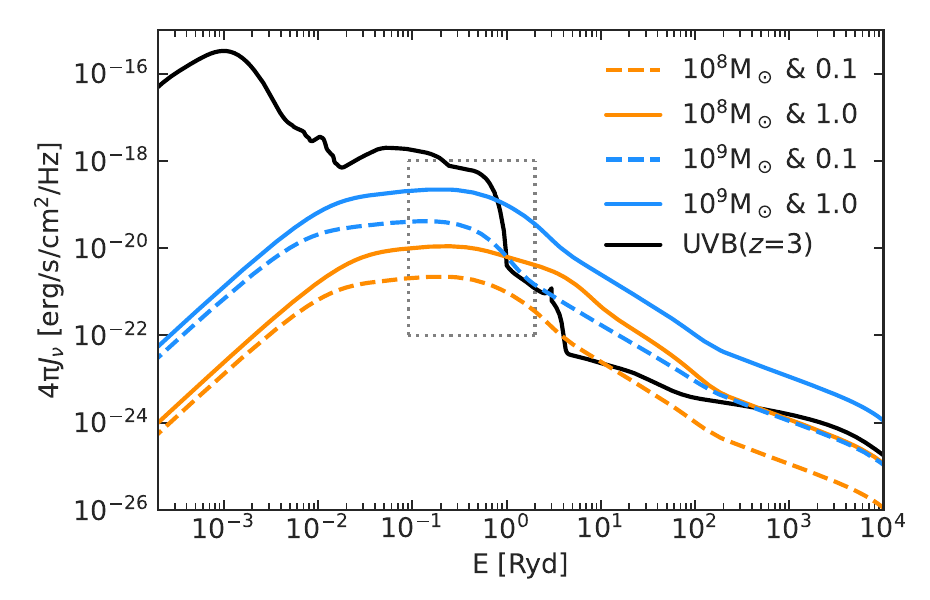}
    \includegraphics[width=0.49\textwidth]{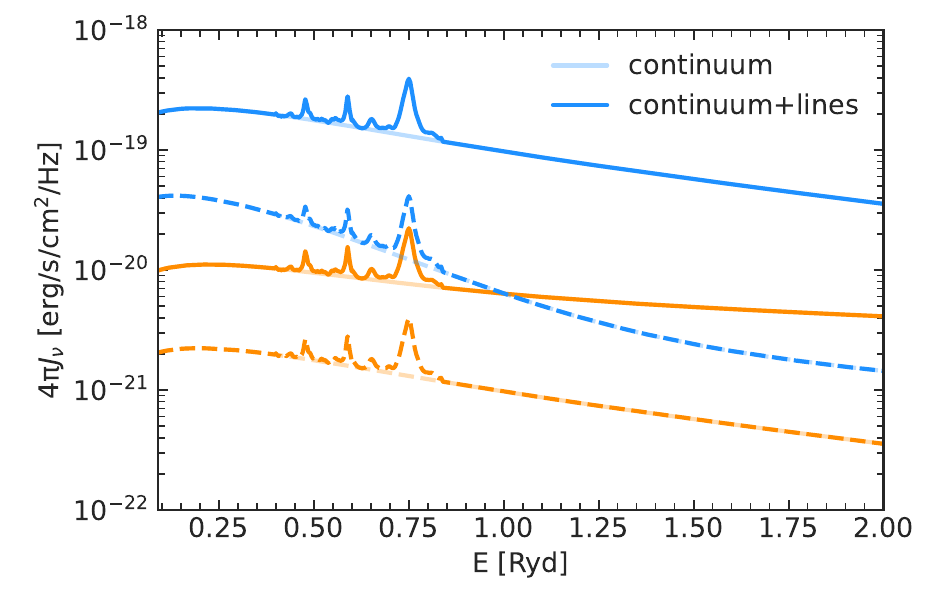}
    \caption{\textbf{Top}: spectra of the photoionization and photoheating sources. The black curve gives the fiducial UVB model of \citet{Khaire:2019} for $z=3$, while the colored ones represent the four \emph{\sc{optxagnf}} AGN spectra \citep{Done:2012} that we use in this work as seen from a distance of 1~Mpc from the quasar, with orange/blue for a SMBH of mass 10$^{\rm 8}$/10$^{\rm 9}$M$_{\rm\odot}$, and solid/dashed for an Eddington accretion ratio of 1.0/0.1. \textbf{Bottom}: Zoom-in on the dotted gray square marked in the top panel, showing also the AGN continua plus the stacked line emission sprectrum of \citet{Lusso:2015}. The most prominent line is the Ly$\alpha$ at 1215.67\AA\ ($\approx$0.75~Ryd), and the second most prominent the \ion{C}{iv} doublet at 1548.19\AA, 1550.78\AA\ ($\approx$0.59~Ryd). Both Ly$\alpha$ and \ion{C}{iv} are resonant transitions.}
    \label{fig:spectra}
\end{figure}

\citet{Jin:2012} used the {\sc optxagnf} model to fit a sample of 51 low redshift Seyfert 1 AGNs. In this sample, 12 objects are classified as Narrow Lines Seyfert 1 (NLS1), while the rest are Broad Lines Seyfert 1 (BLS1). The BLS1 objects are the closest we can get to the intrinsic SED of AGNs, which is precisely what we need in the simulations. 
Hence, we choose the SED parameters based on the corresponding distributions for the BLS1 sample. To construct our template SED models we fixed the BH spin to $a=0$ and the outer disk radius to $R_{\rm out}=10^{\rm 4}R_{\rm g}$, following \citet{Jin:2012}. For the remaining parameters that our target zoom-in cosmological simulations can not resolve, we used the median values over the BLS1 sample. A few of the AGN in this sample have coronal radii at the upper boundary for $R_{\rm out}$, $R_{\rm out}=100R_{\rm g}$. For this reason we excluded these AGNs when computing the median $R_{\rm cor}$. In summary, based on the BLS1 sample of \citet{Jin:2012}, our assumed AGN SEDs have seven of nine parameters fixed to the following values: 

\begin{equation}
\begin{aligned}
  a           &= 0\\
  R_{\rm out} &= 10^{\rm 4}R_{\rm g}\\
  R_{\rm cor} &= 26R_{\rm g}\\
  \Gamma      &= 1.81\\
  f_{\rm PL}  &= 0.25\\
  T_{\rm e}   &= 0.29\, {\rm keV}\\
  \tau        &= 14.8\\
\end{aligned} 
\end{equation}

With all these parameters fixed, our template AGN spectra depend only on the black hole mass $M_{\rm\bullet}$ and on the Eddington accretion ratio $\lambda=L/L_{\rm Edd}$. 

High redshift quasars are hosted by dark matter halos with masses $10^{\rm 12}\leq M_{\rm 200}\leq10^{\rm 13}$M$_{\rm\odot}$ \citep[e.g.][]{Petter:2023}, and have SMBH to $M_{\rm 200}$ ratios between 10$^{\rm -5}$ and few times 10$^{\rm -3}$ \citep[e.g.][]{Shimasaku:2019}. For a dark matter halo mass (or virial velocity) compatible to the one of our simulation at $z=3$, the SMBH masses in observations cover the range 10$^{\rm 7}\leq M_{\rm\bullet}\leq$10$^{\rm 10}$M$_{\rm\odot}$, but are mostly clustered between 10$^{\rm 8}$M$_{\rm\odot}$ and 10$^{\rm 9}$M$_{\rm\odot}$ \citep[e.g.][]{Shimasaku:2019}. Therefore, for our analysis we focus on two different masses 10$^{\rm 8}$M$_{\rm\odot}$ and 10$^{\rm 9}$M$_{\rm\odot}$, and two extreme cases of accretion rates: $\lambda=0.1$ and $\lambda=1.0$. 

The upper panel of Figure~\ref{fig:spectra} gives the spectra for all cases of radiation sources we consider in this study. Notably, the AGN spectra (blue and orange curves) provide significantly more ionizing photons ($E>1$Ryd) than the UVB even at large distances from the quasar (the AGN spectra in the figure correspond to a distance of 1~Mpc). An AGN also emits large number of photons in UV emission lines.
\citet{Fossati:2021} has shown that the UV line emission spectra of AGN are crucial to explaining the observed line ratios from the CGM of high redshift quasars with reasonable gas metallicity values. For this reason, we added on top of the AGN continuum spectra shown in the upper panel, the  stacked emission line spectrum of 53 luminous quasars at $z\sim2.4$ from \citet{Lusso:2015}. The lower panel of Figure~\ref{fig:spectra} shows the resulting AGN spectra in the UV region marked by the dotted gray rectangle in the upper panel. To obtain only the observed stacked emission lines, we subtracted from the published spectrum the continuum fitted by \citet{Lusso:2015}: $f_{\nu}\propto\nu^{-\Gamma}$ with   $\Gamma=1.70\pm0.61$ at $\lambda\leq912$\AA. This spectral slope is, within its uncertainties, compatible with the one we assumed ($\Gamma=1.81$).     

\subsection{UVB spectrum}
\label{subsec:UVB_spectra}

Our cosmological simulation, like most such simulations in the literature, has been run under the assumption that gas is photoionized by an isotropic and time dependent UVB. Therefore, our default {\sc cloudy} models use the recent UVB spectra of \citet{Khaire:2019}. To create synthesis models of the extragalactic background light, extending from far infra-red to $\gamma$-rays, these authors used up-to-date values for the cosmic SFR density, dust attenuation, quasar emissivity, and neutral hydrogen distribution in the IGM. \citet{Khaire:2019} do a particularly detailed treatment of the extreme ultraviolet background, which determines the ionization and temperature of the IGM across time. One practical advantage of using these particular UVB models over the recent alternatives \citep[e.g.][]{Puchwein:2019,FaucherGiguere:2020}, is that they are already incorporated in the latest version of the spectral synthesis code {\sc cloudy} \citep[version C17.03,][]{Ferland:2017}, which we use to create the ancillary data (e.g. cooling and heating tables) needed for cosmological galaxy formation simulations. It is also important to mention that UVB models are generally converged at redshifts lower than the end of reionization \citep[e.g.][]{Khaire:2019}, and for this reason the mismatch between predictions using our choice of UVB in this work \citep{Khaire:2019} and the UVB embedded in the simulations \citep{Haardt:2012} should be minimal. 

In the following sections we describe both the intrinsic and the observable effects that these types of radiation sources have on our simulated quasar host halo, in post-processing.

\section{Post-processing with Cloudy}
\label{sec:cloudy}

In order to create mock observables from this simulation we run large grids of photoionization models with the version 
C17.03 of {\sc cloudy} \citep{Ferland:2017}, covering the multi-dimensional phase space of the gas in a 4$\times r_{\rm 200}$ box centered on the galaxy g3.16e12. These models are plane-parallel models for fixed: total hydrogen density $n_{\rm H}$, gas temperature $T$, metallicity $Z$, and incident radiation field $J_{\rm\nu}$. The control models are run with the sum of the cosmic microwave background (CMB) and the UVB of \citet{Khaire:2019} at $z=3$. The photoionization models with the AGN spectra of Figure~\ref{fig:spectra} as incident spectra also include the CMB (but not the UVB), and span a range in normalization corresponding to a minimum distance from the quasar of 1~kpc and a maximum distance of 2$\sqrt{\rm 3}r_{\rm 200}$ (the distance from the center of the simulated galaxy to the farthest corner of the selected box region). We do not include molecules, dust and cosmic rays, because the cosmological simulation does not follow any of them. To take into account the self-shielding of the gas from the incident radiation field, we run the photoionization models with a column density stopping criteria, depending on the local Jeans scale \citep[e.g.][]{Schaye:2001}, following the set-up used by \citet{Ploeckinger:2020} (see section 2.1 of their paper). Other approximations for the local column density, e.g. Sobolev-like based on the local density \citep{Krumholz:2011}, would require adding extra-dimensions to the already many gas properties we need to account for. Instead, using a stopping column  density based on the Jeans lenght means that the column depends only on temperature and hydrogen number density.         

The outputs of these photoionization models include: heating and cooling rates, electron densities, ionization fractions for a few species for which column densities can be estimated from observations, and line emissivities for a list of optical and UV lines that have been or could be observed in QSO host halos at $z\sim3$. All these properties are saved for the last {\sc cloudy} zone, and all are functions of ($n_{\rm H}$, $T$, $Z$, $J_{\rm\nu}$). An example of a {\sc cloudy} script used is explained in detail in  Appendix~\ref{appendix:scripts}.  Figure~\ref{fig:spectra}. 

\begin{figure}
    \centering
    \includegraphics[width=0.49\textwidth]{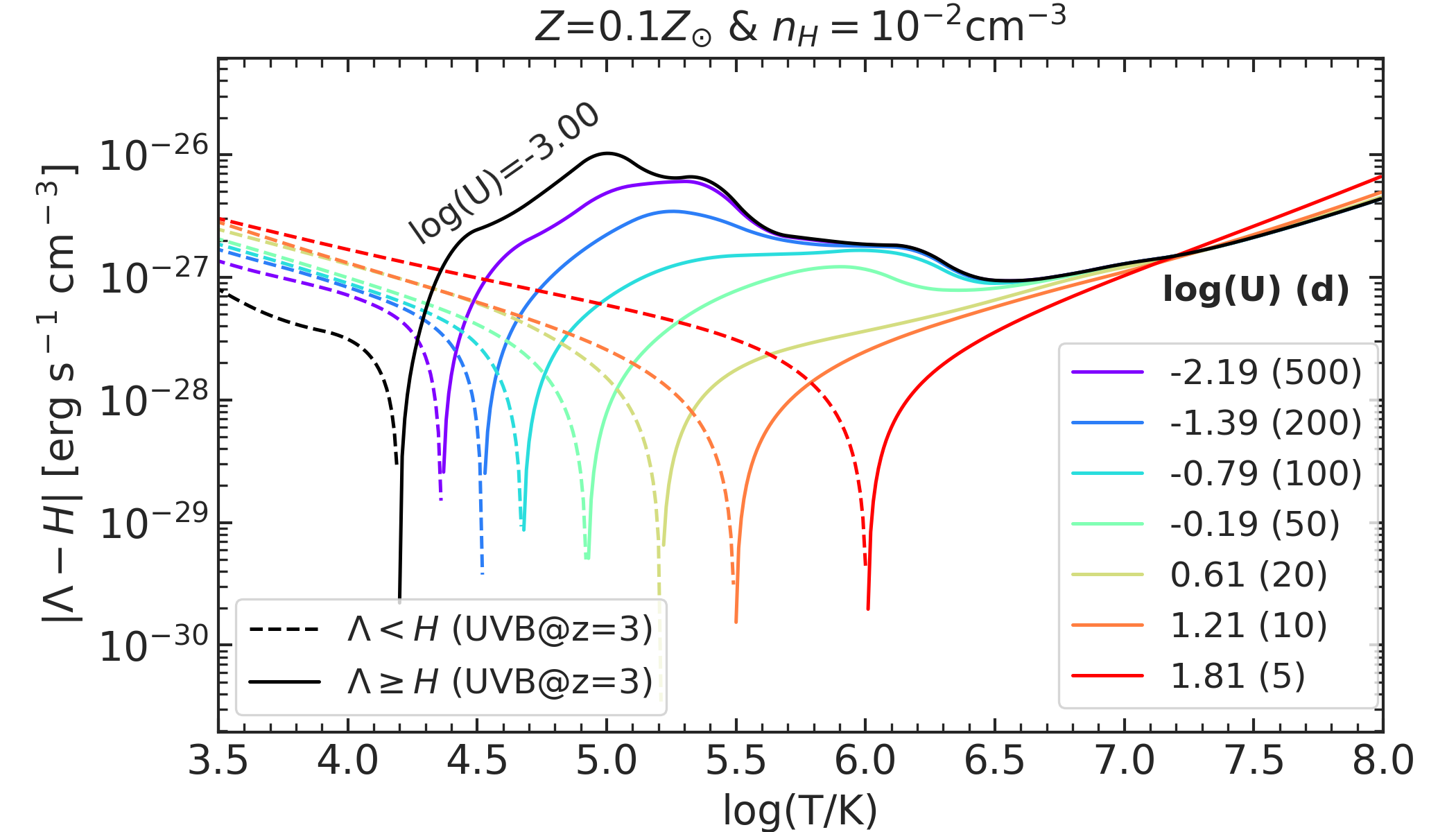}
    \caption{Net cooling rate as a function of temperature for a gas slab with density $n_{\rm H}$=10$^{\rm -2}$cm$^{\rm -3}$ and metallicity 0.1~$Z_{\rm\odot}$, exposed to the $z=3$ UVB (black), or to the spectrum of a 10$^{\rm 9}$~M$_{\rm\odot}$ SMBH accreting at 10 per cent the Eddington rate. Colored curves represent various distances between the radiation source and the gas slab, given in kpc within brackets in the legend. The legend on the right gives also the logarithm of the ionization parameter $U$. The dashed/solid parts of the curves represent the conditions where gas is heating/cooling.}
    \label{fig:netcool_vs_T}
\end{figure}

The main reason for including AGN feedback in simulations of massive galaxies is to mitigate the so-called 'overcooling problem' \citep{Larson:1974,White:1978,Dekel:1986,Navarro:1991}. Thus, we first look at the impact AGN radiation has on the net cooling rate of the gas. From Figure~\ref{fig:g316e12phasespace}, the hot gas reservoir of the CGM is metal enriched and has densities of the order of 10$^{\rm -3}$cm$^{\rm -3}$. In Figure~\ref{fig:netcool_vs_T} we show how the absolute net cooling rate $|\Lambda-H|$, where $\Lambda$ and $H$ are the volumetric cooling and heating rates, for a slab of gas of metallicity 0.1$Z_{\rm\odot}$ and density 10$^{\rm -2}$cm$^{\rm -3}$ varies with the temperature. The black curve in the figure corresponds to the case where the gas is only irradiated by the UVB, while the colored curves show  $|\Lambda-H|$ as a function of $T$ for gas irradiated by a 10$^{\rm 9}$~M$_{\rm\odot}$ SMBH accreting at ten percent the Eddington rate ($\lambda=0.1$) for various distances between the slab of gas and the radiation source (indicated in the legend on the right side). The curves are solid in the region where the gas is cooling ($\Lambda>H$) and dashed where it is heating ($\Lambda<H$). The most notable thing in this figure is how the equilibrium temperature ($T$ for which $\Lambda=H$) moves to higher and higher values as the strength of the incident radiation field increases. While for the UVB case $T_{\rm eq}\approx$10$^{\rm 4.2}$K, gas irradiated by a central quasar and sitting at only 5~kpc (red curve) has $T_{\rm eq}\approx$10$^{\rm 6}$K. 
Based on this figure and the phase space distribution of CGM gas in Figure~\ref{fig:g316e12phasespace}, we can infer that a significant fraction of the CGM gas will not be cooling, but instead will be heating in the presence of a central AGN radiation source.

\section{Geometric effects}
\label{sec:geometry}

As already stated at the beginning of Section~\ref{sec:model_spectra}, the radiation emitted by an AGN does not propagate on halo scales isotropically, but rather within ionization cones, because of obscuring media on small scales. Flavours of the AGN unification model, though very approximate, are quite useful for sub-grid models for galaxy formation simulations, which can not yet resolve the parsec and sub-parsec scales relevant for the SMBH feeding. For example, \citet{Fritz:2006} modeled the dust torus of the SMBH as a flared disk \citep{Efstathiou:1995}, and found that roughly half of their type I AGN and $\sim$70\% of the type II were best fitted with a torus opening angle $\theta\approx$140$^{\rm o}$, while the rest of the sample required $\theta\approx$100$^{\rm o}$. This means that ionization cone opening angles of AGN $\alpha=180^{\rm o}-\theta$ should be mostly restricted to $40^{\rm o}\leq\alpha\leq80^{\rm o}$.

We concentrate in this work on how the ionization cone opening angle affects not only the intrinsic properties of the CGM gas, but also its observables. We define ionization cones by their opening angle $\alpha$ \footnote{$\alpha$ is the full angle at the cone vertex.} and explore the range between 180$^\circ$ (isotropic case) and 0$^\circ$ (no AGN radiation). For each $\alpha$ we choose $N$ random directions passing through the center of the halo to place the cones, and for each of these $N$ directions we place the line of sight (LOS) randomly within the ionization cones. The choice of $N$ is different for the various exercises we do in the following sections.   

First, we investigate how much of the CGM would be catastrophically cooling as a function of opening angle. While the value of the cooling time in itself might not be very informative for $t_{\rm cool}>0$, comparing it with the free-fall time $t_{\rm ff}$ can provide a good idea of how much of the gas is in (quasi)hydrostatic equilibrium, and how much is catastrophically cooling. The free-fall time is only a function of the total average system's density and thus varies with the position in the dark matter halo:
\begin{equation}
    t_{\rm ff}(r) \equiv \sqrt{\frac{3\pi}{32 G \rho_{sys}(<r)}},
    \label{eq:tff}
\end{equation}
where $G$ is the gravitational constant, and $\rho_{sys}(<r)$ is the average mass (dark matter + baryons) density within a sphere of radius $r$: $\rho_{sys}(<r)=3M_{\rm dyn}(<r)/4\pi r^3$. 

\begin{figure}
    \centering
    \includegraphics[width=0.49\textwidth]{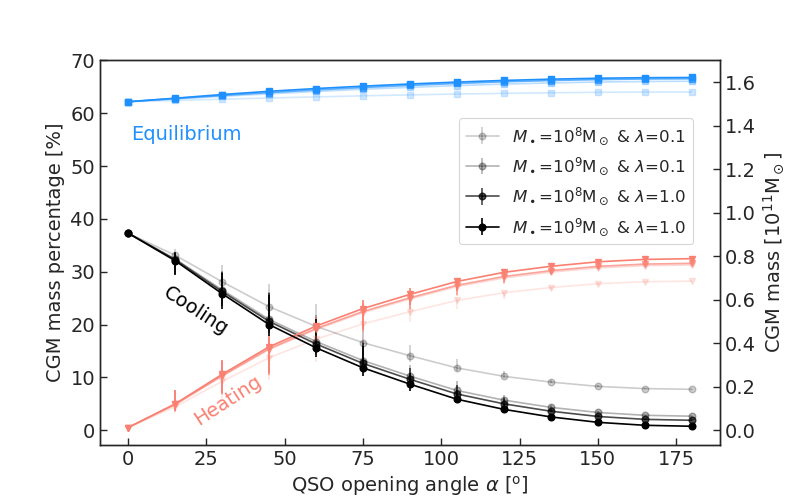}
    \caption{The CGM gas mass budget as a function of the quasar ionization cone's opening angle $\alpha$. The various transparency levels of the curves represent the spectral model for the AGN. The three phases of the CGM represent different thermal states: gas that is heating ($t_{\rm cool}<0$, light-red), gas that is cooling ($0<t_{\rm cool}<t_{\rm ff}$, black), and gas in (quasi)hydrostatic equilibrium ($t_{\rm cool}>t_{\rm ff}$, blue).}
    \label{fig:opening_angle}
\end{figure}

\begin{figure*}
    \centering
    {$\alpha=0^{\rm o}$\hspace{5cm}$\alpha=60^{\rm o}$\hspace{5cm}$\alpha=180^{\rm o}$}\\
    \includegraphics[width=0.33\textwidth]{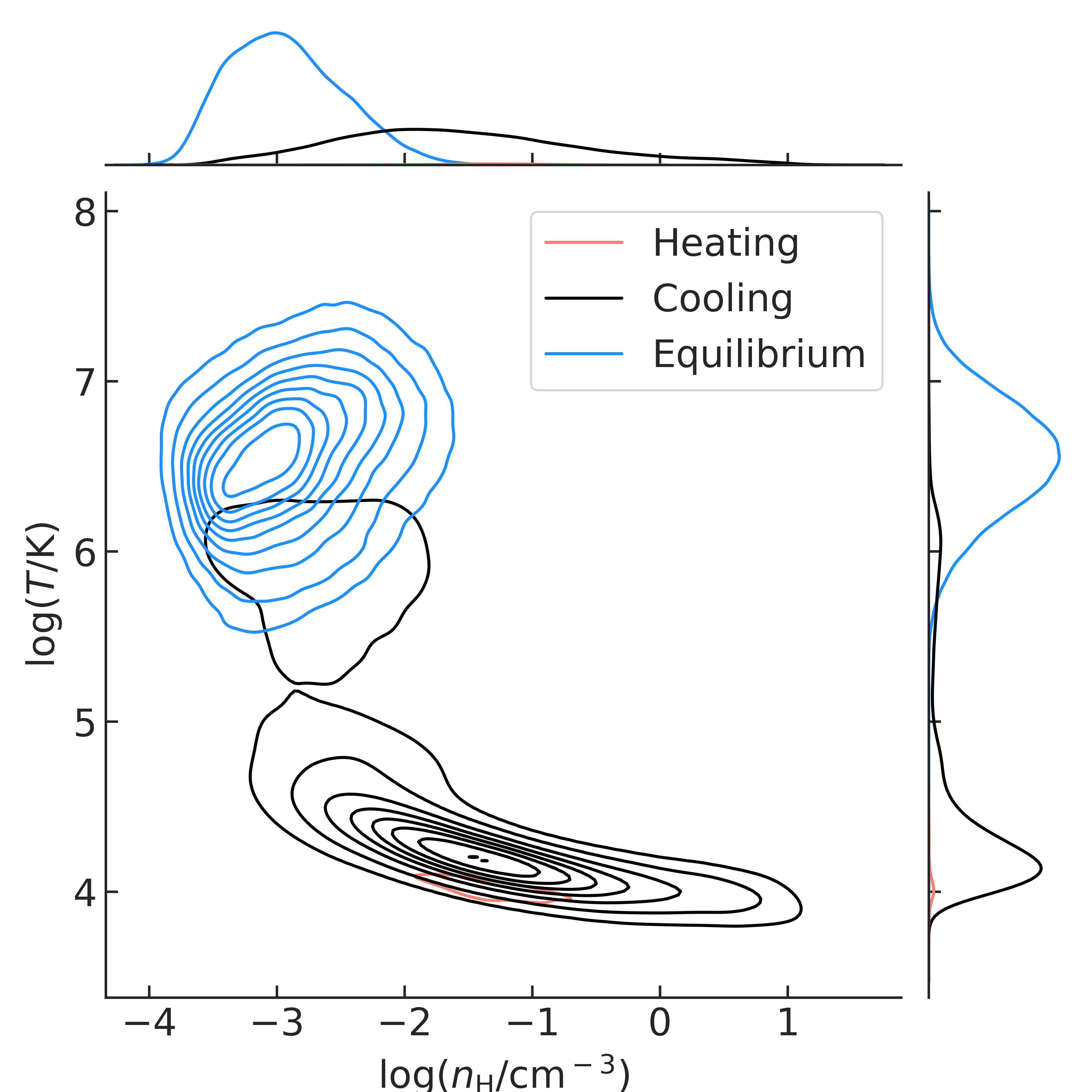}
    \includegraphics[width=0.33\textwidth]{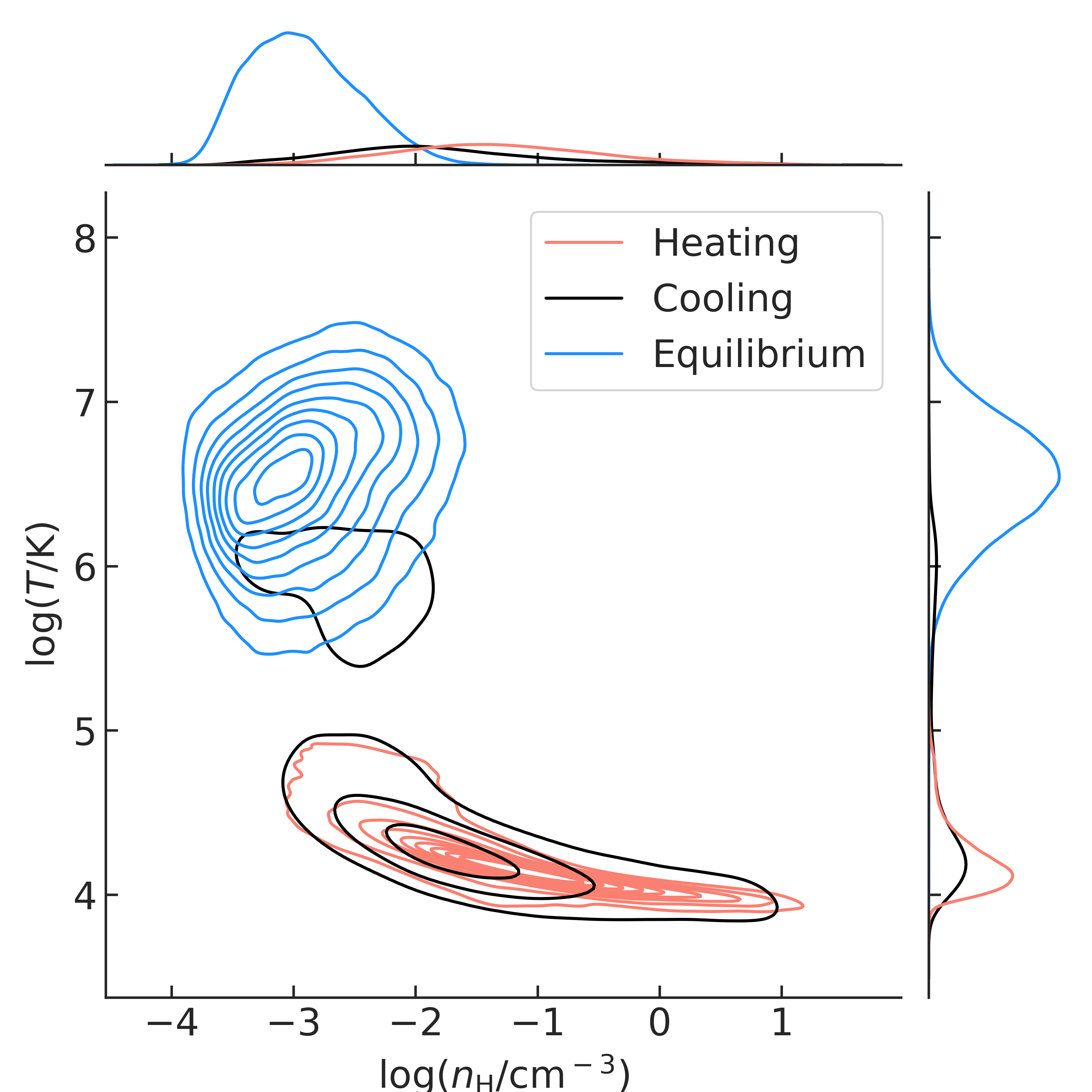}
    \includegraphics[width=0.33\textwidth]{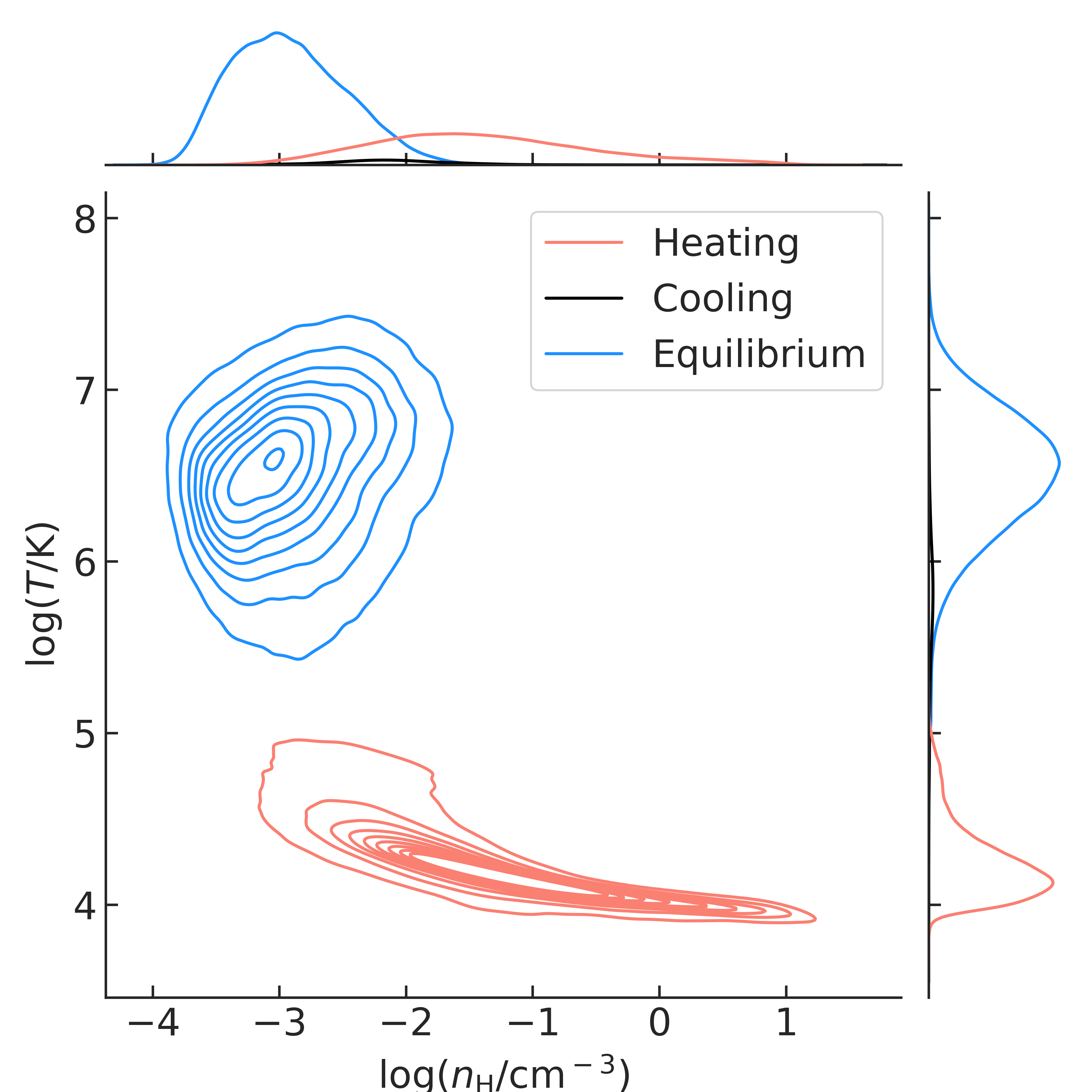}
    \caption{Joint density-temperature phase space distributions for the three thermal states of CGM gas, assuming different quasar ionization cone's opening angles: $\alpha=0^{\rm o}$ representing the UVB only case (left), $\alpha=60^{\rm o}$ (center), and $\alpha=180^{\rm o}$ representing the case where all the CGM gas is impacted by the AGN radiation. The spectral model in all three panels corresponds to an AGN of $M_{\rm\bullet}=10^{\rm 9}$M$_{\rm\odot}$ and a $\lambda=0.1$ Eddington ratio. The contours of the kernel density estimate enclose equally spaced fixed fractions of particles in each of the three thermal states.}
    \label{fig:cgm_phase_space_vs_alpha}
\end{figure*}

Using the definition of Equation~\ref{eq:tff} for the free-fall time, Figure~\ref{fig:opening_angle} quantifies exactly how much of the CGM gas is catastrophically cooling, heating or is in equilibrium, as a function of the quasar spectrum and ionization cone opening angle. The errors on the individual points are the standard deviations of the mass fractions over the $N$=200 random directions chosen to place the ionization cones. In this figure $\alpha=0$ corresponds to the case where all gas sees only the UVB, while increasing values of $\alpha$ correspond to the case where only the gas within the double cone of opening angle $\alpha$ is irradiated by the quasar while the rest of it is irradiated by the UVB. The case $\alpha$=180$^{\rm o}$ means that all gas is directly irradiated by the quasar. Figure~\ref{fig:opening_angle} illustrates nicely the 'overcooling' problem, 37\% of the total CGM gas ( $\sim$2.43$\times$10$^{\rm 11}$M$_{\rm\odot}$) experiencing catastrophic cooling if the only source of radiation is the UVB. As the ionization cone opening angle increases, an increasingly larger fraction of the CGM is heating (salmon curves), while the cooling fraction (black curves) decreases accordingly reaching almost zero ($f_{\rm cooling}\sim$0.7\%) for the extreme $\alpha$=180$^{\rm o}$ and the stronger incident spectra. Only the spectra with $M_{\rm\bullet}$=10$^{\rm 8}$M$_{\rm\odot}$ and $\lambda=0.1$ has a significant $f_{\rm cooling}\sim$8\% for $\alpha$=180$^{\rm o}$. On the other hand, the fraction of CGM gas in (quasi)hydrostatic equilibrium (blue curves) is only slightly increasing with $\alpha$, and varies little with the strength of the quasar spectrum.           

Figure~\ref{fig:cgm_phase_space_vs_alpha} shows the CGM gas in the density--temperature phase--space for three different opening angles. The color coding is the same as in Figure~\ref{fig:opening_angle}, and the SMBH spectra is for $M_{\rm\bullet}$=10$^{\rm 9}$M$_{\rm\odot}$ and $\lambda=0.1$. In the left panel corresponding to the UVB only case, we see that most of the gas with temperatures around 10$^{\rm 4}$K is cooling and, therefore, it is a reservoir for the ISM. The same is true for some of the gas in the hot CGM. For $\alpha$=60$^{\rm o}$ shown in the central panel  \citep[value within the observational limits, e.g. ][]{Fritz:2006}, the fraction of CGM gas which is a reservoir for the ISM is roughly halved, while for the extreme $\alpha$=180$^{\rm o}$ case, virtually all the hot CGM gas is in equilibrium, while the cool CGM is heating under the effect of the central accreting SMBH. Therefore, AGN with large opening angles can completely stall or at least delay the star-formation process as the cool CGM is ultimately its fuel.    

\begin{figure*}
    \centering
    {$M_{\rm\bullet}$=10$^{\rm 8}$M$_{\rm\odot}$ \& $\lambda=0.1$\hspace{3.5cm}$M_{\rm\bullet}$=10$^{\rm 9}$M$_{\rm\odot}$ \& $\lambda=0.1$\hspace{3.5cm}$M_{\rm\bullet}$=10$^{\rm 9}$M$_{\rm\odot}$ \& $\lambda=1.0$}\\
    \includegraphics[width=0.33\textwidth]{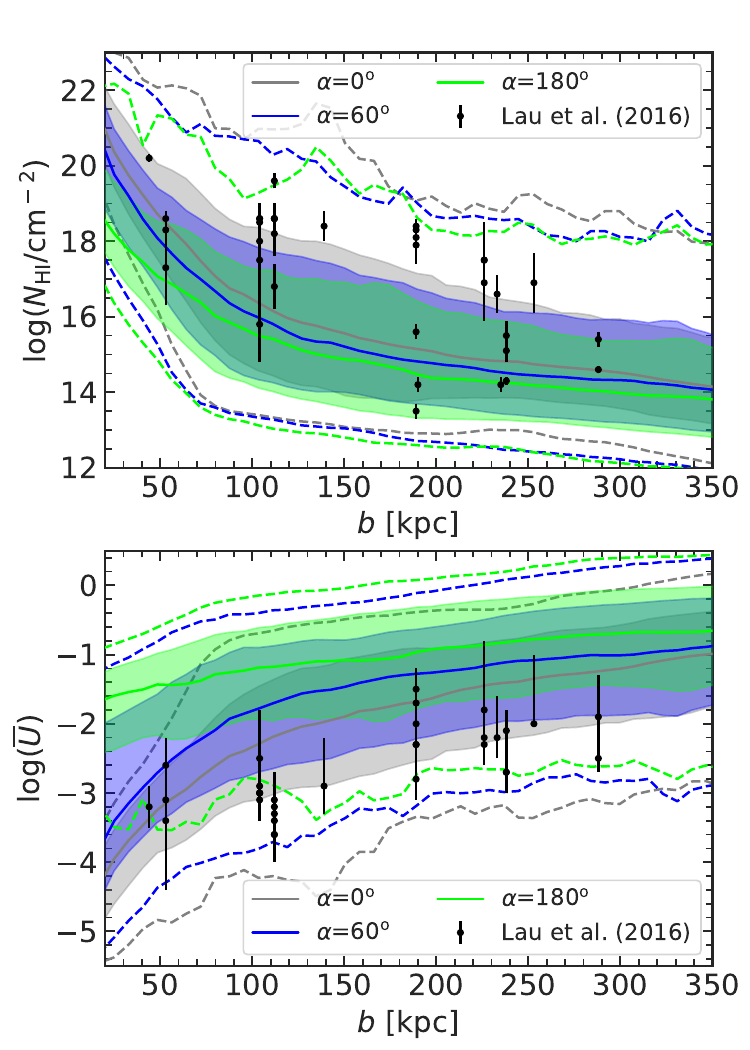}
    \includegraphics[width=0.33\textwidth]{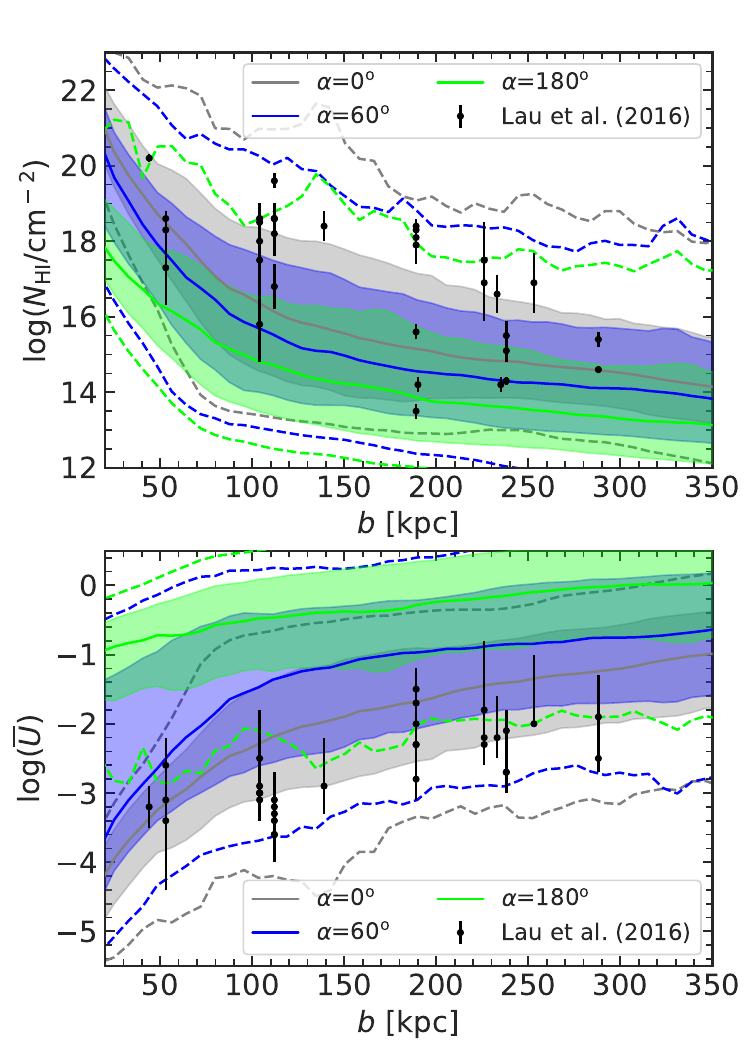}
    \includegraphics[width=0.33\textwidth]{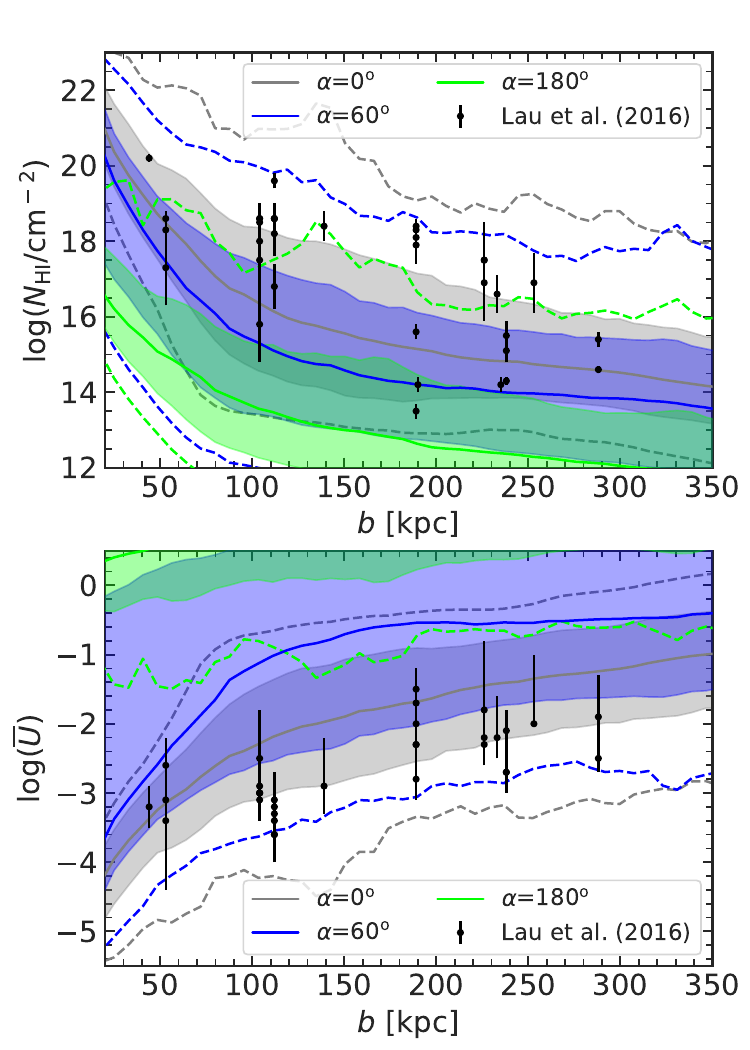}\\
    \includegraphics[width=0.33\textwidth]{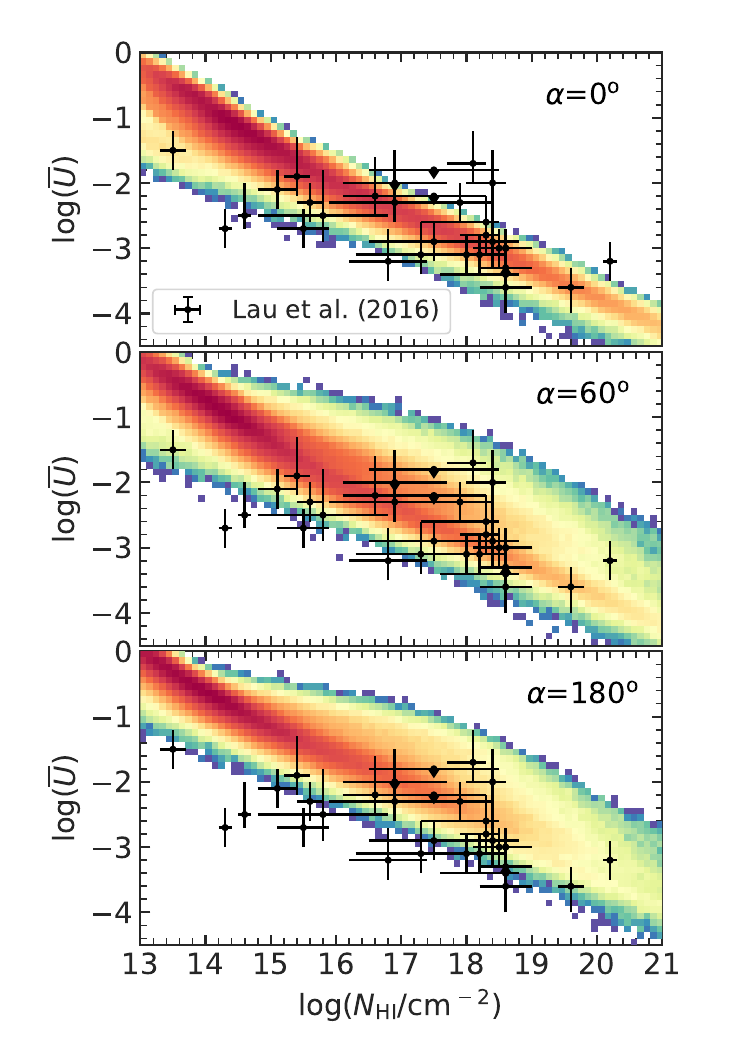}
    \includegraphics[width=0.33\textwidth]{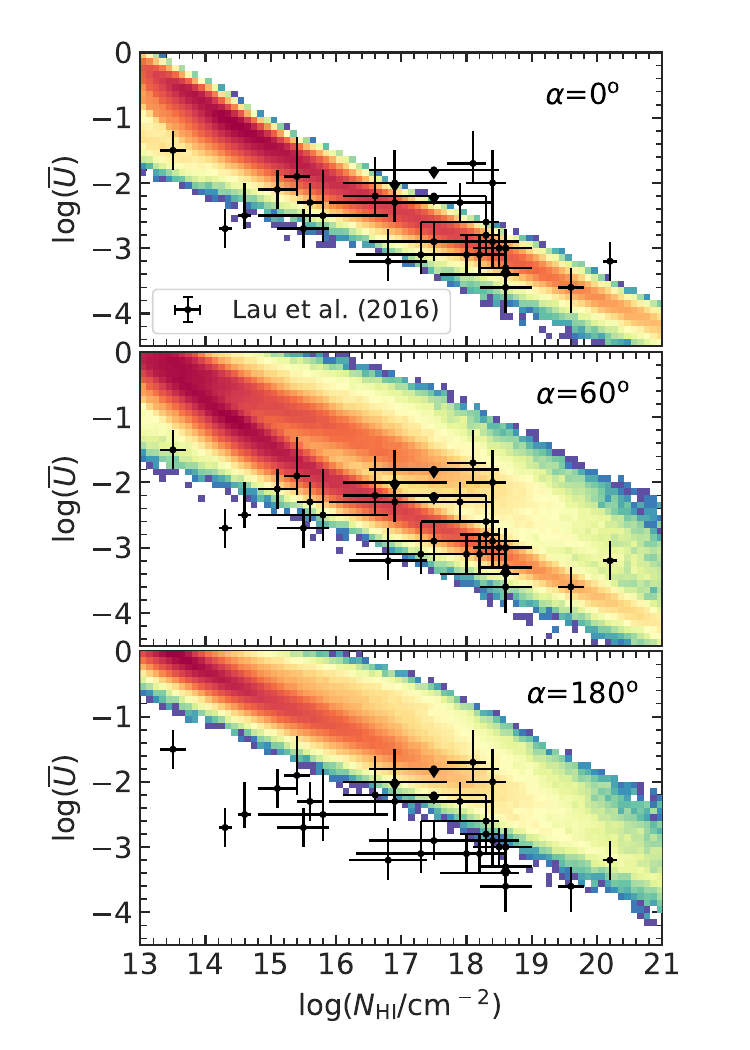}
    \includegraphics[width=0.33\textwidth]{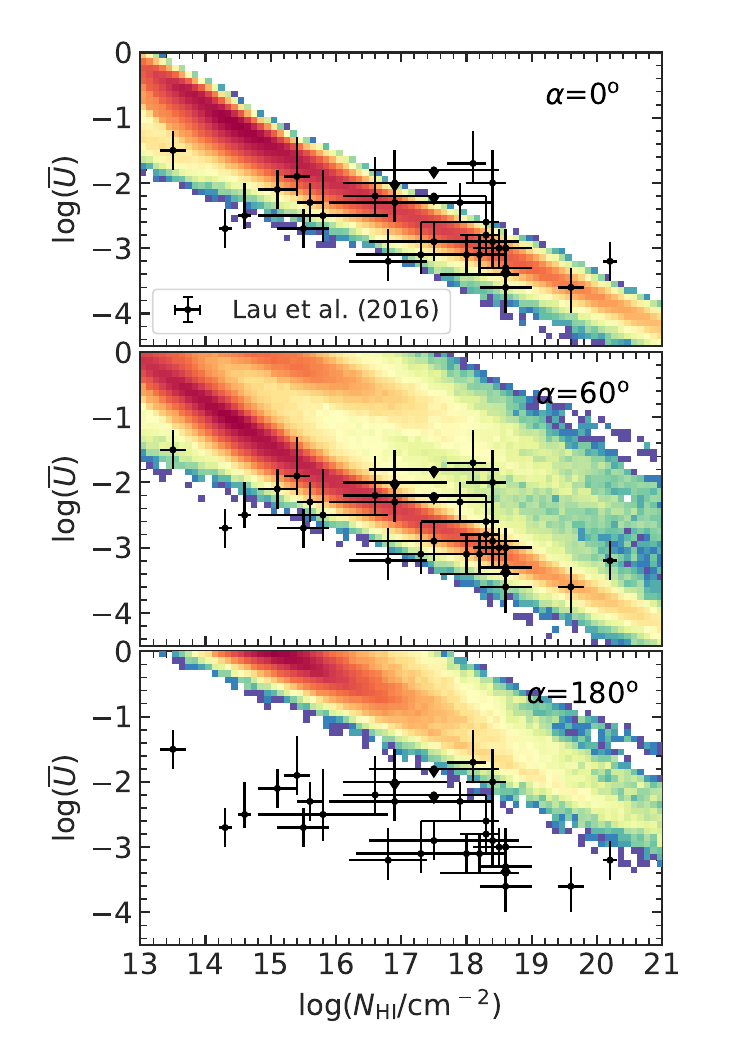}
    \caption{\ion{H}{i} column density (\textbf{top}) and mean ionization parameter $\overline{U}$ (\textbf{center}) as functions of projected distance from the quasar $b$ for three AGN model spectra with increasing number of ionizing photons (left to right; see on top of each column for the values of SMBH mass and Eddington ratio). The solid curves are the 50$^{\rm th}$ percentiles, the shaded areas mark values between the 16$^{\rm th}$ and 84$^{\rm th}$ percentiles, while the lower and upper dashed curves give the 2$^{\rm nd}$ and 98$^{\rm th}$ percentiles, respectively. Colors encode the ionization cone opening angle $\alpha$, as explained in the legends. \textbf{Bottom}: logarithm of the number of sightlines with given \ion{H}{i} column density and mean ionization parameter (colored map) as a function of AGN model (left to right) and ionization cone opening angle (top to bottom). The black points in all panels are the observational estimates of \citet{Lau:2016}.}    
    \label{fig:qpq8_fig13}
\end{figure*}

\begin{figure}
    \centering
    \includegraphics[width=0.45\textwidth]{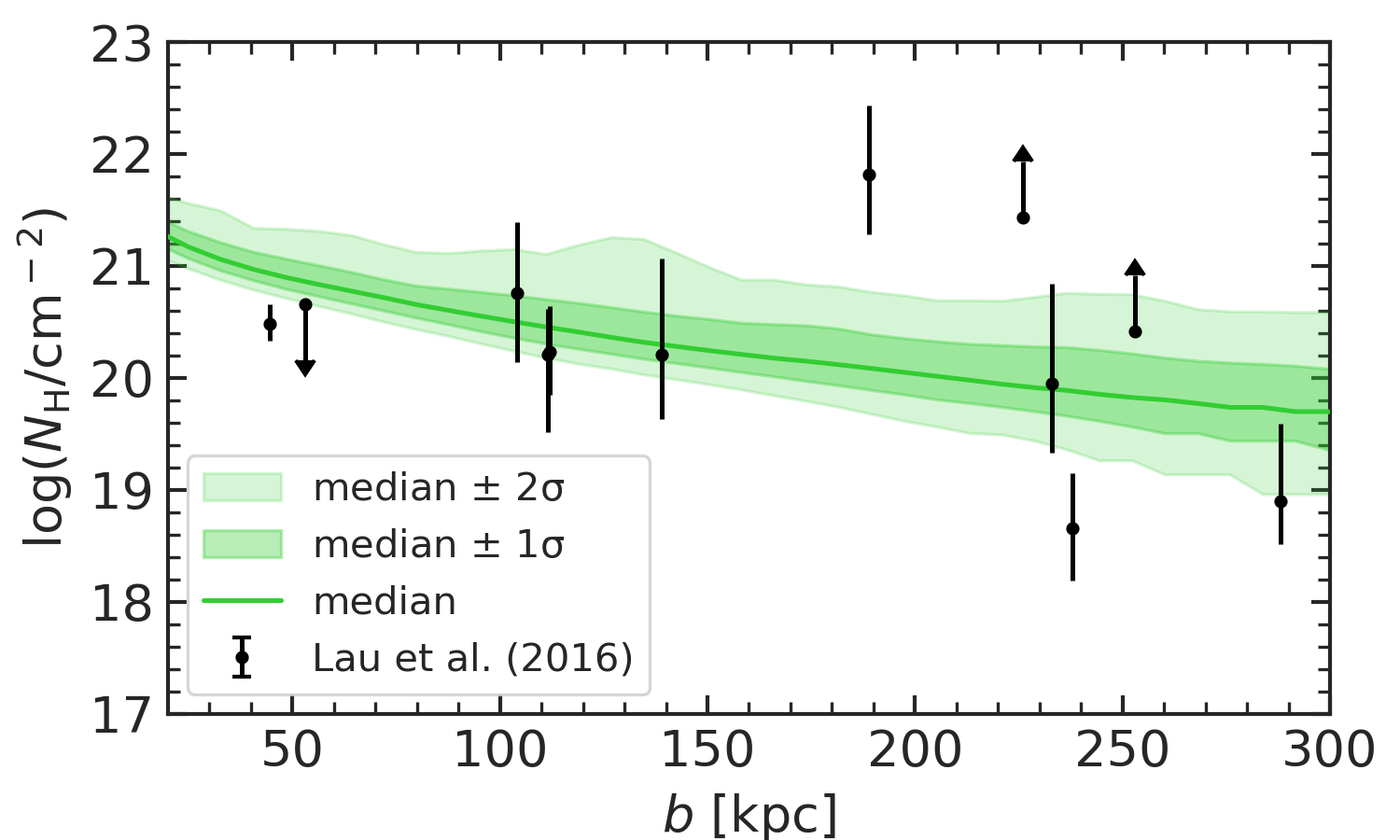}
    \caption{Total hydrogen column density as a function of projected distance from the center of the halo. The median curve and the 1$\rm\sigma$ and 2$\rm\sigma$ shaded areas were obtained by stacking $N$=10 random LOS. The observational data points in black were computed as $N_{\rm H}\equiv N_{\rm \ion{H}{I}}/\mathcal{X}_{\rm \ion{H}{I}}$, where $\mathcal{X}_{\rm \ion{H}{I}}$ is the ionization fraction of hydrogen, estimated via photoionization modeling \citep{Lau:2016}.}
    \label{fig:qpq8_fig15}
\end{figure}

\begin{figure}
    \centering
    \includegraphics[width=0.45\textwidth]{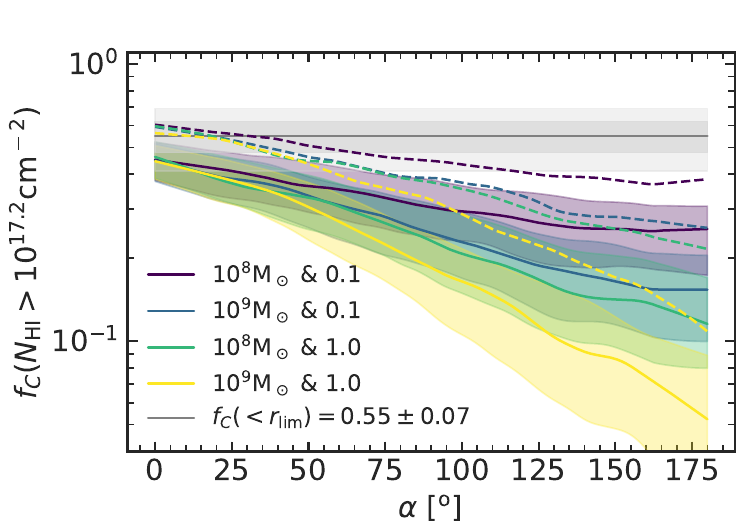}
    \caption{Median covering fractions of \ion{H}{i} with column densities $>$10$^{\rm 17.2}$cm$^{\rm -2}$ for sightlines with projected impact parameters $b\leq$153~kpc, corresponding to a physical radial limit of 200~kpc at the mean redshift of the observational sample in \citet{Prochaska:2013a}, as function of the ionization cone opening angle. Color curves represent the four AGN SED models, shaded areas mark the regions between the 16$^{\rm th}$ and 84$^{\rm th}$ percentiles, and dashed colored curves are the 98$^{\rm th}$ percentiles. The horizontal grey line gives our new estimate of $f_C$ from table 4 of \citet{Prochaska:2013}, using a fixed comoving aperture $r_{\rm lim}$=153(1+3)=612~ckpc, with the two shaded areas marking the 1 and 2$\sigma$ uncertainties.}
    \label{fig:qpq6_fig10}
\end{figure}

\section{CGM observables in absorption}
\label{sec:absorbtion}

For our target galaxies ($z\sim3$ quasar hosts) the best observational data-set in absorption we can compare our simulation with is the one of the project "Quasars probing quasars" (QPQ) introduced by \citet{Hennawi:2006}. This work  pioneered the method of using projected quasar pairs, that are not physically associated with each other but are close on-the-sky, to study quasar environments. Along most sight-lines of the QPQ sample, the measured Lyman-$\alpha$ equivalent width is low $W\approx$1--2\AA, meaning that the \ion{H}{i} column density can not be well estimated because the systems are on the flat part of the curve-of-growth. As a consequence most observations of the QPQ sample can only place limits on $N_{\rm \ion{H}{i}}$ \citep{Prochaska:2013}. \citet{Lau:2016} used only a very small sub-sample of QPQ systems with good measurements of $N_{\rm \ion{H}{i}}$ to also try quantifying the ionization parameter (representing the ratio between the number of ionizing photons and number of hydrogen atoms) at the absorber location. The foreground quasars in this sub-sample have an average redshift $<z>=2.66$, a minimum of 2.00, and a maximum of 4.11. This latter work also provides a formula to approximate the mass of cool gas as a function of projected distance from the quasar. Using their formula we would expect to have $\sim$1.3$\times$10$^{\rm 11}$M$_{\rm\odot}$ within the $r_{\rm 200}$=134~kpc of our simulation, which is larger than our cool ($T<$10$^{\rm 5}$K) CGM mass of $\sim$0.9$\times$10$^{\rm 11}$M$_{\rm\odot}$. As mass fractions, the cool CGM in our simulated galaxy constitutes $\sim1/5$ of the baryons in the halo vs the $\sim1/3$ estimated by \citet{Lau:2016} from their 14 quasar pairs. In the following we provide a first test of how the simulation post-processed with AGN radiation compares to these observations.

First, the top row panels of Figure~\ref{fig:qpq8_fig13} show the comparison between observed and simulated \ion{H}{i} column density as a function of projected distance from the quasar. To compute the percentiles for $N_{\rm\ion{H}{i}}$ for each particular ionization cone opening angle, we proceeded as follows. We generate 10 uniformly distributed directions over the unit sphere, and consider them to give the directions of the ionization cones. Since a quasar is an AGN viewed through its ionization cone, for each $\alpha$ the LOS\footnote{LOS here means the viewing direction of the system from the observer's perspective.} is not centered on the ionization cone axis, but it is chosen randomly among those within the cone\footnote{Drawing more than 10 uniformly distributed directions over the unit sphere, and more LOSs has no impact on our conclusions, but substantially increase the amount of generated data.}. Next, for each LOS we approximate the maps $N_{\rm\ion{H}{i}}(x,y)$ by:
\begin{equation}
\begin{split}
N_{\rm\ion{H}{i}}(x,y) &=  \sum_j N_{\rm j,\ion{H}{i}}(x,y) \\ 
 &= \sum_j m_j(x,y)\mathcal{X}_j(x,y)\mathcal{X}_{j,\ion{H}{i}}(x,y)/\Delta l_j^2,
\end{split}
\label{eq:colden}
\end{equation}
where $N_{\rm j,\ion{H}{i}}(x,y)$ is the column density contribution of particle $j$  with projected position within the pixel $(x,y)$ and position along the LOS, $\rm-500kpc<$$z$$\rm<500kpc$. In Equation~\ref{eq:colden}, $m$ is the gas particle mass, $\mathcal{X}$ is the hydrogen mass fraction,  $\mathcal{X}_{\ion{H}{i}}$ is the ionization fraction, and $\Delta l_j$ is the 'size' of the particle approximated as $\Delta l_j\equiv (m_j/\rho_j)^{1/3}$ with $\rho_j$ the mass density of the particle. To limit the computational time while still having pencil-beam-like information, the maps are done assuming a pixel size of $\Delta$=1.58~kpc ($\sim$0.2~arcsec assuming the halo at $z=3$). While the hydrogen mass fraction $\mathcal{X}$ is given by the simulation, the ionization fraction $\mathcal{X}_{\ion{H}{i}}(n_{\rm H},T,Z,J_{\rm\nu})$ is interpolated from the pre-computed {\sc cloudy} tables. For particles within the ionization cone of the quasar, $J_{\rm\nu}$ has one of the corresponding spectral shapes shown in Figure~\ref{fig:spectra} (blue or orange curves) and a normalization depending on the radial distance to the center of the halo. For particles outside of the ionization cone, $J_{\rm\nu}$ is given by the UVB spectrum at $z=3$ also shown in Figure~\ref{fig:spectra}. Finally, we compute the percentiles in radial bins using the $N_{\rm\ion{H}{i}}(x,y)$ of all 10 LOS together. A more faithful comparison between simulations and observations would require to construct spectra along background sight-lines using a post-processing tool like e.g. {\sc trident} \citep{Hummels:2017}, but this is outside the scope of this paper, and therefore we leave such analysis for a future work.

The radial profiles of $N_{\rm\ion{H}{i}}$ in Figure~\ref{fig:qpq8_fig13} show the simulation to be compatible with the observations if the only source of photoionization is the UVB (all observational data points are between the 2$^{\rm nd}$ and 98$^{\rm th}$ percentiles of the $\alpha=0^{\rm o}$ case given by the dashed grey curves). This result is in line with previous predictions from simulations run only with stellar feedback  \citep[e.g.][]{FaucherGiguere:2016}. As we include ever stronger AGN radiation fields (left to right panels), the simulation is compatible with the observations only for small opening angles: the regions between the 2$^{\rm nd}$ and 98$^{\rm th}$ percentiles of the $\alpha=60^{\rm o}$ case (dashed blue curves) still include all observational data points. The same percentiles for the $\alpha=180^{\rm o}$ case (dashed green curves) include all of the observational data points only for the weakest AGN model (left panel). Both these effects -- with the strength of the radiation field, and with the ionization cone opening angle -- are to be expected, as the ionization parameter $U$ increases with both $J_{\nu}$ and $\alpha$ for fixed number density $n_{\rm H}$. 

It is important to keep in mind that we look at one simulation only, while the observational data points represent the CGM of different systems. Thus, if we consider that our simulation reaches high enough resolution to resolve decently the densities in the CGM \citep[see the effects that resolution can have on $N_{\rm\ion{H}{i}}$ in the CGM, e.g.:][]{Hummels:2019,vandeVoort:2019}, we can interpret the highest values in observations as coming from more massive dark matter hosts (10$^{\rm 12.5}$M$_{\rm\odot}<M_{\rm h}\leq$10$^{\rm 13}$M$_{\rm\odot}$) or from the ISM of satellites (e.g. the bump in the higher percentiles simulated curves at $b\sim$130~kpc is due to the satellite galaxy just entering the virial radius, as can be seen in Figure~\ref{fig:g316e12maps}). For the first possibility, it is instructive to look at the profile of total hydrogen column density $N_{\rm H}$ in comparison with the $N_{\rm H}\equiv N_{\rm \ion{H}{I}}/\mathcal{X}_{\rm \ion{H}{I}}$ reconstructed from the observations, which we show in Figure~\ref{fig:qpq8_fig15}. While most observational estimates are in agreement with the simulation, a few data points are clearly outside the simulation range. The most natural explanation for these data points is that the corresponding absorbers are coming from dark matter halos of different masses. Given that the total gas density in a dark matter halo decreases with radius, we expect the $N_{\rm H}(b)$ of the simulation to also decrease with the impact parameter, as seen in the figure. On the other hand, the total hydrogen content of a halo is proportional to the dark matter mass. Therefore, a more/less massive halo would shift the simulated profile up/down.    

In each of our models, we can also compute the ionization parameter $U_j$ for each particle $j$ in the simulation:
\begin{equation}
 U_j \equiv \frac{\Phi_j(H)}{n_{j,H} c},
\end{equation}
where $c$ is the speed of light, and $\Phi_j(H)$ is the ionizing photon flux ($E>$13.6~eV) at the particle position (see the {\sc cloudy} documentation), which depends on the AGN SED model and distance to the AGN for particles within the ionization cone, and only on the particle's hydrogen number density $n_{\rm j,H}$ for particles outside (as the UVB just gives a constant $\Phi(H)$ for a given redshift). To compute an average ionization parameter $\overline{U}$ we weight each particle by its column $N_{\rm j,\ion{H}{i}}(x,y)$ to construct similar $\overline{U}(x,y)$ maps:
\begin{equation}
 \overline{U}(x,y) \equiv \frac{\sum_j N_{\rm j,\ion{H}{i}}(x,y)U_j(x,y)}{N_{\rm\ion{H}{i}}(x,y)}
\end{equation}

The panels in the second row of Figure~\ref{fig:qpq8_fig13} show the comparison between the radial profiles of $\overline{U}$ and the values estimated by  \citet{Lau:2016} for the absorbers in the 14 foreground quasars of their pairs. 
In their calculation, it is assumed, as in our work, that the ionization structure of the gas is set by photoionization. Specifically, \citet{Lau:2016} modeled each absorber as a plane-parallel slab with its corresponding $N_{\rm HI}$ value and a fixed $n_{\rm H}=0.1$~cm$^{-3}$. The ionization parameter $U$ is determined by varying the metallicity and the input radiation field from the UVB of \citet{Haardt:2012} to higher amplitude assuming a EUV power-law spectrum $f_{\nu}\propto \nu^{-1.57}$ until the results converge on the available multiple ionization states of individual elements. Same as for the top row, the simulations are in good agreement with the observations if the only photoionization source is the UVB, or if the AGN's ionization cone opening angle is small enough. To make the comparison between observations and simulations clearer, the lower panels of Figure~\ref{fig:qpq8_fig13} show the 2D log($N_{\ion{H}{i}}$)--log($\overline{U}$) distribution of sightlines in a logarithmic scale (colored map with red marking the maximum) for three different angles $\alpha$. In these panels it is clear that the observational data favor the AGN models with low accretion rate $\lambda=0.1$ (left and center) and low opening angle $\alpha=60^{\rm o}$. On the other hand the UVB model alone (the panels with $\alpha=0$) results in a very marked anti-correlation between log($\overline{U}$) and log($N_{\ion{H}{i}}$), which, however, is not spread enough in log($\overline{U}$) to cover well the observational data.    

Since \citet{Lau:2016} also uses absorption lines of other ions besides \ion{H}{i} to constrain $U$, for completeness, in Figure~\ref{fig:qpq8_figallions} of Appendix~\ref{appendix:absorbtion} we give our models predictions for various ion columns as a function of impact parameter in comparison to their results. Specifically, we show predictions for column densities of \ion{Si}{ii} (16.3~eV), \ion{Si}{iv} (45.1~eV), \ion{C}{iv} (64.5~eV), \ion{N}{v} (97.9~eV) and \ion{O}{vi} (138.1~eV) as a function of impact parameter. Given that most of the data are limits, we cannot draw definite conclusions from these plots, i.e. none of the models stand out as preferred. The most important points are (i) the agreement with observations of the predictions for the aforementioned favored model with low accretion rate $\lambda=0.1$ and $\alpha=60^{\rm o}$ (central column), and (ii) the exclusion of the model with the highest radiation input ($M_{\bullet}=10^9$~M$_{\odot}$, $\lambda=1.0$) injected isotropically because of the tension with the Silicon columns (top right panels). 

To conclude this section, we condense the behavior of our AGN models into one parameter: the covering fraction $f_C$ of optically thick absorbers ($N_{\rm\ion{H}{i}}>10^{\rm 17.2}$cm$^{\rm -2}$) within the projected CGM region, and compare it with QPQ estimates \citep{Prochaska:2013a,Prochaska:2013}. In a work using only the QPQ spectra with signal-to-noise ratios S/N>9.5 around the Ly$\alpha$ restframe wavelength of the absorbers, \citet{Prochaska:2013a} concluded that $f_C$($b<200$kpc)=0.64$^{\rm +0.06}_{\rm -0.07}$. This estimate is based on 50 sightlines (32 of which are optically thick), with the foreground quasars having redshifts between 1.63 and 2.97, and a mean redshift $<z>_{\rm QPQ5}=2.06$. Even assuming that quasars sit in halos of fixed mass at all epochs, as usually done, the limit radius of 200~physical kpc will cover different fractions of the projected virial radius at the redshifts of the sample. To correct for such aperture effects given the large $z$ range of the QPQ sample, we take a fixed comoving aperture set as $r_{\rm lim}=(1+<z>_{\rm QPQ5})200$kpc=612~ckpc and recompute the observational $f_C$ based on the table 4 of \citet{Prochaska:2013}. We get that 51 absorbers with S/N>9.5 have $b(1+z)\leq r_{\rm lim}$, and 28 of them are catalogued as optically thick. This translates into a new estimate of $f_C(<r_{\rm lim})=0.55\pm0.07$, with the uncertainty given by the binomial distribution. Figure~\ref{fig:qpq6_fig10} nicely shows how only small ionization cone opening angles are compatible with observations quantified by $f_C(<r_{\rm lim})=0.55\pm0.07$, and that the stronger the AGN's SED, the smaller the allowed $\alpha$ values \footnote{To compute $f_C(<r_{\rm lim})$ for the simulation, we use the same 10 random LOS for each $\alpha$, and for each LOS we randomly pick 51 positions with $b\leq r_{\rm lim}$ 10 times, counting how many of them have $N_{\rm\ion{H}{i}}>10^{\rm 17.2}$cm$^{\rm -2}$.}. The upper limits on $\alpha$ given in Table~\ref{tab:alpha} are computed as the intersections between the upper 2$\sigma$ percentile curves of the four AGN SED models (dashed colored curves) and the lower 2$\sigma$ limit on the observationally estimated $f_C(<r_{\rm lim})$ of 0.41. 

It is important, though, not to over interpret these upper limits for various reasons. First of all, we use only one simulation, and because it's a zoom-in we can only consider a limited high-resolution region around the main galaxy. Second, most QPQ foreground quasars have smaller redshifts than the simulation, and it is expected that $f_C$ increases with redshift \citep[e.g.][]{Rahmati:2015}. However, the low number statistics behind the observational estimate of $f_C$ makes  measuring a $z$-trend unreliable. Third, the most generous ranges for dark matter halos hosting quasars span from 10$^{\rm 12}$ to 10$^{\rm 13}$M$_{\rm\odot}$. This mass uncertainty coupled with the wide redshift range of observations means that the probed quasar halos can be in significantly different evolutionary stages, with vastly different amounts of cool gas in the CGM \citep[e.g.][]{Dekel:2006}. Finally, the uncertainties in the quasars' redshifts impose lower limits on the wavelength (velocity) window used to model the Ly$\alpha$-line in absorption, which corresponds to physical depths a few times larger than the viral radii of the dark matter host halos. This latter limitation biases high the observationally derived $f_C(<r_{\rm lim})$. All these caveats are discussed by \citet{Prochaska:2013a} and \citet{Prochaska:2013}, and quantified in some of the  numerical works using large-box simulations, like the one of \citet{Rahmati:2015}.

\section{CGM observables in emission}
\label{sec:emission}

\begin{figure}
    \centering
    \includegraphics[width=0.15\textwidth]{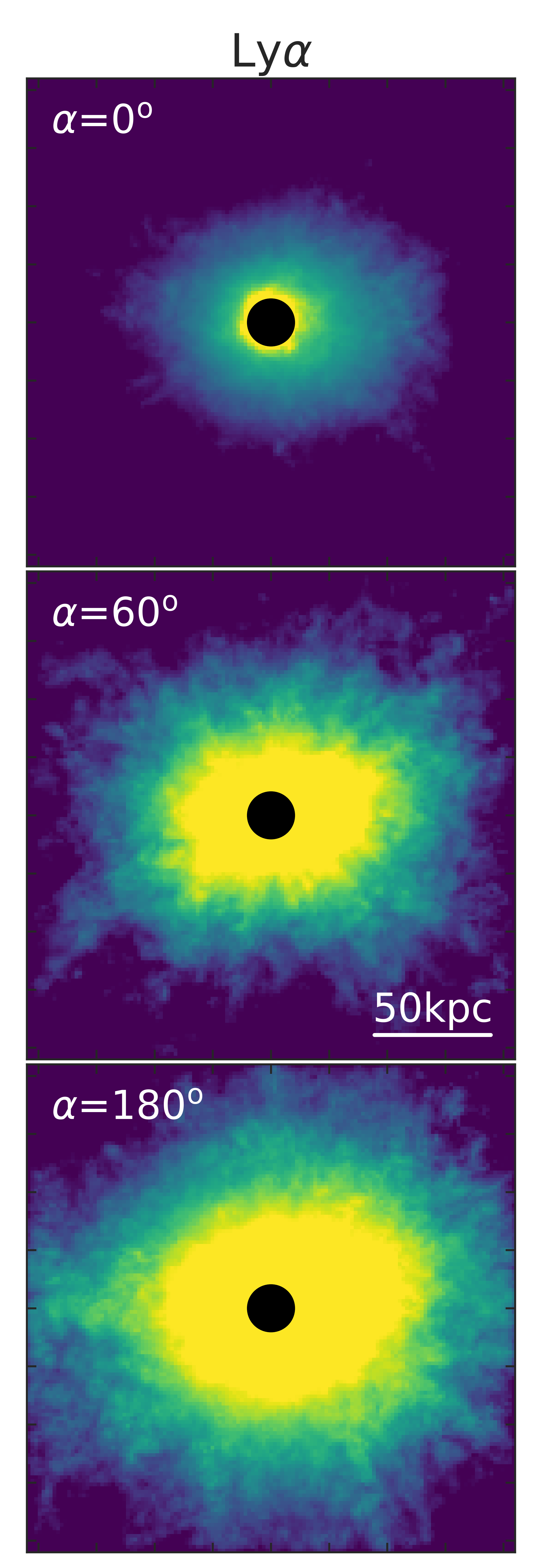}
    \includegraphics[width=0.15\textwidth]{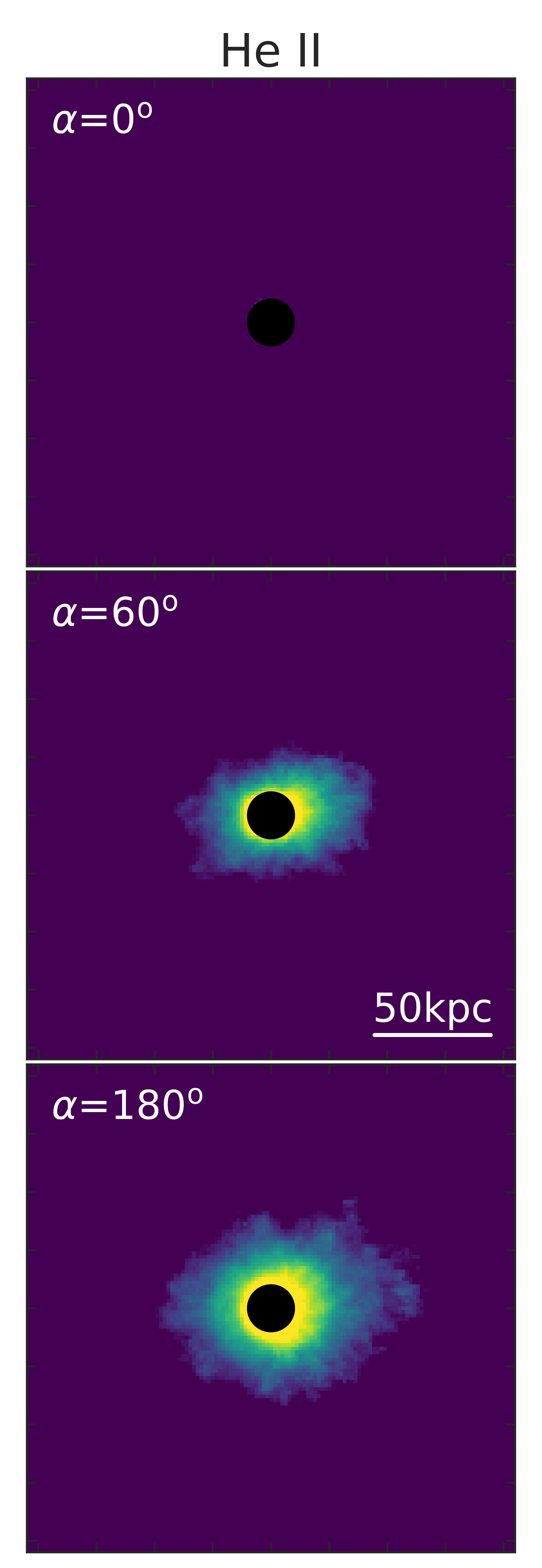}
    \includegraphics[width=0.15\textwidth]{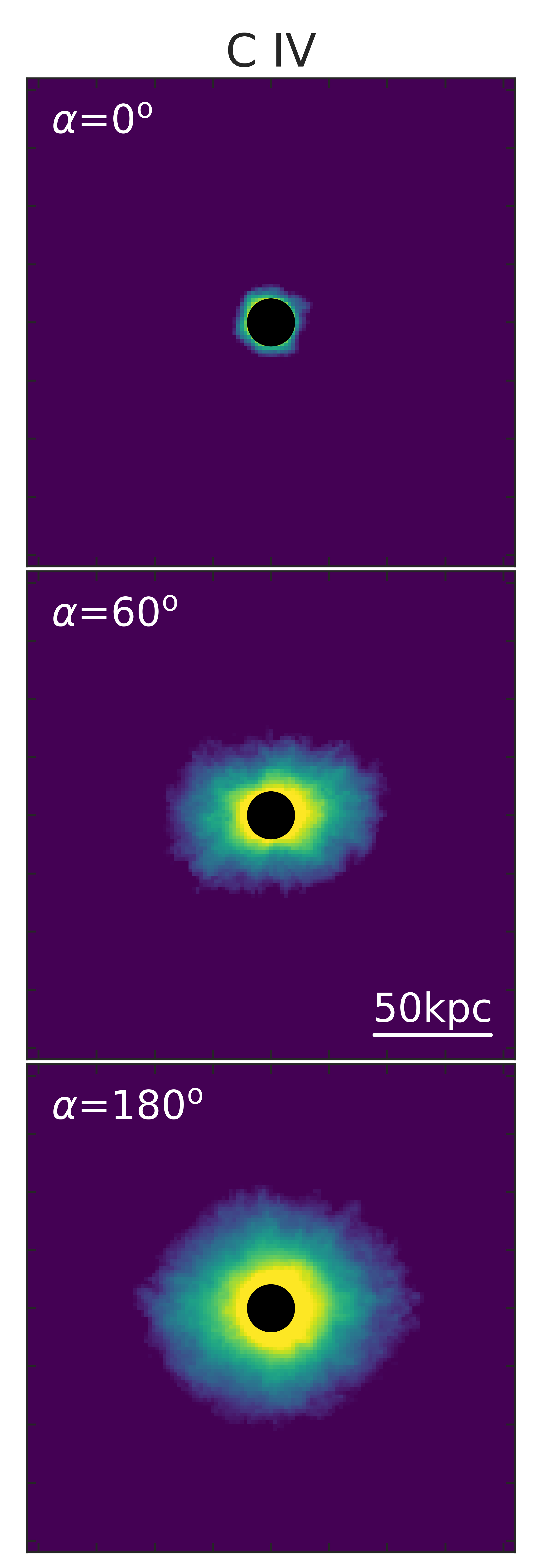}\\
    \includegraphics[width=0.45\textwidth]{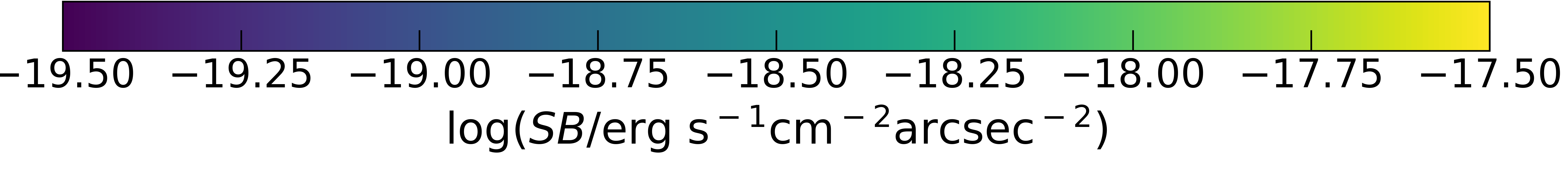}
    \caption{Median stacked surface brightness maps for Ly$\alpha$ (left), \ion{He}{ii}1640\AA (center), and \ion{C}{iv}1549\AA (right) for the AGN model $M_{\rm\bullet}$=10$^{\rm 9}$M$_{\rm\odot}$ \& $\lambda=0.1$ and three different ionization cone opening angles (top row gives the UVB only case, bottom row the AGN only, and the middle row an intermediate $\alpha$). The black dots mask the central regions (radius of 10~kpc) where the observational stacks of \citet{Fossati:2021} are dominated by the point spread function of the quasar. The images span 210$\times$210~kpc$^{\rm 2}$ and the emission is computed as coming from a halo placed at the median redshift of the observations $z_{\rm med} = 3.75$. The images are smoothed with a Gaussian kernel of FWHM$=0.65$~arcsec, corresponding to the average FWHM in \citet{Fossati:2021}.}    
    \label{fig:fossati_fig6}
\end{figure}

\begin{figure*}
    \centering
    \includegraphics[width=0.40\textwidth]{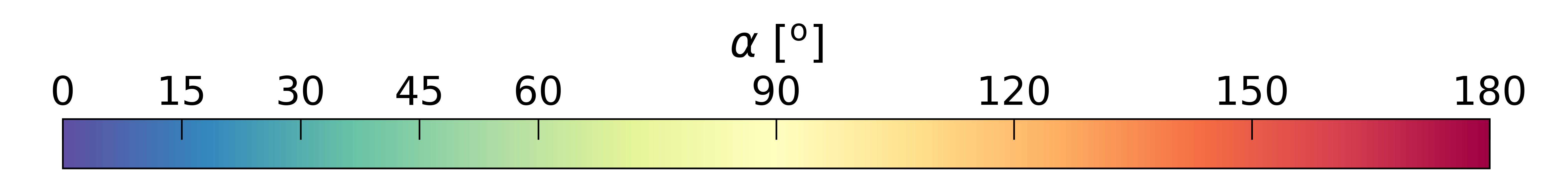}\\
    \includegraphics[width=0.33\textwidth]{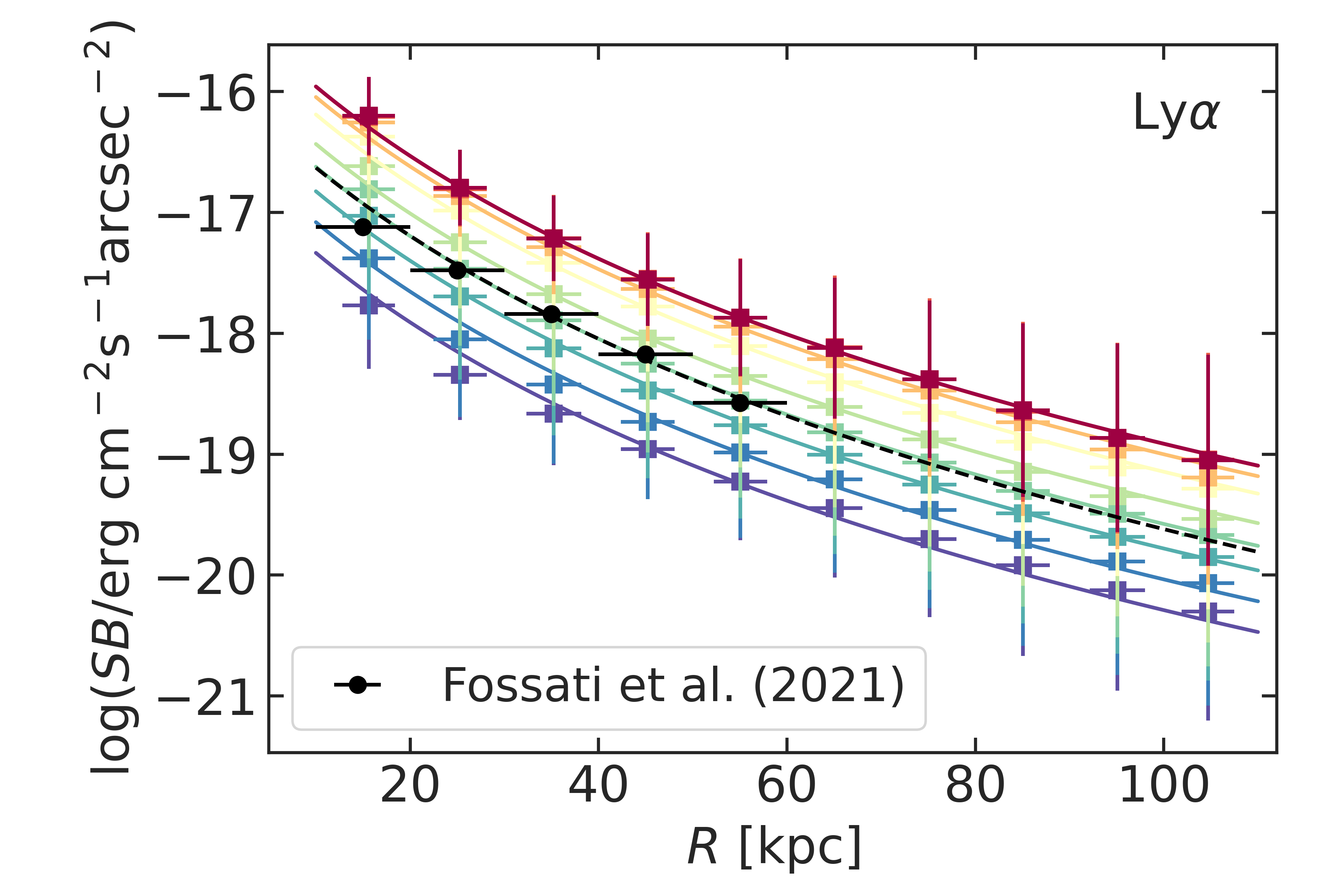}
    \includegraphics[width=0.33\textwidth]{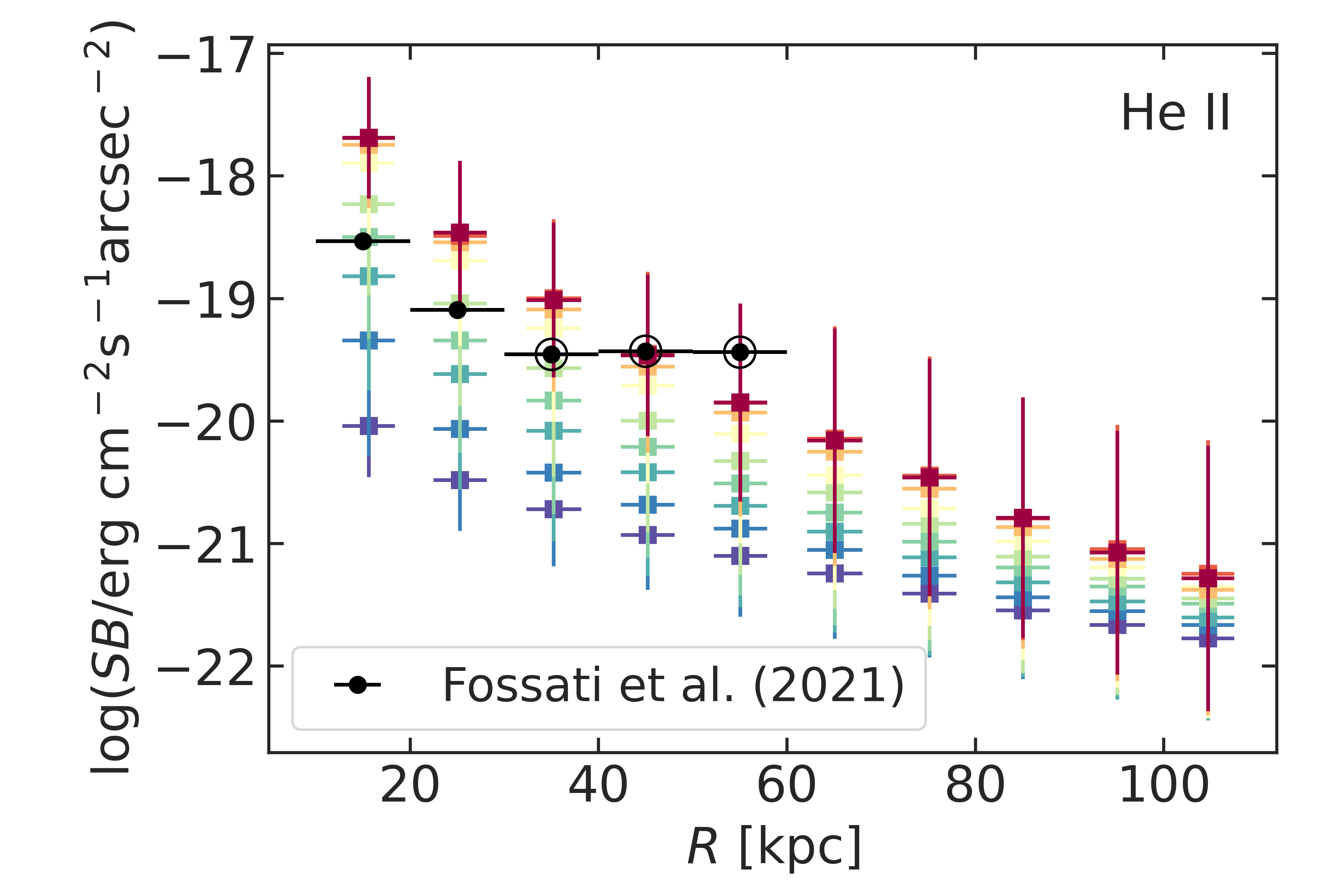}
    \includegraphics[width=0.33\textwidth]{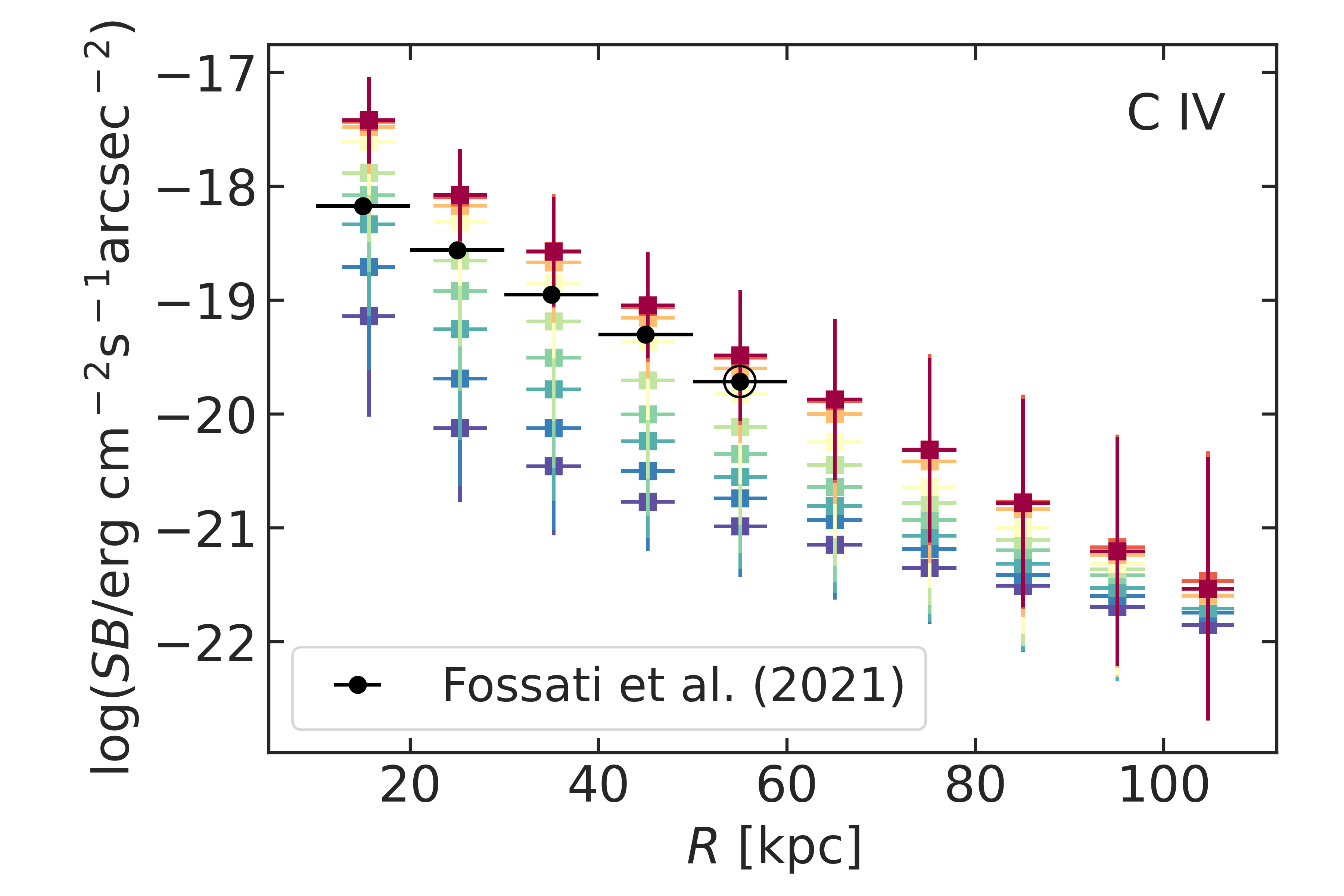}
    \caption{Averaged radial surface brightness profiles for Ly$\alpha$ (left), \ion{He}{ii}1640\AA (center), and \ion{C}{iv}1549\AA (right) for the AGN model $M_{\rm\bullet}$=10$^{\rm 9}$M$_{\rm\odot}$ \& $\lambda=0.1$ for different values of opening angle $\alpha$ corresponding to the ticks of the colorbar. The models are compared with the observations of \citet{Fossati:2021}. The emission is computed as coming from a halo placed at the median redshift of the observations $z_{\rm med} = 3.75$. The observational data points marked with an empty black circle are at the noise level or below. The curves in the left panel have the functional form SB$_{\rm fit}$=SB$_0$(1+$R$/$R_s$)$^{-\beta}$, with $R_s$=26.8~kpc and $\beta$=5.5, while SB$_0$ is computed averaging the normalization over all data points along a curve: log(SB$_0$)=$\langle$log(SB$_{\rm data}$)+$\beta$(1+$R$/$R_s$)$\rangle$. The normalization of the black dashed curve was computed from the observed data points, and has exactly the same value as the one of the $\alpha$=45$^{\rm o}$ model: SB$_0$=10$^{\rm -15.86}$erg~s$^{\rm -1}$cm$^{\rm -2}$arcsec$^{\rm -2}$.}    
    \label{fig:fossati_fig7} 
\end{figure*}

\begin{figure}
    \centering
    \includegraphics[width=0.4\textwidth]{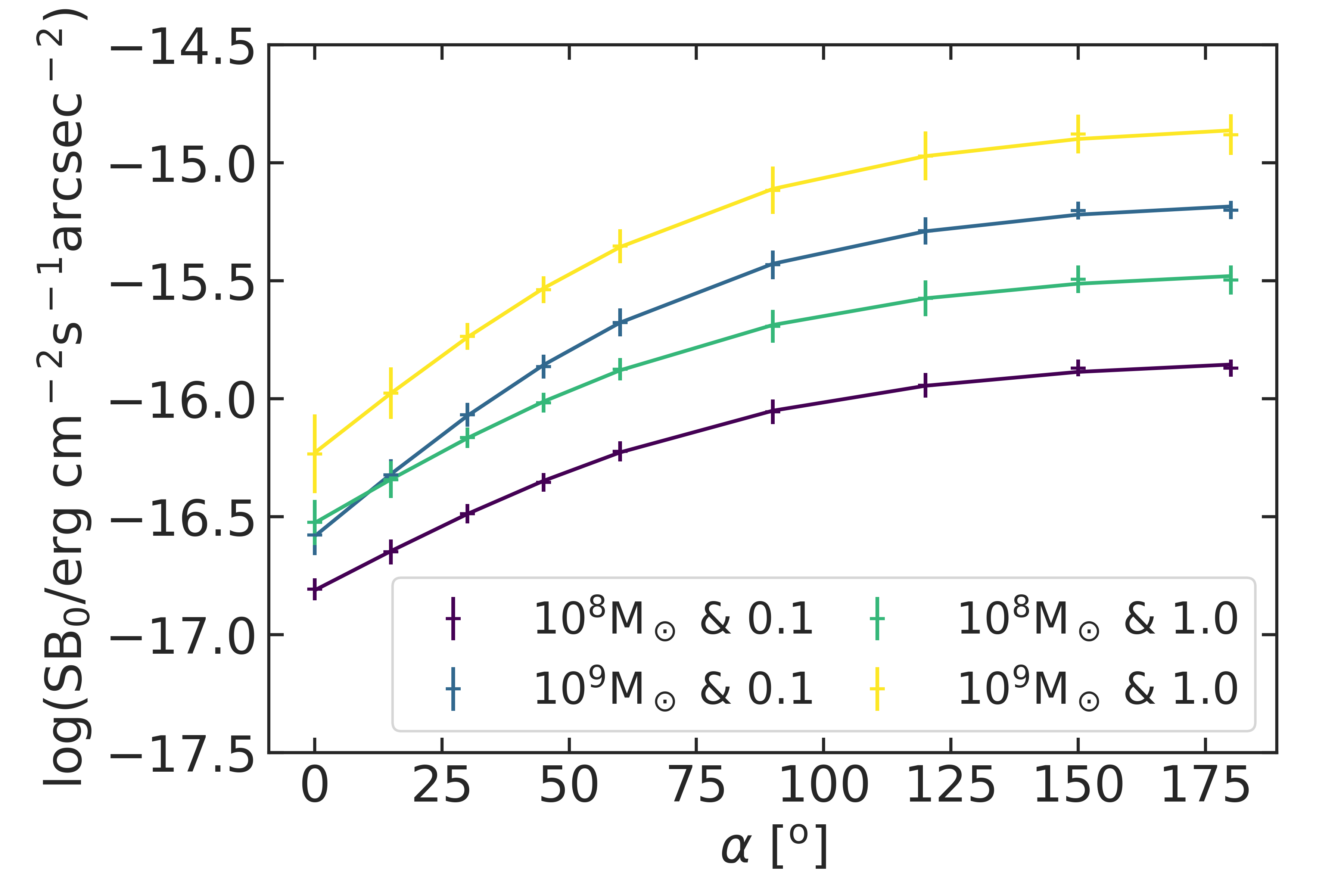}\\
    \caption{The normalization variations of the Ly$\alpha$ profiles (colored points) with the angle $\alpha$ for the four AGN SED models (different colors). The curves follow Equation~\ref{eq:lya_normfit}, with the maximum likelihood parameters of Table~\ref{tab:Lya_fits}.}    
    \label{fig:lya_profile_norm} 
\end{figure}

\begin{figure*}
    \centering
    \includegraphics[width=0.33\textwidth]{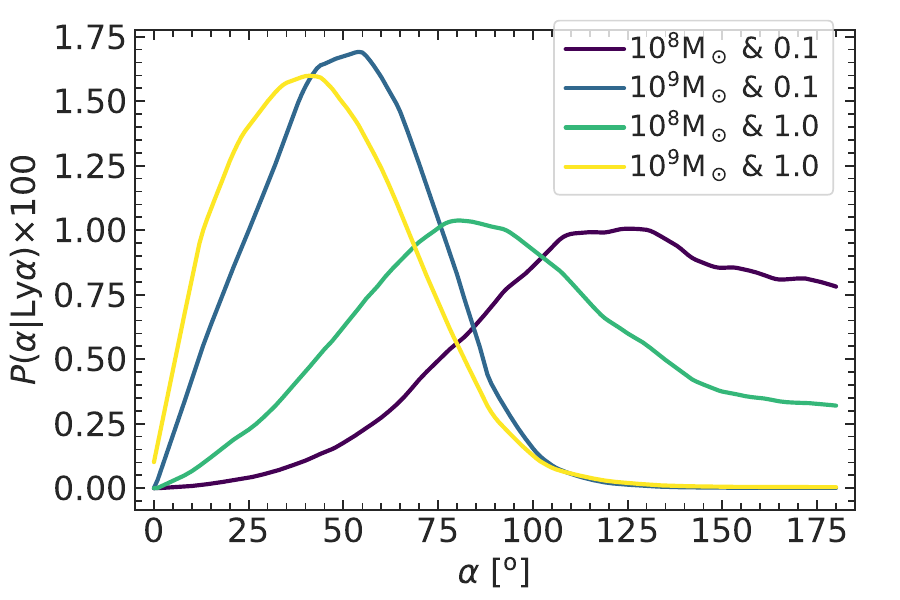}
    \includegraphics[width=0.33\textwidth]{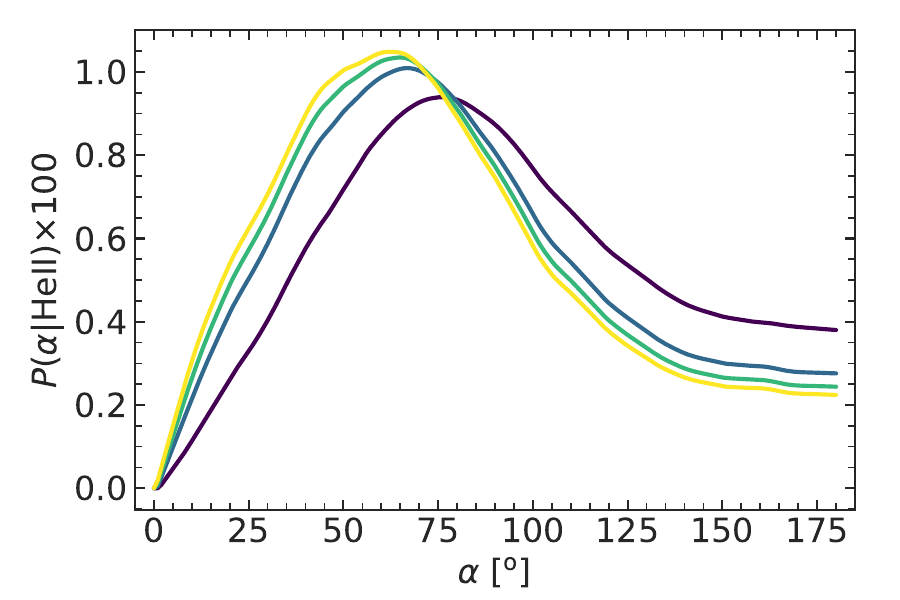}
    \includegraphics[width=0.33\textwidth]{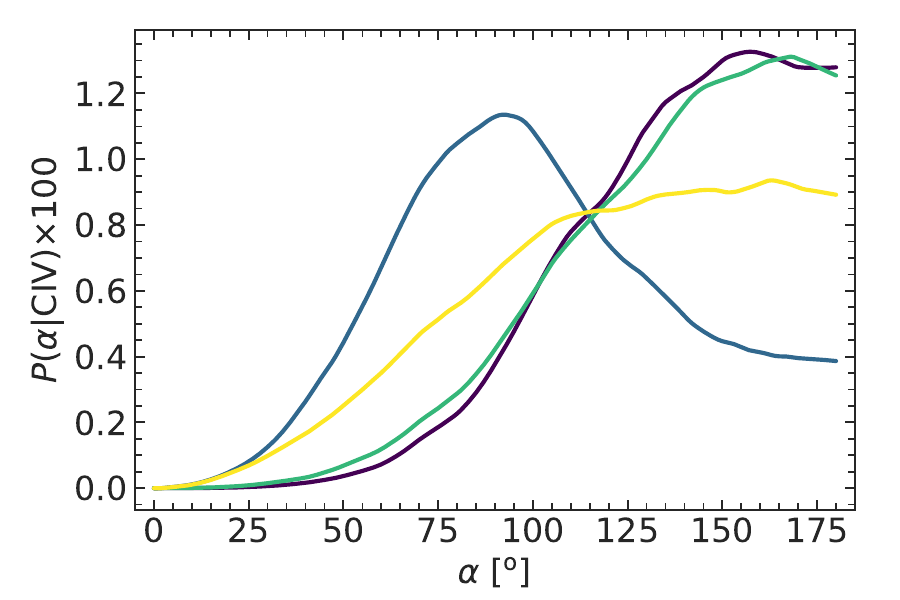}
    \caption{Posterior probability distribution of $\alpha$ for the four AGN SED models (different colors), as constrained by the stacked emission profiles above the noise level for the Ly$\alpha$ (left), \ion{He}{ii} (center), and \ion{C}{iv} (right)}    
    \label{fig:model_comparison_emission}
\end{figure*}

A few previous works reproduced the physical scales and level of emission in Ly$\alpha$ seen in high-$z$ observations using a variety of simulations and modeling approaches, from dealing with radiation transfer effects only in post-processing \citep[e.g.][]{Cantalupo:2014,Gronke:2017,Byrohl:2021} to running also the simulations with radiation-hydrodynamics codes \citep[e.g.][]{Mitchell:2021,Costa:2022}.

In this work, we compare our simulation with the observations of \citet{Fossati:2021}, who targeted 27 quasars at a median redshift $z_{\rm med}=3.75$ with MUSE, and derived average stacked emission line profiles not only for Ly$\alpha$, but also for the rest frame UV lines \ion{He}{ii}1640\AA~and \ion{C}{iv}1549\AA. \ion{He}{ii} is a recombination line which requires a hard ionization spectrum to excite the gas ($E\geq 54.4$eV~$=4$~Ryd), and therefore can be a good tracer of AGN radiation effects. Ly$\alpha$ and \ion{C}{iv} are resonant lines, whose modeling is much more complicated \citep[e.g.][]{Dijkstra:2019}. The \ion{C}{iv} emission, also requiring high energies ($E>4$~Ryd), encodes information on the metallicity of the gas and therefore its extent can provide an estimate of the scale out to which the halo is metal-enriched. 

To obtain mock emission line maps we proceed as follows. We make the simplifying assumption that a gas particle of density $n_{H}$, temperature $T$ and metallicity $Z$ exposed to a radiation field $J_{\nu}$ has an emergent emissivity at a particular line frequency $\nu_0$:
\begin{equation}
 \varepsilon_{\nu_0,{\rm emergent}} =  \varepsilon_{\nu_0,{\rm intrinsic}}e^{-\tau_{\nu_0}},
 \label{eq:emissivity}
\end{equation}
where $\varepsilon_{\nu_0,{\rm intrinsic}}$ is the intrinsic line emissivity in units of erg~s$^{\rm -1}$~cm$^{\rm -3}$ computed with {\sc cloudy}, and $\tau_{\nu_0}$ the local optical depth at the line frequency. As our photoionization models have a hydrogen column stopping criteria, and we consider particles to be represented by the physical properties of the last {\sc cloudy} zone, we compute the local optical depth at the line frequency as:
\begin{equation}
 \tau_{\nu_0} = N_{\nu_0}\sigma_{\nu_0} = n_{\nu_0}d_{\rm zone}\sigma_{\nu_0}, 
 \label{eq:tau}
\end{equation}
where $N_{\nu_0}$ and $n_{\nu_0}$ are the column and volume number density of the ionic species producing the line (e.g. \ion{H}{i} for Ly$\alpha$), $\sigma_{\nu_0}$ is the absorption cross-section at line center, and $d_{\rm zone}$ is the corresponding depth of the last {\sc cloudy} zone, which is also a function of $(n_{\rm H},T,Z,J_{\nu})$. The Ly$\alpha$ absorption cross-section at line center is a function of temperature only \citep[e.g.][]{Dijkstra:2019}:
\begin{equation}
 \sigma_{\rm Ly\alpha} = 5.9\times10^{-14} {\rm cm}^2 \sqrt{\frac{10^4{\rm K}}{T}},
\end{equation}
and $\tau_{Ly\alpha}= \sigma_{\rm Ly\alpha}n_{\ion{H}{i}}d_{\rm zone}$, while $n_{\ion{H}{i}}$ is computed as described in the previous section. 

The other emission lines we are interested in are \ion{He}{ii}$\lambda$1640\AA, \ion{C}{iv}$\lambda$1549\AA, and H$\alpha$. To compute the absorption cross-sections for these other lines, we use the fact that the line center optical depth is proportional to the oscillator strength $f$\footnote{All the values of atomic physics used in this manuscript are taken from the NIST database \url{https://www.nist.gov/pml/atomic-spectra-database}}. Thus, for H$\alpha$ we have 
\begin{equation}
\sigma_{\rm H\alpha} = \sigma_{\rm Ly\alpha} \frac{f_{\rm H\alpha}}{f_{\rm Ly\alpha}} \frac{g_2}{g_1} exp\left(\frac{-10.2{\rm eV}}{k_BT}\right), 
\end{equation}
with $f_{\rm Ly\alpha}=0.416$ and $f_{\rm H\alpha}=0.6958$, the ratio of the statistical weights of the first two levels of hydrogen $g_2/g_1=2/8$, $k_B$ the Boltzmann constant, and 10.2~eV the difference in energy between the two levels. The relevant column density is the same as in the Ly$\alpha$ case: $N_{\ion{H}{i}}=n_{\ion{H}{i}}d_{\rm zone}$.

For the \ion{He}{ii} line we have to also consider the difference in atomic mass between helium and hydrogen, such that:
\begin{equation}
\tau_{\ion{He}{ii}} = \tau_{\rm Ly\alpha} \frac{n_{\ion{He}{ii}}}{n_{\ion{H}{i}}} \frac{g_2}{g_1} \frac{f_{\ion{He}{ii}}}{f_{\rm Ly\alpha}}\sqrt{\frac{A_{\rm He}}{A_{\rm H}}} exp\left(\frac{-40.8{\rm eV}}{k_BT}\right), 
\end{equation} 
where $g_2/g_1 = 8/2$, $f_{\ion{He}{ii}}= 0.6958$ and $A_{\rm He}=4$ and $A_{\rm H}=1$ are the two atomic masses. In this case the difference in energy levels is 40.8~eV.

The \ion{C}{iv} line is like the Ly$\alpha$ line, but for the triply-ionized carbon atom, such that:
\begin{equation}
 \tau_{\ion{C}{iv}} = \tau_{\rm Ly\alpha} \frac{n_{\ion{C}{iv}}}{n_{\ion{H}{i}}} \frac{f_{\ion{C}{iv}}}{f_{\rm Ly\alpha}}\sqrt{\frac{A_{\rm C}}{A_{\rm H}}}, 
\end{equation}
with the oscillator strength $f_{\ion{C}{iv}}=0.19$ and $A_{\rm C}=12$. For this line we have to make a further assumption, as the {\sc Gasoline2} version we used to run this simulation does not trace the carbon enrichment explicitly. Therefore, we assume that particles have the same carbon to oxygen number density ratio as the Sun's photosphere \citep{Asplund:2009}, and use the oxygen abundance of each particle, which is traced by the simulation code, to compute $n_{\ion{C}{iv}}$ taking into account the different atomic masses of the two elements.   

Once we have the emergent emissivity for each line, we compute for each particle the luminosity at the particular line frequency $\nu_0$ as:
\begin{equation}
 L_{\nu_0} = \varepsilon_{\nu_0,{\rm emergent}}(n_{\rm H},T,Z,J_{\nu})\frac{m}{\rho},
\end{equation}
where $m$ and $\rho$ are the mass and mass volume density of the particle. Finally, the observed flux from a particle becomes:
\begin{equation}
 F_{\nu_0} = \frac{L_{\nu_0}}{4\pi D_L^2},
 \label{eq:flux}
\end{equation}
where $D_L$ is the luminosity distance. For $z=3$ and our assumed cosmology $D_L$=26070~Mpc.

We stress that for the resonant transitions Ly$\alpha$ and \ion{C}{iv}, we are not performing a full radiative transfer calculation. Such a calculation would require an even higher resolution in the CGM and a realistic dust distribution to converge \citep[e.g.][]{Camps:2021}. Nevertheless, our approach takes into account this process in an approximated way. Following the results presented in \citet{Costa:2022} to reproduce extended emission around high-z quasars, we assume that Ly$\alpha$ and \ion{C}{iv} photons are able to escape to CGM scales through paths of least resistance (the ionization cones in this work), and are able to resonantly scatter in the CGM (inclusion of line emissions in the quasar input spectrum and continuum pumping in the {\sc Cloudy} calculation). Finally, our calculation of the emergent emissivity (Equation~\ref{eq:emissivity}) for the resonant lines is conservative as it assumes the cross-section at line center, which maximizes absorption. In reality, local absorption of resonant photons is quickly followed by the re-emission of another photon. These re-emitted photons could escape towards the observer in one or more scatterings. We stress that the local absorption correction of Equation~\ref{eq:emissivity} is important when the photoionization source is the UVB, as this spectrum results in larger local column densities of neutral hydrogen than the AGN ones. In fact, in the CGM region, the local absorption correction is minimal if the photoionization source is an AGN, as demonstrated by Figure~\ref{fig:NH_and_tau_lya} of Appendix~\ref{appendix:nolines}. Lastly, no spatial redistribution of resonant photons is taken into account. However such phenomenon should not affect strongly the observables used in this work (i.e. stacked images and the SB profiles) as its effects are (i) washed out by the stacking of several objects (resulting in a symmetric glow) and (ii) stronger on scales similar to the radial bins used in the following analysis. 

The sample of nebulae observed by \citet{Fossati:2021} comprises 27 quasar fields, and these authors stack all of them together in order to increase the signal-to-noise (S/N) of the Ly$\alpha$, \ion{He}{ii} and \ion{C}{iv} emission lines (see their Figure 6 with the median stacked surface brightness maps). To mimic their observations, and since we have only one simulation, we pick 27 random directions through the halo center to place the AGN ionization bi-cone, and for each of them we randomly select as LOS of our mock observations, a direction within this bi-cone. We then redshift the simulation to the median redshift of the observed quasar sample $z_{\rm med} = 3.75$ and chose for our mock images the same pixel scale as in observations (0.2 arcsec). For each pixel in one of these individual mock observations we sum the line flux contribution (Equation~\ref{eq:flux}) from all the particles with 2D projected positions within the pixel, and along a LOS length of 210~kpc (-105~kpc to +105~kpc). The obtained flux in each pixel is then divided by the pixel area to obtain surface-brightness (SB) maps. Each of the 27 maps for a particular emission line is then smoothed with a Gaussian kernel with FWHM~$=0.65$~arcsec (or $ \approx5$~kpc) to take into account the average image quality of the data in \citet{Fossati:2021}. Finally the 27 SB maps have been median stacked to compare them to the observations. 

Figure~\ref{fig:fossati_fig6} shows how the obtained median stacked SB maps change with the ionization cone opening angle for the AGN SED model with $M_{\rm\bullet}$=10$^{\rm 9}$~M$_{\rm\odot}$ and $\lambda=0.1$. The top panels give the case where only the UVB acts as photoionization source, while the bottom one the extreme case where the AGN illuminates all the CGM gas. The color bar in all panels is the same as the one in Figure 6 of \citet{Fossati:2021}. In all three lines, the emission is brighter for larger $\alpha$, while for a fixed $\alpha$ the Ly$\alpha$ SB is larger than that of \ion{C}{iv}, which in its turn is larger than that of \ion{He}{ii}. Interestingly, at the sensitivity of Fossati et al.'s observations, a Ly$\alpha$ glow would be observed even if the AGN has a minimal impact on the CGM gas ($\alpha=0$), but there would be no detectable emission in \ion{He}{ii} and \ion{C}{iv}.

To compare in a quantitative manner the simulation with the observations, Figure~\ref{fig:fossati_fig7} shows the SB profiles as a function of projected radius $R$ for the same  AGN SED model with $M_{\rm\bullet}$=10$^{\rm 9}$M$_{\rm\odot}$ and $\lambda=0.1$, as functions of $\alpha$ (shown only nine profiles with $\alpha$ corresponding to the values of the colorbar ticks). For the Ly$\alpha$ (left panel), the simulation reproduces well the shape of the observed stacked profile (black points). The normalization of the simulated profiles scales with the $\alpha$ angle, with $\alpha\sim45^{\rm o}$ falling on top of the observations. For the \ion{He}{ii} and \ion{C}{iv} (center and right panels), the simulated profiles change not only their normalization, but also their shape with varying $\alpha$. In the case of \ion{He}{ii}, for which only the first two radial bins are above the noise level in the stacked profile of Fossati et al., it seems that also the model with $\alpha\sim45^{\rm o}$ agrees best with observations. The observed stacked profile of \ion{C}{iv} seems to require a larger $\alpha$ than the one of Ly$\alpha$, for a fixed AGN model. 

We fit the Ly$\alpha$ profiles with the following type of power-law:
\begin{equation}
    {\rm SB}_{\rm fit}={\rm SB}_0(1+R/R_s)^{-\beta},
\label{eq:lya_fit}
\end{equation}
where SB$_0$ is the normalization, $R_s$ the scale-length, and $\beta$ the exponent. In a first step, we fitted the Ly$\alpha$ profile of each AGN--$\alpha$ model using Equation~\ref{eq:lya_fit}. Since the profiles for various $\alpha$ angles are approximately self-similar for a given AGN SED model, we fixed the scale-length and the exponent to their mean values over all $\alpha$s (see values in Table~\ref{tab:Lya_fits}). Then, for each curve we computed the normalization as a simple average log(SB$_0$)=$\langle$log(SB$_{\rm data}$)+$\beta$(1+$R$/$R_s$)$\rangle$. These normalizations vary smoothly with the angle $\alpha$ following: 
\begin{equation}
 {\rm log}({\rm SB}_0) = {\rm log}({\rm SB}_1) + a*{\rm tanh}(\alpha/\alpha_s),
 \label{eq:lya_normfit}
\end{equation}
as shown in Figure~\ref{fig:lya_profile_norm}. The parameters of equations~\ref{eq:lya_fit} and \ref{eq:lya_normfit}, for all four AGN SED models, are given in Table~\ref{tab:Lya_fits}. 
In the literature, the Ly$\alpha$ profiles are usually fit with exponentials or power-laws \citep[e.g.][]{ArrigoniBattaia:2019,Farina:2019,Fossati:2021}. We have tried the already published
functional forms, but Equation~\ref{eq:lya_fit} is performing better for our simulated profiles, and it is quite a good approximation for the observational profile of Fossati et al. (dashed black curve in Figure~\ref{fig:fossati_fig7}). In section~\ref{subsec:HeII} we will discuss in more detail the simulated \ion{He}{ii} profiles, which are not as self-similar as the Ly$\alpha$ ones.     

A tension between observations and the simulated profiles is visible for the \ion{C}{iv} emission, with the latter having steeper profiles than the former. While the CGM metal enrichment of the simulation is compatible with absorption studies \citep[$Z_{\rm CGM}/Z_{\rm\odot}\geq 0.1$, e.g. ][and bottom panel of Figure~\ref{fig:radial_profiles}]{Prochaska:2009,Lau:2016}, the metallicity gradient is probably too steep in comparison with observations (right panel of Figure~\ref{fig:fossati_fig7}). This is a possible indication of a missing mechanism to push metals far away enough from the ISM. Thermal AGN feedback can produce such a strong enrichment of the CGM \citep[e.g.][]{Sanchez:2019,Nelson:2019}. One should also keep in mind that the details of metal diffusion in numerical experiments are far from being settled \citep[e.g.][]{Wadsley:2008}.  Also, as we discussed previously, this simulation does not trace explicitly the carbon enrichment. Running it with the new chemical enrichment implementation of \citet{Buck:2021} will remove this uncertainty. Lastly, the steeper simulated profiles could be also due to the fact that we do not take into account in our modeling the spatial diffusion due to resonant scattering. However, giving that the models reproduce the Ly$\alpha$ profile, which is expected to be even more affected by resonant scattering, indicates that this effect should be marginal for \ion{C}{iv}. Because of these uncertainties, we do not provide in this work fits to the \ion{C}{iv} profiles.

We can, however, quantitatively compare the simulated profiles with the observed ones, without any fitting, and constrain the ionization cone opening angle for each AGN model. To find the optimal angle $\hat{\alpha}$ we compute the posterior probability $P(\alpha|{\rm data})\propto P({\rm data}|\alpha)P(\alpha)$, where 'data' are, one at a time, the line emission stack profiles of \citet{Fossati:2021} for Ly$\alpha$, \ion{He}{ii}, and \ion{C}{iv}, and we assume a flat prior $P(\alpha)=1/180^{\rm o}$. The likelihood of the observational data given the model is then: 
\begin{equation}
 P({\rm data}|\alpha) = \Pi_i P({\rm log(SB_{i,obs})}|\mathcal{N}({\rm log}({\rm SB}_{i,{\rm model}}),\sigma_{i,{\rm model}})),
\end{equation}
where $\mathcal{N}$ is the normal distribution\footnote{The uncertainties of the points along the simulated profile are not truly independent, so assuming a normal distribution for the simulation is just an approximation.}, ${\rm SB}_{i,{\rm model}}$ the model prediction (simulation $\otimes$ AGN SED) at the impact parameter $R_i$ of the observed surface brightness ${\rm SB}_{i,{\rm obs}}$, and $\sigma_{i,{\rm model}}$ the corresponding dispersion of the model in logarithm. In Figure~\ref{fig:model_comparison_emission} we show the posterior probability distribution of $\alpha$ for the four AGN SED models (different colors), as constrained by the stacked emission profiles above the noise level for the Ly$\alpha$ (left), \ion{He}{ii} (center), and \ion{C}{iv} (right). 

Table~\ref{tab:alpha} gives $\hat{\alpha}$ with its confidence interval constrained for all combinations of SED and observed emission line profiles. Given the resonant scattering effects that can affect the Ly$\alpha$ and \ion{C}{iv} lines, in a best case scenario one should trust more the constraints provided by recombination lines like \ion{He}{ii}. However, \ion{He}{ii} has only been detected in some of the Ly$\alpha$ nebulae, and typically on much smaller scales. For this reason and also because of the uncertainties on the CGM metal enrichment in our simulation (which affect the predicted \ion{C}{iv} profiles) we decided to give separately the constraints for each observed emission line. In an ideal case, maybe feasible in the future, one would combine all the observational constraints into one single likelihood. All these caveats aside, one of the four SED models tested ($M_{\rm\bullet}$=10$^{\rm 9}$M$_{\rm\odot}$ and $\lambda=0.1$, highlighted in grey in the table) leads to compatible $\hat{\alpha}$ among all three emission lines, and, within their uncertainties, these values are less than the upper limit provided by the covering fraction of optically thick absorbers. Thus, for this best AGN model we can conclude that $\hat{\alpha}\sim$60$^{\rm o}$ is a value in good agreement with observational studies in emission, and it is also within the limits given by observations in absorption (see Section~\ref{sec:absorbtion}). E.g. \citet{Trainor:2013} used Ly$\alpha$ emitters in the vicinity of hyperluminous QSOs to constrain the activity period to 1~Myr$\leq t_{\rm Q}\leq$20~Myr and the ionization cone opening angle to $\alpha\geq$60$^{\rm o}$. Using a combination of absorption and emission measurements of the QPQ sample, \citet{Hennawi:2013} find that $\alpha\leq$90$^{\rm o}$. The constraints on $\alpha$ of both these works overlap in a narrow range with each other as well as with the older study of \citet{Fritz:2006}.

\newcolumntype{g}{>{\columncolor{Gray}}c}
\begin{table}
 \caption{Ionization cone opening angle constraints $\hat{\alpha}$[$^{\rm o}$]. The first three lines give $\hat{\alpha}$ in units of degrees estimated from fits to the stacked SB profiles of \citet{Fossati:2021}, while the last line gives the upper limit of $\alpha$ that agrees with the covering fraction of optically thick absorbers of \citet{Prochaska:2013}.} 
 \centering
 \label{tab:alpha}
 \begin{tabular}{l|c|g|c|c}
  \hline
& $M_{\rm\bullet}$=10$^{\rm 8}$M$_{\rm\odot}$ & $M_{\rm\bullet}$=10$^{\rm 9}$M$_{\rm\odot}$ & $M_{\rm\bullet}$=10$^{\rm 8}$M$_{\rm\odot}$ & $M_{\rm\bullet}$=10$^{\rm 9}$M$_{\rm\odot}$\\ 
Obs  & $\lambda=0.1$ & $\lambda=0.1$ & $\lambda=1.0$ & $\lambda=1.0$\\
  \hline
  ${\rm SB}$(Ly$\alpha$) & 125$^{\rm+37}_{\rm-37}$ & 54$^{\rm+25}_{\rm-21}$ & 80$^{\rm+38}_{\rm-46}$ & 41$^{\rm+25}_{\rm-23}$\\
  ${\rm SB}$(\ion{He}{ii}) & 76$^{\rm+44}_{\rm-48}$ & 67$^{\rm+40}_{\rm-41}$ & 65$^{\rm+42}_{\rm-38}$ & 62$^{\rm+40}_{\rm-37}$\\
  ${\rm SB}$(\ion{C}{iv}) & 157$^{\rm+23}_{\rm-27}$ & 92$^{\rm+37}_{\rm-35}$ & 168$^{\rm+12}_{\rm-27}$ & 163$^{\rm+17}_{\rm-38}$\\
  $f_{\rm C}$ & $<$117 & $<$72 & $<$71 & $<$57\\
  \hline
 \end{tabular}
\end{table}

\begin{table}
 \caption{Best fit parameters of Equations~\ref{eq:lya_fit} and \ref{eq:lya_normfit} applied to the simulated Ly$\alpha$ profiles as function of AGN SED model and ionization cone opening angle.
 The units of SB$_1$ are erg~s$^{\rm -1}$cm$^{\rm -2}$arcsec$^{\rm -2}$. The last three parameters were obtained with the \emph{emcee} package \citep{2013PASP..125..306F}, but only $\alpha_s$ has an asymmetric marginalized posterior distribution.}
 \centering
 \label{tab:Lya_fits}
 \begin{tabular}{l|c|c|c|c}
  \hline
  Param & $M_{\rm\bullet}$=10$^{\rm 8}$M$_{\rm\odot}$ & $M_{\rm\bullet}$=10$^{\rm 9}$M$_{\rm\odot}$ & $M_{\rm\bullet}$=10$^{\rm 8}$M$_{\rm\odot}$ & $M_{\rm\bullet}$=10$^{\rm 9}$M$_{\rm\odot}$\\ 
  & $\lambda=0.1$ & $\lambda=0.1$ & $\lambda=1.0$ & $\lambda=1.0$\\
  \hline
  $R_s$ [kpc] & 31.2$\pm$6.0 & 26.8$\pm$6.5 & 25.9$\pm$5.2 & 23.8$\pm$7.2\\
  $\beta$ & 5.5$\pm$0.5 & 5.5$\pm$0.6 & 5.5$\pm$0.5 & 5.8$\pm$0.8\\
  log(SB$_1$) & -16.81$\pm$0.04 & -16.58$\pm$0.06 & -16.51$\pm$0.07 & -16.20$\pm$0.11\\
  $a$ & 0.99$\pm$0.05 & 1.43$\pm$0.06 & 1.09$\pm$0.07 & 1.41$\pm$0.10\\
  $\alpha_s$ [$^{\rm o}$] & 89.0$^{\rm+12.7}_{\rm-13.2}$ & 80.4$^{\rm+9.1}_{\rm-9.6}$ & 85.2$^{\rm+13.0}_{\rm-24.5}$ & 81.0$^{\rm+14.4}_{\rm-19.7}$\\
  \hline
 \end{tabular}
\end{table}

\begin{figure}
 \centering
 \includegraphics[width=0.45\textwidth]{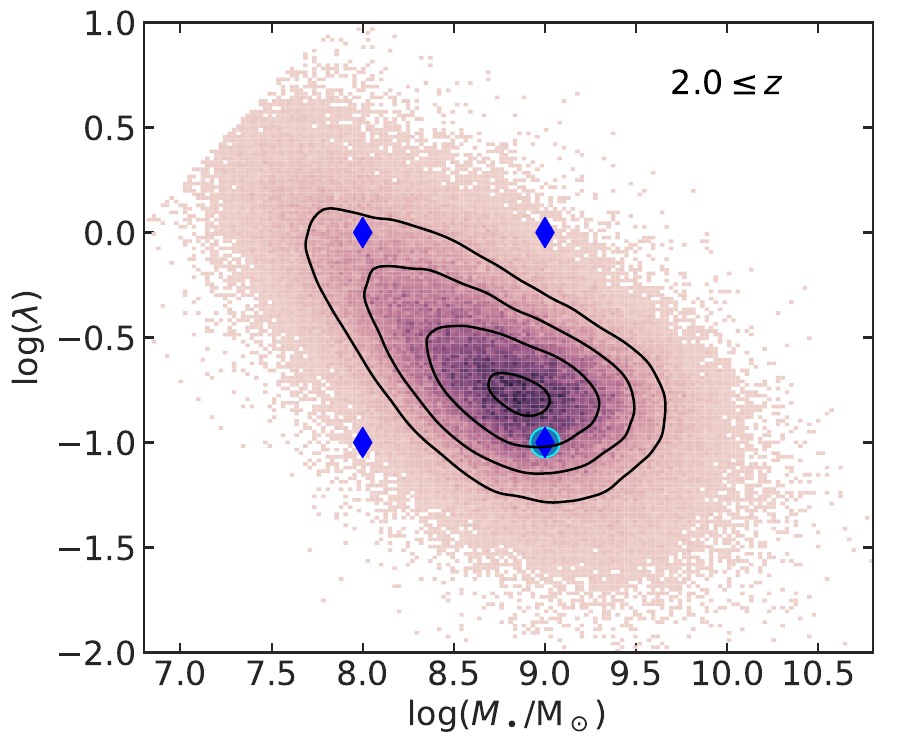}
 \caption{Distribution of high-$z$ ($z\geq2.0$) SDSS quasars \citep{Rakshit:2020} in the parameter space of black hole mass and accretion rate. The black contours are kernel density estimates for the 20, 40, 68 and 95 percentiles, while the four black diamonds are the positions of the AGN SED models we tested in this work. The model favored by our analysis is highlighted in cyan.}
 \label{fig:sdss}
\end{figure}

It is important to mention that using only the AGN continuum, that is the SED without the contribution from the stacked line emission spectrum of \citet{Lusso:2015}, leads to lower SB levels in  Ly$\alpha$ and \ion{C}{iv}. In its turn, this implies that larger ionization cone opening angles are required to match the observations, but these larger $\alpha$s are in tension with absorption studies. Figure~\ref{fig:nolines_Lya} in Appendix~\ref{appendix:nolines} illustrates the differences between SB$_{\rm Ly\alpha}$ for the AGN SED with and without the stacked emission lines. Thus, we can conclude that resonant pumping of the AGN spectrum is essential to explain the observed SB levels in the resonant lines, and their ratios with the HeII recombination line, similar to the conclusion reached by \citet{Fossati:2021}. 

To conclude this section we also look at how representative are the four AGN SED models for observed quasars. Figure~\ref{fig:sdss} gives the distribution of SDSS quasars from the catalog published by \citet{Rakshit:2020} in the parameter space log($M_{\rm\bullet}$)--log($\lambda$). To construct the distribution we use only the quasars with the best quality flag (0), and restrict the redshift to $z\geq2.0$, which is the range that covers the observational samples of \citet{Lau:2016} and \citet{Fossati:2021}. These criteria restrict the observational quasar sample to 191073 objects. The four AGN SED models we tested in this work are marked with blue diamonds, and we can notice that our preferred model (log($M_{\rm\bullet}$/M$_{\rm\odot}$)=9.0 and log($\lambda$)=-1.0) is close to the 2D peak of the SDSS distribution, meaning that it is also the most representative model for observed quasars at $z\geq2$.

\begin{figure*}
 \centering
 \includegraphics[width=0.33\textwidth]{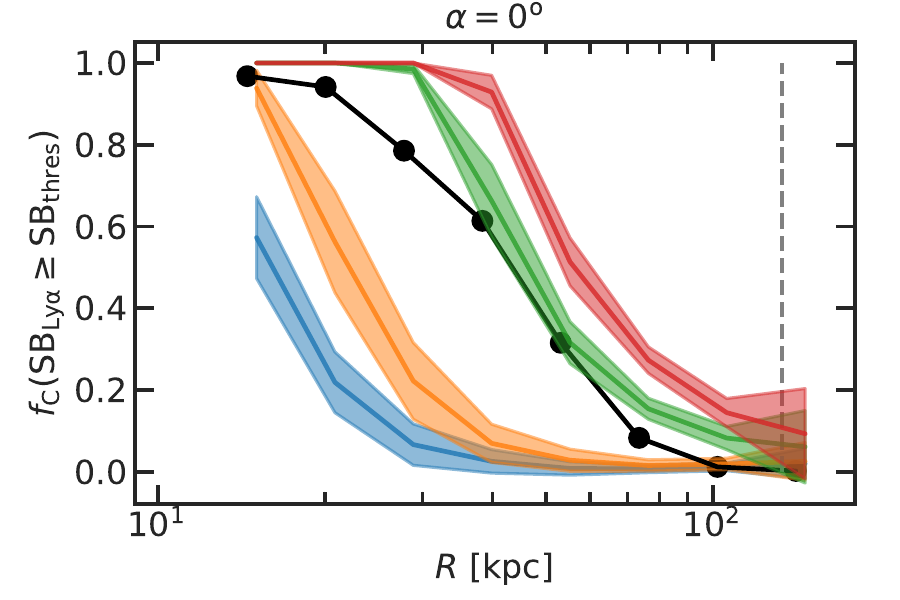}
  \includegraphics[width=0.33\textwidth]{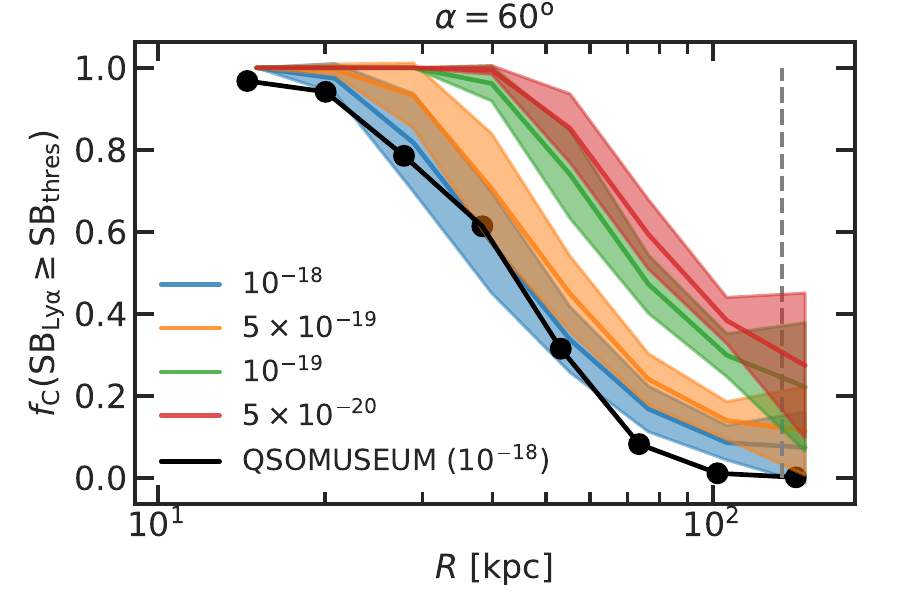}
  \includegraphics[width=0.33\textwidth]{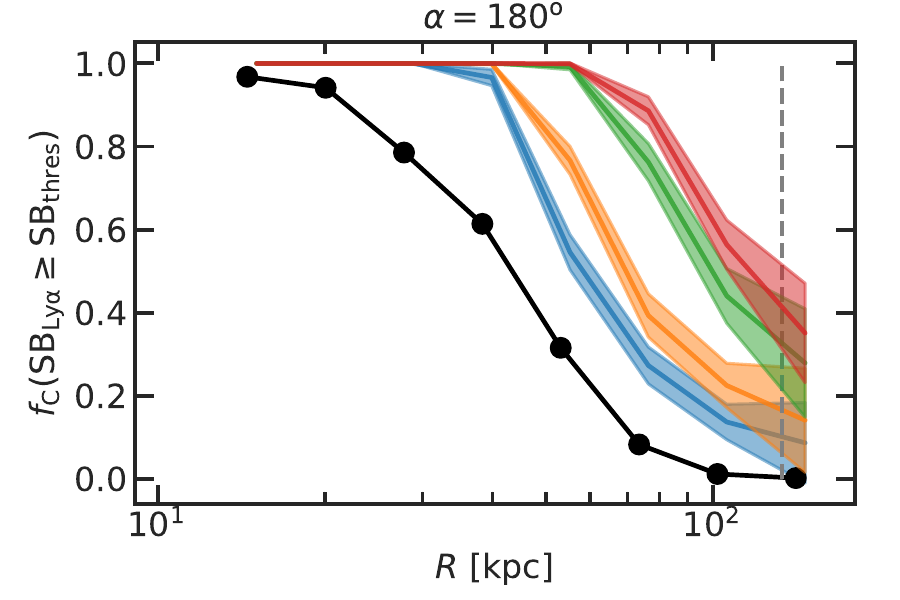}
 \caption{Covering fraction of Ly$\alpha$ emission as function of projected radius for various SB thresholds (indicated in the legend in units of erg~s$^{\rm -1}$~cm$^{\rm -2}$~arcsec$^{\rm -2}$) for our preferred AGN SED model ($M_{\rm\bullet}$=10$^{\rm 9}$M$_{\rm\odot}$ and $\lambda=0.1$) and three ionization cone opening angles: $\alpha=0^{\rm o}$ (left), $\alpha=60^{\rm o}$ (center), and $\alpha=180^{\rm o}$ (right). Colored curves in all three panels are the predictions from the simulation, while the black one is the average covering fraction profile for the QSO MUSEUM sample with a threshold of 10$^{\rm -18}$~erg~s$^{\rm -1}$~cm$^{\rm -2}$~arcsec$^{\rm -2}$ \citep{ArrigoniBattaia:2019}. The vertical grey dashed line in all panels marks the virial radius $r_{\rm 200}$.}
 \label{fig:Lya_coverf}
\end{figure*}

\begin{figure*}
    \centering
    \includegraphics[width=0.40\textwidth]{figures/emission_profiles_cbar.png}\\
    \includegraphics[width=0.33\textwidth]{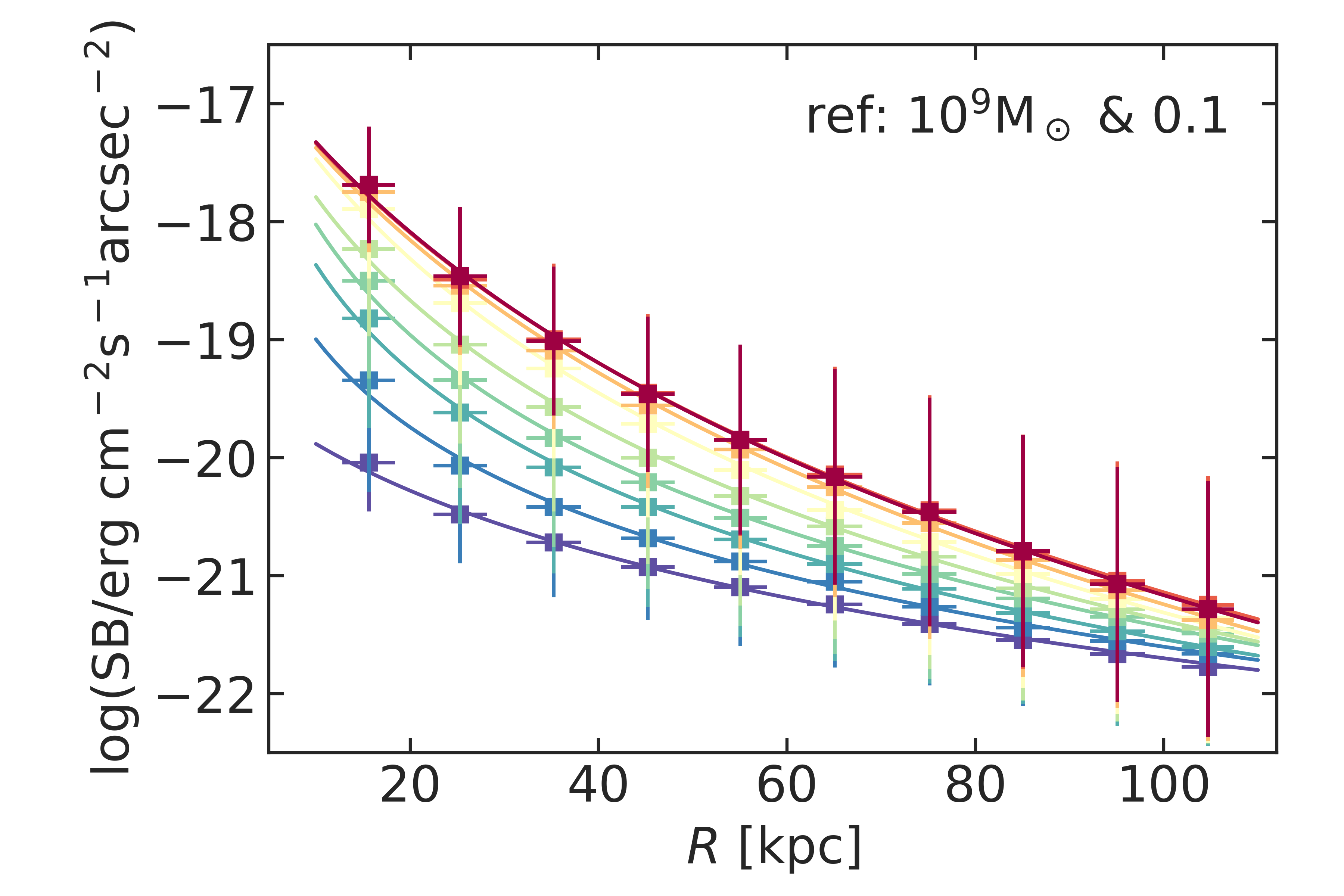}
    \includegraphics[width=0.33\textwidth]{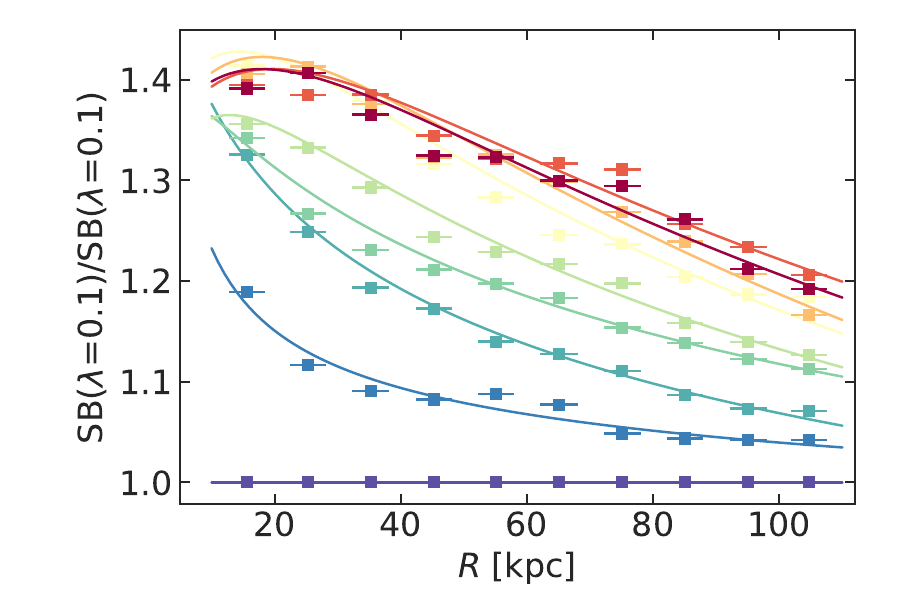}
    \includegraphics[width=0.33\textwidth]{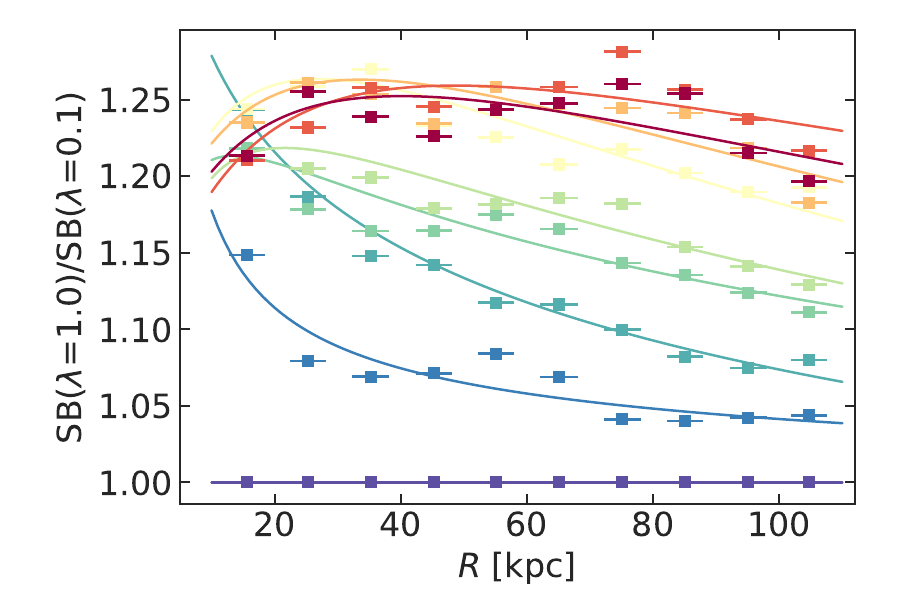}
    \caption{\textbf{Left}: predicted SB profiles for \ion{He}{ii}1640\AA~for our reference AGN SED model $M_{\rm\bullet}$=10$^{\rm 9}$M$_{\rm\odot}$ and $\lambda=0.1$ (colored points). \textbf{Center}: ratios of \ion{He}{ii} profiles at fixed accretion rate $\lambda$=0.1. \textbf{Right}: ratios of \ion{He}{ii} profiles at fixed black hole mass $M_{\rm\bullet}$=10$^{\rm 9}$M$_{\rm\odot}$. In all three panels, colored curves represent the best ML fits using Equation~\ref{eq:lya_fit} (left), or ratios of such fits (center and right). For easier visualisation, we do not show the uncertainties in SB ratios on the individual points.}    
    \label{fig:HeII_profile} 
\end{figure*}

\begin{table*}
 \caption{Maximum likelihood fit parameters of Equation~\ref{eq:lya_fit} applied to the simulated \ion{He}{ii} profiles as function of AGN SED model and ionization cone opening angle.
 The units of SB$_0$ are erg~s$^{\rm -1}$~cm$^{\rm -2}$~arcsec$^{\rm -2}$.}
 \centering
 \label{tab:HeII_fits}
\begin{tabular}{c|c|ccccccccc}
  \hline
Model & Param $\backslash$ $\alpha$ & 0$^{\rm o}$ & 15$^{\rm o}$ & 30$^{\rm o}$ & 45$^{\rm o}$ & 60$^{\rm o}$ & 90$^{\rm o}$ & 120$^{\rm o}$ & 150$^{\rm o}$ & 180$^{\rm o}$ \\
\hline
& log(SB$_0$) & -19.29 & -16.99 & -15.94 & -15.77 & -16.29 & -16.30 & -16.41 & -16.42 & -16.42 \\
$M_{\rm\bullet}$=10$^{\rm 8}$M$_{\rm\odot}$ & $R_s$ [kpc] & 15.82 & 2.07 & 2.01 & 2.92 & 7.58 & 13.64 & 20.12 & 21.65 & 22.34 \\
$\lambda$=0.1 & $\beta$ & 2.79 & 2.73 & 3.30 & 3.69 & 4.46 & 5.52 & 6.32 & 6.41 & 6.54 \\
& $rmsd$ & 0.07 & 0.06 & 0.05 & 0.05 & 0.05 & 0.06 & 0.07 & 0.08 & 0.08 \\
\hline
& log(SB$_0$) & -19.29 & -16.64 & -15.75 & -15.68 & -16.22 & -16.19 & -16.30 & -16.31 & -16.30 \\
$M_{\rm\bullet}$=10$^{\rm 9}$M$_{\rm\odot}$ & $R_s$ [kpc] & 15.82 & 1.64 & 2.09 & 3.22 & 8.60 & 15.26 & 22.54 & 23.91 & 24.59 \\
$\lambda$=0.1 & $\beta$ & 2.79 & 2.77 & 3.42 & 3.82 & 4.68 & 5.83 & 6.72 & 6.76 & 6.90 \\
& $rmsd$ & 0.07 & 0.06 & 0.04 & 0.05 & 0.05 & 0.06 & 0.07 & 0.08 & 0.07 \\
\hline
& log(SB$_0$) & -19.29 & -16.52 & -15.71 & -15.70 & -16.22 & -16.17 & -16.28 & -16.29 & -16.28 \\
$M_{\rm\bullet}$=10$^{\rm 8}$M$_{\rm\odot}$ & $R_s$ [kpc] & 15.82 & 1.54 & 2.18 & 3.49 & 9.20 & 16.22 & 23.93 & 25.33 & 26.00 \\
$\lambda$=1.0 & $\beta$ & 2.79 & 2.79 & 3.48 & 3.88 & 4.77 & 5.97 & 6.89 & 6.91 & 7.06 \\
& $rmsd$ & 0.07 & 0.06 & 0.04 & 0.05 & 0.05 & 0.05 & 0.07 & 0.07 & 0.07 \\
\hline
& log(SB$_0$) & -19.29 & -16.27 & -15.58 & -15.68 & -16.20 & -16.15 & -16.25 & -16.27 & -16.25 \\
$M_{\rm\bullet}$=10$^{\rm 9}$M$_{\rm\odot}$ & $R_s$ [kpc] & 15.82 & 1.25 & 2.07 & 3.57 & 9.39 & 16.58 & 24.17 & 25.58 & 26.19 \\
$\lambda$=1.0 & $\beta$ & 2.79 & 2.79 & 3.50 & 3.90 & 4.81 & 6.01 & 6.92 & 6.92 & 7.07 \\
& $rmsd$ & 0.07 & 0.06 & 0.04 & 0.05 & 0.05 & 0.05 & 0.06 & 0.07 & 0.06 \\
\hline
\end{tabular}
\end{table*}

\begin{figure*}
    \centering
    \includegraphics[width=0.95\textwidth]{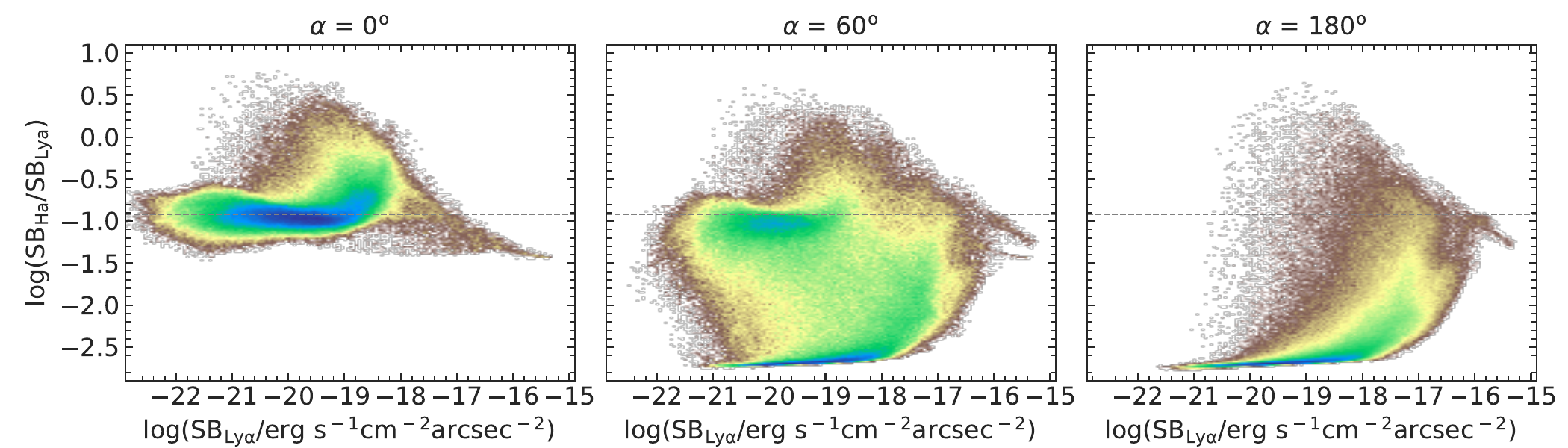}\\
    \includegraphics[width=0.95\textwidth]{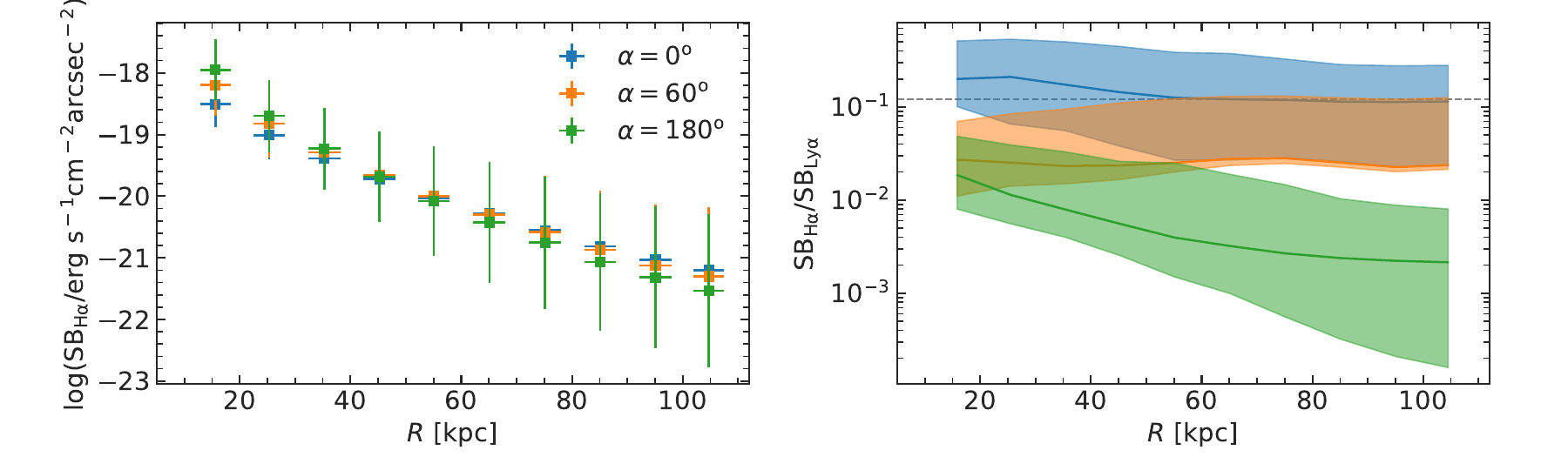}\\
    \caption{\textbf{Top}: 2D distributions of 0.2$\times$0.2~arcsec$^{\rm2}$ CGM (20$<r<$134~kpc) pixels in the space of Ly$\alpha$ surface brightness and SB ratios of H$\alpha$ to Ly$\alpha$ for three different ionization cone opening angles (left to right) for our preferred AGN SED model ($M_{\rm\bullet}$=10$^{\rm 9}$M$_{\rm\odot}$ \& $\lambda=0.1$). 
    In each panel, the horizontal dashed grey line gives the ratio value for Case B recombination SB$_{\rm H\alpha}$/SB$_{\rm Ly\alpha}$=0.12 for $T$=10$^{\rm 4}$K and typical CGM metallicities of 0.5$Z_{\rm\odot}$. At fixed angle $\alpha$ we collapsed together all the pixels of 27 images, representing 27 random orientations of the ionizing cone, and implicitly the LOS. \textbf{Bottom}: average H$\alpha$ SB profile (left), and median SB$_{\rm H\alpha}$/SB$_{\rm Ly\alpha}$ profiles (right) for the same AGN SED model and different $\alpha$ angles. The shaded areas in the right panel enclose the regions between the 16$^{\rm th}$ and 84$^{\rm th}$ percentiles of ratios, while the horizontal dashed grey line gives the same Case B recombination ratio as in the top panels.}    
    \label{fig:Ha_predictions} 
\end{figure*}

\section{Discussion}
\label{sec:discussion}

\subsection{Predictions on the covering fraction of Ly$\alpha$ emitting gas around quasars}
\label{subsec:LyaCovF}

In Section~\ref{sec:emission} we used the observed emission profiles to constrain the central quasar properties and opening angle, favoring an AGN model with $M_{\rm \bullet}=10^9$~M$_\odot$, $\lambda=0.1$ and $\hat{\alpha}\approx60$ degrees. Here, we compute the covering fraction of the Ly$\alpha$ emitting gas and compare it to observational estimates. \citet{ArrigoniBattaia:2019} computed the average covering factor profile $f_{\rm C}$($R$) of each Ly$\alpha$ nebula in their sample of 61 $z\sim3$ quasars' fields in logarithmic radial bins, defining $f_{\rm C}$  as the ratio between the area occupied by Ly$\alpha$ emission above a SB threshold of $\approx10^{-18}$~erg~s$^{-1}$~cm$^{-2}$~arcsec$^{-2}$ (S/N$=2$ in those observations) and the total area of each radial bin. Assuming equal weights for the profile of each individual target, they derived an average covering factor profile. We apply an identical procedure to the same 27 random orientations constructed for the simulated Ly$\alpha$ SB profile in Section~\ref{sec:emission} assuming an AGN SED with $M_{\rm \bullet}=10^9$~M$_\odot$, $\lambda=0.1$,  and repeat the calculation for three critical opening angles $0^{\rm o}$, $60^{\rm o}$, $180^{\rm o}$, corresponding to no quasar ionization, the favored opening angle, and isotropic AGN illumination, respectively.

Figure~\ref{fig:Lya_coverf} shows the results of this exercise. The predicted covering factor profile for Ly$\alpha$ emission above $10^{-18}$~erg~s$^{-1}$~cm$^{-2}$~arcsec$^{-2}$ (blue) is much lower, in agreement, and much higher with respect to the observations (black) for an opening angle of $0^{\rm o}$ (left panel), $60^{\rm o}$ (central panel), and $180^{\rm o}$ (right panel), respectively. The agreement between the observed profile and the simulation for an opening angle of 60$^{\rm o}$ further confirms the analysis in Section~\ref{sec:emission}, i.e. the quasars illuminate on average a relatively small volume fraction of their halos ($2\Omega/4\pi=sin^2(\alpha/4)\approx0.13$), and dominate the powering of the Ly$\alpha$ emission. 

By construction, the covering factor profile depends on the depth of the observational data. It is therefore interesting to make predictions on how such a profile change for more sensitive data sets. We provide forecasts for the average covering factor of Ly$\alpha$ emitting gas as a function of SB threshold. Specifically, Figure~\ref{fig:Lya_coverf} shows how the covering factor profile varies by lowering the Ly$\alpha$ SB threshold to $5\times10^{-19}$ (orange), $10^{-19}$ (green), $5\times10^{-20}$ erg~s$^{-1}$~cm$^{-2}$~arcsec$^{-2}$ (red). Even ultra-deep observations, reaching SB levels of $5\times10^{-20}$ erg~s$^{-1}$~cm$^{-2}$~arcsec$^{-2}$ (e.g., \citealt{Bacon:2021}), would result in at most an average $f_{\rm C}\sim0.4$ close to the expected virial radius of the DM halo (vertical dashed line). Importantly, these predictions take into account only the radiation from the central quasar. Additional sources would contribute to the extended Ly$\alpha$ emission resulting in higher covering factors at larger radii (e.g., see Figure 7 in \citealt{ArrigoniBattaia:2019}).

\subsection{QSO properties from \ion{He}{ii} extended emission}
\label{subsec:HeII}

The importance of detecting \ion{He}{ii} emission in extended quasars' Ly$\alpha$ nebulae and extended Ly$\alpha$ nebulae in general has been always emphasized since their discovery (e.g., \citealt{Heckman:1991b,Villar-Martin:2003,Yang:2006,Prescott:2015,ArrigoniBattaia:2015b,Herenz:2020}). \ion{He}{ii} is a non-resonant line and it is conveniently located in the observed optical range for high-redshift sources. Because of this, \ion{He}{ii} can be used to: assess how strong are the resonant scattering effects on the Ly$\alpha$ line, constrain the ionization state of the gas and hardness of the impinging ionization spectrum, and study the gas kinematics. In this work, we have shown that \ion{He}{ii} emission is the cleaner tracer of the AGN ionization effects on the surrounding CGM, and that together with other emission and absorbtion lines, it can provide estimates for the central AGN's parameters (SMBH mass, Eddington ratio, and ionization cone opening angle). This first estimate showcases the potential in the use of CGM observables to measure the properties of the central engine and could provide independent and complementary constraints on AGN properties (with respect to the usual methods using quasars emission lines, e.g., \citealt{Farina:2022}), especially at the earliest epochs where these values are most needed to understand the rapid formation of SMBHs (\citealt{Review_FBS:2022} and references therein). As the \ion{He}{ii} emission is very faint, and hard to detect in individual objects \citep[e.g.][]{Guo:2020,Fossati:2021}, we provide fits to its SB profile as predicted by our modeling, and quantify how much better we could do with current instruments.

Figure~\ref{fig:HeII_profile} shows the simulated \ion{He}{ii} profile for the favored AGN SED model, this time together with the maximum likelihood fits using Equation~\ref{eq:lya_fit}. Contrary to the Ly$\alpha$ profiles, the \ion{He}{ii} requires varying all three parameters (normalization, scale-length and exponent). Table~\ref{tab:HeII_fits} gives all these parameters for all four AGN SED models, and the root-mean-squared deviation $rmsd$ of the simulated data points from the fitted function, to quantify the goodness of fit. 

The center and right panels of Figure~\ref{fig:HeII_profile} show the profiles of SB ratios at fixed accretion rate ($\lambda$=0.1, center) to gauge the effect of changing the black hole mass, and the profiles of SB ratios at fixed black hole mass ($M_{\rm\bullet}$=10$^{\rm 9}$M$_{\rm\odot}$, right) to gauge the effect of changing the accretion rate. We can notice that changing $\lambda$ at fixed $M_{\rm\bullet}$ has only a mild effect on the SB$_{\rm\ion{He}{ii}}$ with an increase of 20--25\% when increasing $\lambda$ from 0.1 to 1.0 for the larger ionization cone opening angles ($\alpha\geq90^{\rm o}$ right panel of´ Figure~\ref{fig:HeII_profile}). Increasing $M_{\rm\bullet}$ at fixed $\lambda$ by an order of magnitude has a larger effect on SB$_{\rm\ion{He}{ii}}$ (center panel). The SB ratios for reasonable ionization cone opening angles ($\alpha\sim60^{\rm o}$) are very sensitive to the projected radius at which they are measured. All this, together with the comparison between our simulated profiles and the stacked data of \citet{Fossati:2021}, proves that we need more sensitive observations in order to use CGM data for constraining firmly the AGN properties. For example, to push the observations to a 2$\sigma$ (3$\sigma$) significance at the SB level of $10^{-20}$~erg~s$^{-1}$~cm$^{-2}$~arcsec$^{-2}$ for the \ion{He}{ii} transition, one would need to target additional 315 (756) objects to the sample of \citet{Fossati:2021} with $\sim4$~hours MUSE observations, following their quoted sensitivity and assuming their redshift distribution\footnote{This experiment could be eased by targeting sources with \ion{He}{ii} in wavelength ranges with less contamination from sky emission.}. Such a large sample of sources should also be able to unveil the targets with the brightest \ion{He}{ii} nebulae (e.g., \citealt{Guo:2020}) which could then be used to study the kinematics of the quasar's CGM (e.g., \citealt{Zhang:2023}) and obtain estimates for individual quasar properties (e.g., ionization cone opening angle).

\subsection{Constraining the Ly$\alpha$ emission processes with H$\alpha$ glows}

Observing the H$\alpha \lambda$6563\AA\ transition from the quasar's CGM can provide similar information as \ion{He}{ii}. In particular, H$\alpha$ emission should be mainly produced through recombination and should not be subject to resonant scattering effects under most astrophysical conditions. For this reason, it can be used to pin down the powering mechanism of Ly$\alpha$ emission, constrain the gas physical properties (e.g., density) and the gas kinematics. Unlike \ion{He}{ii}, however, the detection of H$\alpha$ emission is not possible from the ground for $z \gtrsim 2.5$ due to the Earth atmosphere (e.g., \citealt{Hayes:2019}), and space telescopes such as the James Webb Space Telescope (JWST, \citealt{Gardner:2006}) are needed to overcome this obstacle. To our knowledge, there are yet no published constraints on H$\alpha$ emission for the CGM (on scales $>20$~kpc as in this study) of $z\sim3$ quasars. Therefore, we provide predictions for the average H$\alpha$ SB levels expected in the model favored by our analysis, i.e. a CGM illuminated by an AGN SED with $M_{\rm \bullet}=10^9$~M$_\odot$ and $\lambda=0.1$. 

We construct H$\alpha$ SB maps as done for the other transitions and assuming three different opening angles $\alpha= 0^{\rm o}$, $60^{\rm o}$, and $180^{\rm o}$. As for the other emission lines, we pick 27 random directions of the ionization cones and for each of those a random line of sight within the cone. Importantly, these directions are the same as those used for the construction of the Ly$\alpha$ SB maps, so that we can now compare the emission levels of the two transitions. First, we focus on the distribution of  SB$_{\rm H\alpha}$/SB$_{\rm Ly\alpha}$ ratios as a function of SB$_{\rm Ly\alpha}$ on CGM scales (radii $R>20$~kpc). The top panels in Figure~\ref{fig:Ha_predictions} show the distribution of the ratio computed for CGM regions (pixels with area of $0.2\times0.2$~arcsec$^2$) for the three $\alpha$s, 0$^{\rm o}$ (left), $60^{\rm o}$ (center), and $180^{\rm o}$ (right). For reference, we indicate the Case B recombination value SB$_{\rm H\alpha}$/SB$_{\rm Ly\alpha}=0.12$ for $T=10^4$~K and a typical CGM metallicity 0.5~Z$_{\odot}$ computed with {\sc cloudy}\footnote{We tested all case B implementations available in {\sc cloudy} (\citealt{HS:1987,SH:1995,PS:1964}) and found no variation of the SB$_{\rm H\alpha}$/SB$_{\rm Ly\alpha}=0.12$ for $T=10^4$~K and densities in the range 
$ -2\leq {\rm log(}n_{\rm H}{\rm /cm}^{\rm -3}{\rm)} \leq 2$.}. This approximated regime occurs when the gas is optically thick to Lyman-line photons, which are expected to undergo several scatterings and be converted into Balmer line photons and Ly$\alpha$ or two-photon emission (e.g., \citealt{Ferland:1999}). 

We find that for all angles $\alpha$ the predicted SB$_{\rm H\alpha}$/SB$_{\rm Ly\alpha}$ ratios do not follow uniquely the frequently invoked Case B value. For $\alpha=0^{\rm o}$, i.e. the UVB-only case, most of the CGM follows roughly case B, but there are regions where the ratio is lower and much higher. The regions with a lower ratio are in between case B and case A (according to {\sc cloudy} at $T=10^4$~K case A gives log(SB$_{\rm H\alpha}$/SB$_{\rm Ly\alpha})=-1.8$), while the regions with higher ratios are patches where the local absorption correction is the strongest (Figure~\ref{fig:NH_and_tau_lya}). Introducing the ionization cones of the quasar as described in Section~\ref{sec:geometry} widens the distribution of ratios towards lower values because continuum pumping increases the Ly$\alpha$ SB. For the extreme case where the AGN shines on all the CGM most of the regions have values SB$_{\rm H\alpha}$/SB$_{\rm Ly\alpha}\lesssim$0.01 (top right panel of Figure~\ref{fig:Ha_predictions}). Removing the quasar's emission lines from the input spectrum (Section~\ref{sec:geometry}) would result in higher ratios as the continuum pumping of the resonant lines would be less significant (Figure~\ref{fig:Ha_predictions_woLusso} in Appendix~\ref{appendix:nolines}). 

The bottom panels of Figure~\ref{fig:Ha_predictions} give the predicted average H$\alpha$ SB (left) and the median SB$_{\rm H\alpha}$/SB$_{\rm Ly\alpha}$ (right) as functions of radius for the three $\alpha$ angles. The shaded areas in the bottom right panel mark the 16$^{\rm th}$ to 84$^{\rm th}$ ratio percentiles. Overall, considering the full distribution of allowed ratios for $\alpha=60^{\rm o}$ and the fact that small opening angles could be in place (Table~\ref{tab:alpha}), our models suggest that the easiest H$\alpha$ detections would be those corresponding to SB$_{\rm H\alpha}$/SB$_{\rm Ly\alpha}\sim0.1$, close to a case B recombination scenario. However, statistical samples of observations should result in lower ratios.

With the advent of the JWST Near Infrared Spectrograph (NIRSpec) observations can test our models and hence obtain firmer constraints on the physical properties of the $z\sim3$ quasars' CGM and on the physical properties of the quasars themselves (i.e., $M_{\rm \bullet}$, $\lambda$, and opening angle). While observations of individual systems are necessary as pilot studies, large samples of sources are needed to really pin down the physics of the cool CGM. First results on this front are expected for quasars at the highest redshifts, $z\sim6$, with known Ly$\alpha$ nebulae (e.g., \citealt{Farina:2019}) which have been targeted with JWST observations during cycle 1, and will also be targeted during cycle 2.

At lower redshifts, for which both Ly$\alpha$ and H$\alpha$ are accessible from the ground, a recent work targeted a $z\sim2.26$ quasar and reported a detection for both transitions (\citealt{Langen:2023}). These authors used the IFU KCWI and a longslit spectrum with the MOSFIRE instrument on the Keck telescope to target Ly$\alpha$ and H$\alpha$, respectively. A MOSFIRE $1\arcsec$ slit has been placed on the brightest portion of the Ly$\alpha$ nebula, which is also the brightest in their sample. The H$\alpha$ detection extends for $2.5\arcsec$ (or $\sim20$~kpc) on one side of the quasar. After extracting a pseudoslit from the KCWI data corresponding to the MOSFIRE one, they report a ratio of SB$_{\rm H\alpha}$/SB$_{\rm Ly\alpha}=0.27$, in a region of $1\arcsec\times5.5\arcsec$. If this value would be present throughout the whole Ly$\alpha$ nebula, it would be in tension with our predictions. However, we find surprising their quoted value of the Ly$\alpha$ flux $F_{\rm Ly\alpha} = 3.5\times10^{-17}$~erg~s$^{-1}$~cm$^{-2}$ within the constructed pseudoslit, because it corresponds to only 7\% of the total Ly$\alpha$ flux of the nebula. This flux clearly does not agree with the SB map presented in their Figure~2: only the $1\arcsec \times 1\arcsec$ region at the high SB peak should already have at least a factor of two higher flux than what they report for the slit. A rough estimate of the flux within the drawn pseudoslit on the same image suggests instead that the Ly$\alpha$ flux in that $1\arcsec\times5.5\arcsec$ region should be $\sim20$\% of the total. This would make the ratio SB$_{\rm H\alpha}$/SB$_{\rm Ly\alpha}\sim0.09$, a value more consistent with our modeling. We stress that $z\sim2$ quasar's Ly$\alpha$ nebulae are known to be on average dimmer than those at $z\sim3$ (e.g., \citealt{Cai:2019}), so the conditions in the CGM of quasars at different redshifts could be different (\citealt{ArrigoniBattaia:2019}). 

Once again, the discussion of this example showcases how critical it is to acquire a large statistical sample of observations targeting the H$\alpha$ transition from the high-redshift quasar's CGM. Importantly, we need observations of the entirety of the Ly$\alpha$ nebulae to constrain radiative transfer effects well. Data achieving SB levels of $\sim10^{-19}$~erg~s$^{-1}$~cm$^{-2}$~arcsec$^{-2}$ should be able to detect on average H$\alpha$ emission out to $\sim30$~kpc (Figure~\ref{fig:Ha_predictions}), if the densities in the simulation are a good representation of reality (Section~\ref{sec:caveat_densities}).

\subsection{Density distributions of the cool CGM emitting Ly$\alpha$}
\label{sec:caveat_densities}

The SB levels of quasars' Ly$\alpha$ nebulae require the emitting gas to have high densities ($\gtrsim 1$~cm$^{-3}$; e.g., \citealt{Heckman:1991,Cantalupo:2014,ArrigoniBattaia:2015,Hennawi:2015}), or very broad density distributions (e.g., \citealt{Langen:2023} and references therein) for a pure recombination scenario. Testing for the presence of such CGM densities can be done in two ways: via sub-grid prescriptions for the unresolved dense cool CGM gas in cosmological simulations (e.g., \citealt{Hummels:2019}), or via very high-resolution simulations that can trace the survival of such dense cool structures in a hot medium (e.g., \citealt{McCourt:2018,Gronke:2022}). On the other hand, if quasar's photons contribute to the emission in resonant lines (Ly$\alpha$, \ion{C}{iv}), the required densities would be lowered. Using a suite of high-resolution (maximum spatial resolution $\sim$80~pc), radiation-hydrodynamic cosmological simulations targeting $z\sim6$ quasar host halos and post-processed with a Ly$\alpha$ radiative transfer code, \citet{Costa:2022} was able to match the observed Ly$\alpha$ SB profile. Importantly, this work showed that quasar's photons can efficiently scatter in the quasar's CGM indicating that extreme densities may not be necessary to match observations. Motivated by this work, in our analysis we used a quasar spectrum including the emission lines to allow continuum pumping to be effective (Section~\ref{sec:geometry}), and we show that the predicted observables (column densities and SB profiles) match current observations. Therefore, it is interesting to look at what kind of densities could be recovered from the mock observations in comparison to the ground truth of the simulation. 

\begin{figure}
    \centering
    \includegraphics[width=0.48\textwidth]{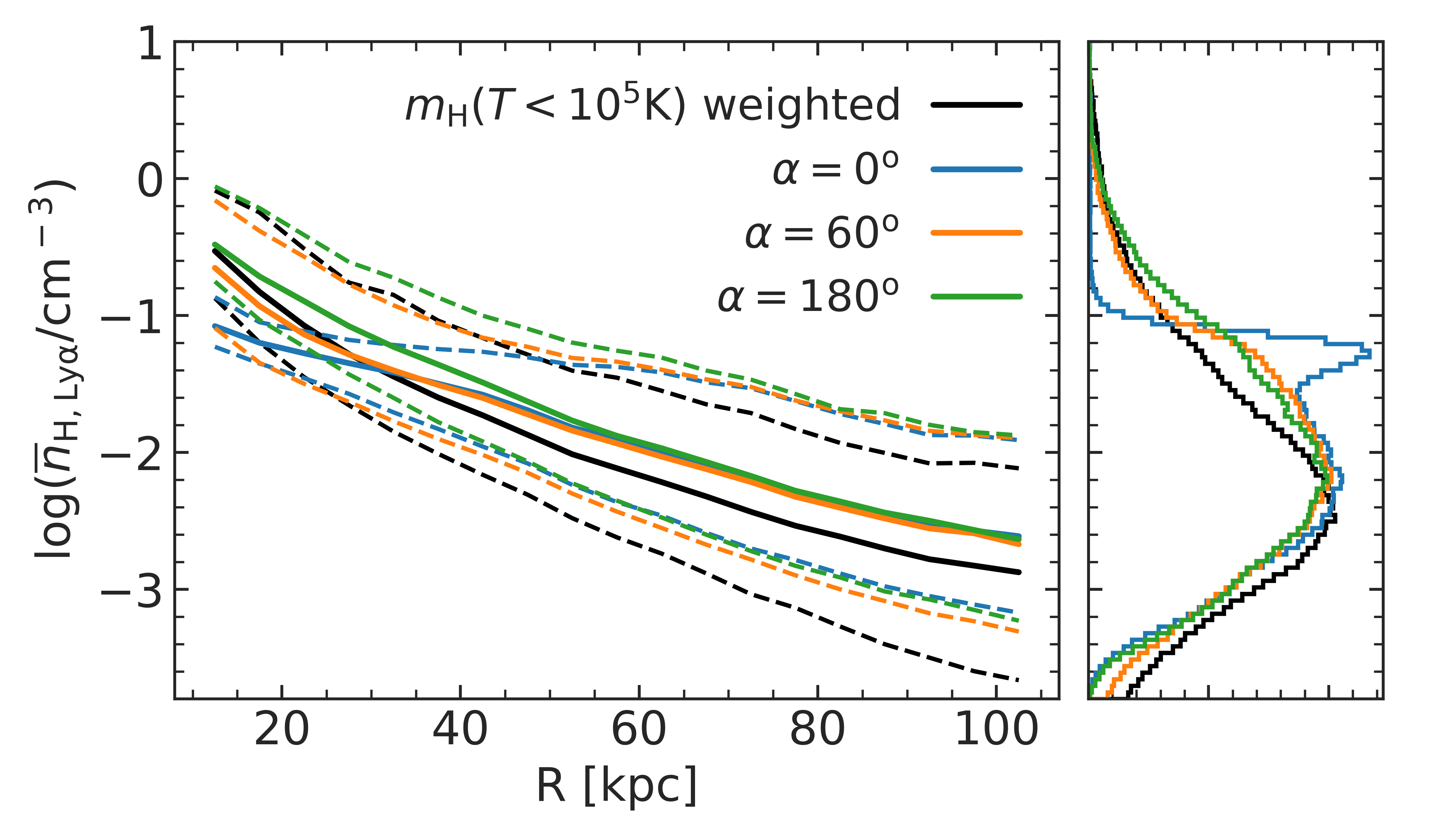}
    \caption{Median (solid curves), and 16$^{\rm th}$ (lower dashed curves) and 84$^{\rm th}$ (upper dashed curves) volume number density of hydrogen weighted with the Ly$\alpha$ luminosity of each particle (colored curves), or with the hydrogen mass of cool ($T<$10$^{\rm 5}$K) particles (black curves), as functions of projected distance from the center (left panel). The smaller right panel shows the corresponding $n_{\rm H}$ distributions over the full projected radial range of the left panel. The AGN SED considered is the one for our preferred model, and the three colors represent three different ionization cone opening angles.}
    \label{fig:Lya_weigthed_densities}
\end{figure}

We constructed the hydrogen number densities radial distributions in Figure~\ref{fig:Lya_weigthed_densities} by creating 27 random orientation maps (as done for all the results in Section~\ref{sec:emission}) with the line-of-sight mass (black curve) or Ly$\alpha$ luminosity (colored curves) weighted $n_{\rm H}$. For the mass-weighted $n_{\rm H}$ we used only the cool gas particles ($T<$10$^{\rm 5}$K). The estimates for $\overline{n}_{\rm H,Ly\alpha}$ should mimic hydrogen number densities that could be extracted from observations. While most of the CGM emitting Ly$\alpha$ is characterized by densities $\sim10^{-2}$~cm$^{-3}$, the distributions in Figure~\ref{fig:Lya_weigthed_densities} show tails at high densities, up to $\sim10$~cm$^{-2}$ for the models with an AGN. These high densities are responsible for the brightest Ly$\alpha$ emission in the inner CGM. On the contrary, in the model including only the UVB as photoionization source there is almost no contribution for $n_{\rm H}>0.1$~~cm$^{-3}$ as those regions are mostly neutral with such a faint impinging continuum. The $\overline{n}_{\rm H,Ly\alpha}$ distributions for the two cases where the AGN is on follow better the intrinsic cool gas mass-weighted density distribution than the UVB case. At large radii the intrinsic densities are lower than the luminosity-weighted ones.      

Re-running the simulation at higher resolution will allow resolving higher densities in the CGM, and likely increase the \ion{H}{i} column densities by a few tens of percent \citep[e.g.][]{Hummels:2019,vandeVoort:2019,Peeples:2019,Suresh:2019,Bennett:2020}. As a consequence, the ionization cone opening angles permitted by the covering fraction of optically thick absorbers would increase. Simultaneously, higher densities would affect the emission levels, requiring smaller angles to match observations and possibly a smaller contribution from continuum pumping. Nevertheless, we should also keep in mind that running the same simulation, but with this new AGN EM feedback model on-the-fly might change the phase space distribution of CGM gas, and therefore, also the distribution of densities. Finally, we should test our predictions against radiation transfer post-processing, which might naturally provide the means to bring in complete agreement the $\alpha$ angles required by the Ly$\alpha$ and \ion{He}{ii} emission lines.

\section{Summary}
\label{sec:summary}

We used a high resolution (gas particle mass resolution of 3.6$\times$10$^{\rm 4}$M$_{\rm\odot}$, and gravitational softening down to 72~pc) zoom-in cosmological simulation of a massive (10$^{\rm 12.5}$M$_{\rm\odot}$) halo at $z=3$ to study in post-processing the direct effects of the radiation field generated by a central accreting black hole on its CGM. The simulation has been run with the galaxy formation model used for the NIHAO project \citep{Wang:2015}, without any AGN feedback. For the AGN SED we used the physically consistent continuum model of \citet{Done:2012}, with all but two parameters fixed to the median values  over a sample of type I AGN \citep{Jin:2012}, to which we added the stacked quasars line emission spectrum of \citet{Lusso:2015}. The two free parameters of the AGN SED are the SMBH mass and the Eddington accretion ratio, and for both we explored two different values (10$^{\rm 8}$ and 10$^{\rm 9}$M$_{\rm\odot}$ masses, and 0.1 and 1.0 accretion ratios), which together bracket the estimates for a large sample of SDSS quasars at $z>2$ \citep{Rakshit:2020}. To model the gas response to the radiation field we run large grids of {\sc cloudy} models with the incident radiation fields given by our choice of AGN SEDs and distances covering all the CGM of the halo, while the control models were run with the $z=3$ UVB of \citet{Khaire:2019}. 

The radiation from AGN is not emitted isotropically, and to study this anisotropy we positioned bi-cones of given opening angle $\alpha$ at various randomly placed sightlines through the centre of the halo, and stacked any intrinsic or observational CGM property of interest over these sightlines. We consider only the particles within the bi-cones to be illuminated by the AGN, while the ones outside are illuminated by the UVB. Under the radiation of an AGN, a significant fraction of the cool ($T<10^{\rm 5}$K) CGM will not cool catastrophically, but instead heat up. This fraction steadily increases with the angle $\alpha$ reaching almost 100\% for the maximum $\alpha=180^{\rm o}$ -- when all gas is irradiated by the AGN -- and for all but the weakest AGN SED studied ($M_{\rm\bullet}=10^8$M$_{\rm\odot}$ and $\lambda=0.1$), see Figure~\ref{fig:opening_angle}. For angles $\alpha=50^{\rm o}$--$60^{\rm o}$, half of the cool CGM is heating, while the other half is cooling catastrophically. From this exercise, it becomes clear that AGN radiation can stop or at least stall the cooling of CGM gas in massive halos \citep[e.g.][]{Ciotti:1997,Ciotti:2007}, and hence provide a viable negative (or preventive) feedback for star formation. 

We then turn to observables.
In particular, at high redshift, there are CGM observations for quasar host halos both in absorption \citep[e.g. the QPQ survey][]{Hennawi:2006,Hennawi:2013,Prochaska:2009,Prochaska:2013a,Prochaska:2013,Prochaska:2014,Lau:2016,Lau:2018}, and in rest-frame UV emission lines  \citep[e.g.][]{ArrigoniBattaia:2019,Fossati:2021}. To compare with these observations, we 
constructed mock column density and ionization parameter profiles, and surface brightness profiles from SB maps in three emission lines (Ly$\alpha$, \ion{He}{ii}, \ion{C}{iv}). Similar to previous studies using simulations run without AGN feedback \citep[e.g.][]{FaucherGiguere:2016}, we found \ion{H}{i} column densities as high as those inferred from the QPQ sample \citep[][]{Lau:2016} if the CGM is only photoionized by the UVB. A similar agreement can be found, also if we do consider a central AGN source, but restrict the ionization cone opening angle $\alpha$ to low values (see Table~\ref{tab:alpha}). \citet{Lau:2016} also estimated ionization parameters for their absorbers, so we can use the parameter space $U$ vs $N_{\rm\ion{H}{i}}$ to check which combination of photoionization sources better explains the observations. In this case, we find that only AGN models with low $\alpha$ are capable of simultaneously explaining the spread in 'observed' $U$ and $N_{\rm\ion{H}{i}}$ values (Figure~\ref{fig:qpq8_fig13}). The UVB only model instead leads to a very marked anti-correlation between $U$ and $N_{\rm\ion{H}{i}}$, with a scatter too small to encompass all observations.    

Firmer constraints on the AGN model come from comparing the simulation with emission nebulae around quasars. We created mock SB maps for the rest-frame UV lines Ly$\alpha$, \ion{He}{ii}, \ion{C}{iv}, and from them stacked average radial profiles that can be compared to the stack of 27 quasar fields observed by \citet{Fossati:2021}. The simulated average profile of Ly$\alpha$ has the same functional form as the one in observations (Figure~\ref{fig:fossati_fig7} and Equation~\ref{eq:lya_fit}) with scale-length $R_s=26.8$kpc and power $\beta=5.5$, while its normalization depends on the AGN model and angle $\alpha$ (Table~\ref{tab:Lya_fits}). This allowed us to use the observed profile for constructing the posterior probability of $\alpha$ for each of the four AGN models. Doing similar comparisons observation--simulation for the other two emission lines we can thus see which AGN model results in compatible $\alpha$ angle for all three emission lines. In our case, the model with $M_{\rm\bullet}=$10$^{\rm 9}$M$_{\rm\odot}$ and $\lambda=0.1$ has the best compatibility between the three $\alpha$ values (see Table~\ref{tab:alpha}), with $\alpha=60^{\rm o}$ reproducing well the Ly$\alpha$ and \ion{He}{ii} observations, and falling within 1$\rm\sigma$ of the angle required by \ion{C}{iv}. The tension with the $\alpha$ required by \ion{C}{iv} arises from the fact that the average stacked simulated \ion{C}{iv} profile is steeper than the observational one, likely indicating that our simulation does not push enough metals beyond the inner CGM region. Rerunning the simulation with this EM AGN feedback on-the-fly will clarify if the CGM can be enriched to larger radii, or if a more explosive type of feedback is needed.

The comparison between the simulation and the CGM observations revealed a few very interesting facts: 
\begin{itemize}
    \item It demonstrates how CGM observables can be used to pinpoint properties of the central AGN engine. In this respect, we think that deeper observations targeting non-resonant lines (i.e. \ion{He}{ii}, H$\alpha$) will have the greatest power in constraining AGN properties (e.g. mass, accretion rate, opening angle; see Section~\ref{subsec:HeII}).
    \item The very good match between the brightness (Figure~\ref{fig:fossati_fig7}) and morphology (Figure~\ref{fig:Lya_coverf}) of simulated and observed Ly$\alpha$ nebulae indicates that the resolution of this simulation is likely enough to numerically model these types of CGM emission (see also Section~\ref{sec:caveat_densities} and \citealt{Costa:2022}). The absorption observations, instead, seem to require only slightly higher resolving power for the cool CGM.
    \item Continuum pumping of the quasar's emission lines is needed to bring in agreement the predictions from emission and absorption observations (Figure~\ref{fig:nolines_Lya}).
    \item Detection of \ion{C}{iv} nebulae around individual quasars at high significance could trace the metal enrichment to larger distances from the galaxy, and consequently help understand if a realistic CGM enrichment around single quasars can be achieved with only a gentle EM AGN feedback, or if explosive feedback is needed.
    \item Among the four AGN models, the one resulting in the best agreement with all the observations we considered ($M_{\rm\bullet}=$10$^{\rm 9}$M$_{\rm\odot}$ and $\lambda=0.1$) is also the most representative of the wider high-$z$ population of quasars at the targeted redshifts (Figure~\ref{fig:sdss}).
    \item This favored model implies a balance between the CGM reservoir undergoing heating and cooling (Figure~\ref{fig:opening_angle}), hinting to the importance of self-regulation around AGNs also for the EM feedback, as it has been shown for other AGN feedback mechanisms (e.g., \citealt{Gaspari:2011}).
\end{itemize}

To validate the exercise we made in this paper, we plan to: implement this EM AGN feedback in the {\sc gasoline2} code, run more similarly massive and slightly higher resolution simulations, both without and with this new implementation of feedback, and apply in post-processing a radiation transfer code that can follow also resonant scattering effects. Building up statistically significant samples of high resolution massive galaxies, run with different AGN implementations, is needed in order to fully exploit the potential of CGM observations for constraining not only feedback models, but also the properties of individual observed quasars (e.g. AGN mass, accretion rate, ionization cone opening angle). Future works have to understand whether also the properties of the simulated galaxies match observations, when the simulations are run with a milder or absent AGN ejective feedback together with EM feedback.

\section*{Acknowledgements}

The authors would like to thank Simeon Bird for providing a timely and insightful report that helped improve the clarity of the paper. The authors would also like to thank Seok-Jun Chang for fruitful discussions on the line center optical depths used in Section~\ref{sec:emission}. A.O. is funded by the Deutsche Forschungsgemeinschaft (DFG, German Research Foundation) -- 443044596. T.B.'s contribution to this project was made possible by funding from the Carl Zeiss Foundation. The simulation and subsequent analysis were carried out on the High Performance Computing resources at New York University Abu Dhabi. We are indebted to the developers of the following open-source {\sc python} libraries: {\sc pynbody} \citep{pynbody}, {\sc numpy} \citep{numpy}, {\sc scipy} \citep{scipy}, {\sc matplotlib} \citep{matplotlib}, {\sc pandas} \citep{pandas}, {\sc seaborn} \citep{seaborn}, {\sc statsmodels} \citep{statsmodels}, and {\sc astropy} \citep{2018AJ....156..123A}.

\section*{Data Availability}

The simulation used in this work is available upon request to the corresponding author. All the tables generated with {\sc cloudy}, containing the heating and cooling, ionization fraction for the various species, and emissivities for various rest-frame UV and optical lines as functions of ($n_{\rm H}$,$T$,$Z$,$J_{\nu}$($M_{\rm\bullet}$,$\lambda$) || $J_{\nu}$(UVB)) are also available upon request to the corresponding author. These tables will be made open-access once we have completed to run all the {\sc cloudy} models covering the full ranges of SMBH masses and accretion rates.  


\bibliographystyle{mnras}
\bibliography{biblio} 


\appendix

\section{Example of {\sc cloudy} input}
\label{appendix:scripts}

The lines starting with \verb|##| are comments, while the lines starting with only one hashtag \verb|#| are alternative options that can be de-commented. 

The UVB models of \citet{Khaire:2019} are included in the current version of {\sc cloudy} (C17.03). Therefore, in principle, one would load them with the command \verb|table KS18 redshift 3.00|, for a particular redshift of interest. However, using this default command implies that resonant pumping of the UVB continuum will be on by default, leading to SB in Ly$\alpha$ and \ion{C}{iv} much higher that what is seen in observations. One way to avoid this anomalous  behavior for the UVB is to pass it as a custom made SED file using the \verb|table| command (similar to how we pass the AGN SED) and put to 0 the spectrum at the wavelengths of the resonant lines \citep[see also][]{ArrigoniBattaia:2015}. The various options of {\sc cloudy} to suppress resonant pumping of the continuum can have far reaching implications (see section 12.4 of the Hazy1 documentation). For this reason, we prefer to mask the Ly$\alpha$ and \ion{C}{iv} wavelengths in the UVB spectrum manually. 

For both the UVB and the AGN cases, we set the normalization of the SED using the \verb|phi(H) [value]| command, where \verb|[value]| is the log of the ionizing photon flux $\Phi(H)$ in units of cm$^{\rm-2}$s$^{\rm-1}$. For the UVB we use 5.48 (value computed for the fiducial $z=3$ model of \citet{Khaire:2019}), while for the AGN models we compute a grid of values representing different distances from the source, with a lower limit of 1~kpc and an upper limit given by the distance from the galaxy center to a corner of the selected 534~kpc box.   

\begin{verbatim}
constant density
hden [value] log
constant temperature, t = [value] K log
abundances "solar_GASS10.abn"
metals [value] log
no molecules
## set the incident radiation fields 
#table SED "[name_UVBsed_file]"
#phi(H) [value]
table SED "[name_AGNsed_file]"
phi(H) [value]
CMB redshift 3.00
# set the stopping criteria
stop temperature off
stop column [value] log
## if t>5.0 stop after 1 zone and fix its thickness 
#set dr 20.0
#stop zone 1
## set the output
set save prefix "[name_input_file]"
save heating each last ".heating"
save cooling each last ".cooling"
save element hydrogen last ".H_ionf"
save element helium last ".He_ionf"
save element oxygen last ".O_ionf"
save element carbon last ".C_ionf"
save element nitrogen last ".N_ionf"
save element silicon last ".Si_ionf"
save lines emissivity emergent last ".emissivities"
H  1 1215.67A
He 2 1640.43A
C  4 1548.19A
C  4 1550.78A
Blnd 1549.00A
H  1 6562.81A
end of lines
iterate to convergence
\end{verbatim}

\section{Absorption columns of higher energy ions}
\label{appendix:absorbtion}

\begin{figure*}
    \centering
    {$M_{\rm\bullet}$=10$^{\rm 8}$M$_{\rm\odot}$ \& $\lambda=0.1$\hspace{3.5cm}$M_{\rm\bullet}$=10$^{\rm 9}$M$_{\rm\odot}$ \& $\lambda=0.1$\hspace{3.5cm}$M_{\rm\bullet}$=10$^{\rm 9}$M$_{\rm\odot}$ \& $\lambda=1.0$}\\
    \includegraphics[width=0.33\textwidth]{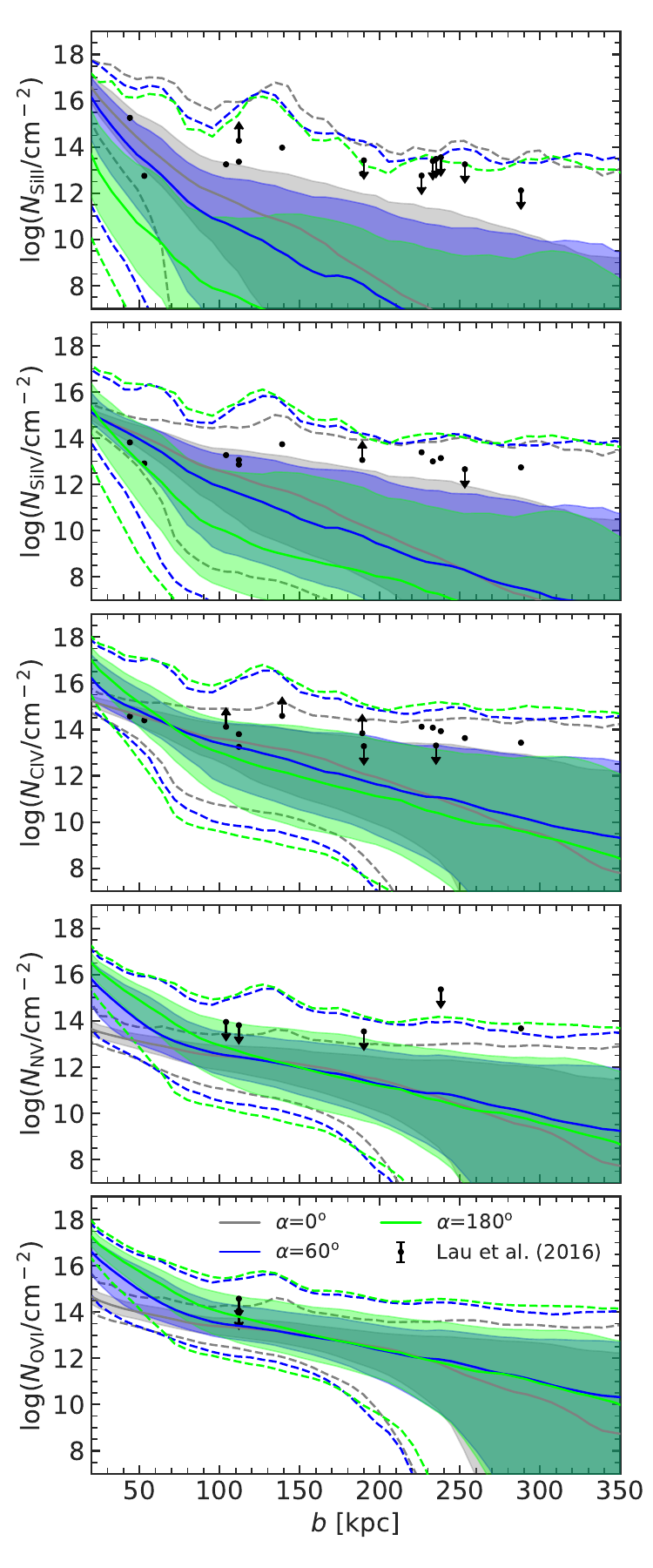}
    \includegraphics[width=0.33\textwidth]{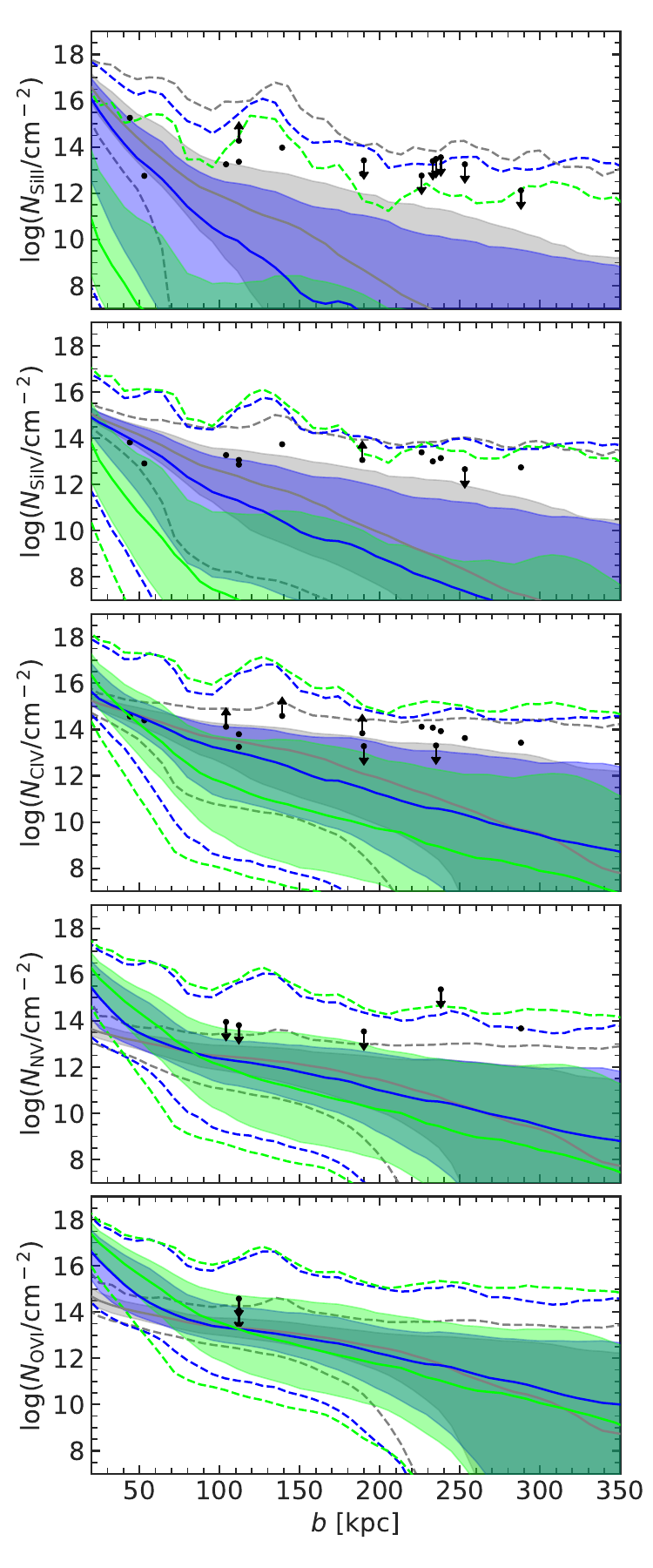}
    \includegraphics[width=0.33\textwidth]{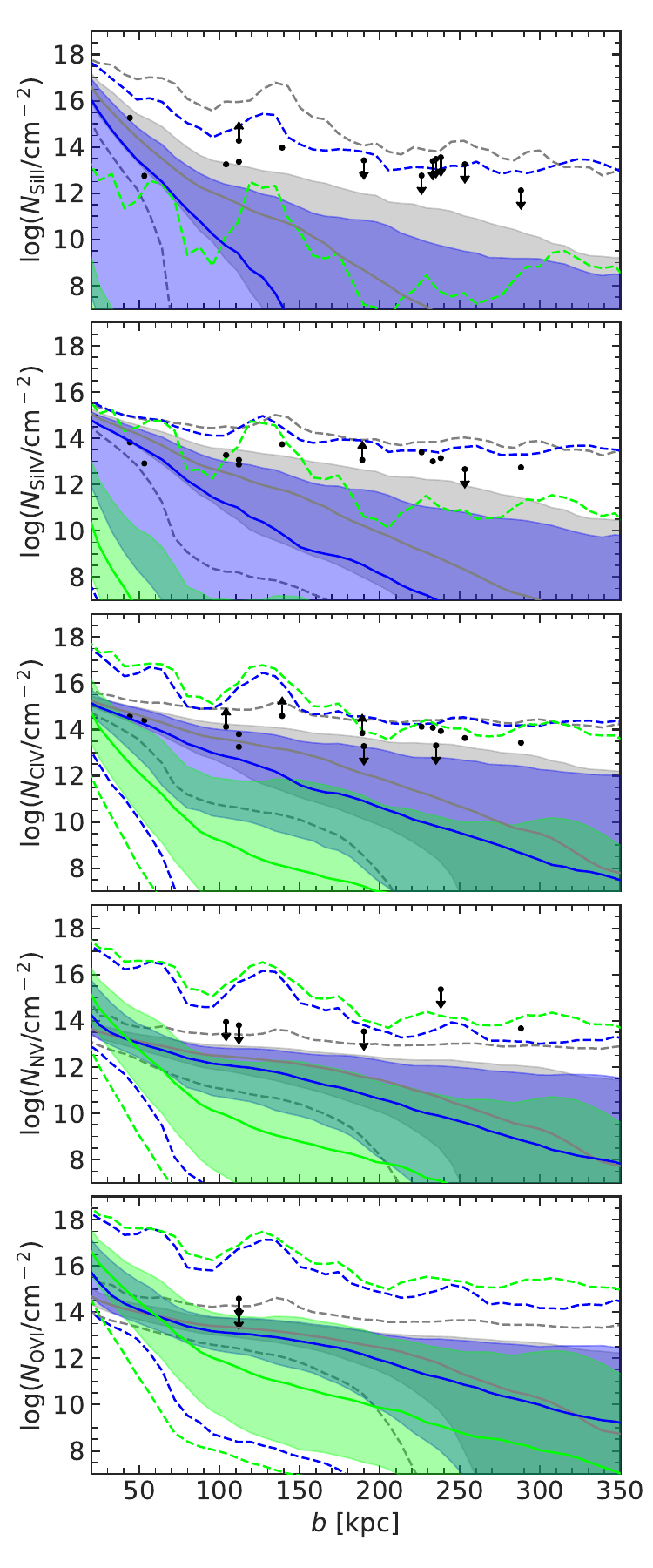}    
    \caption{Column densities for five ionic species of increasing ionization energy as a function of projected distance from the quasar for three AGN spectra with increasing number of ionizing photons from left to right. From top to bottom we show the column densities of \ion{Si}{ii} (16.3~eV), \ion{Si}{iv} (45.1~eV), \ion{C}{iv} (64.5~eV), \ion{N}{v} (97.9~eV) and \ion{O}{vi} (138.1~eV). The black points are the observational estimates of \citet{Lau:2016} foreground quasars. Each percentile curve was obtained from the stacking of 10 different directions to the simulated halo (each within a randomly placed ionization cone). 
    The line styles and colors represent the same percentiles and $\alpha$s as in Figure~\ref{fig:qpq8_fig13}.}
    \label{fig:qpq8_figallions}
\end{figure*}

In this appendix we show, for completeness, the predictions for a set of 
ion column densities as a function of impact parameter. The ions covered are
\ion{Si}{ii} (16.3~eV), \ion{Si}{iv} (45.1~eV), \ion{C}{iv} (64.5~eV), \ion{N}{v} (97.9~eV) and \ion{O}{vi} (138.1~eV), which are those used by \citet{Lau:2016} in combination
with the \ion{H}{i} absorption to constrain the ionization parameters $U$ around their sample of quasars. Figure~\ref{fig:qpq8_figallions} compares our predictions
with the data available from that work. The main points of this comparison are summarized in Section~\ref{sec:absorbtion}.

\section{Effects of local absorption correction and AGN emission lines on CGM observables}
\label{appendix:nolines}

\begin{figure}
     \centering
     \includegraphics[width=0.48\textwidth]{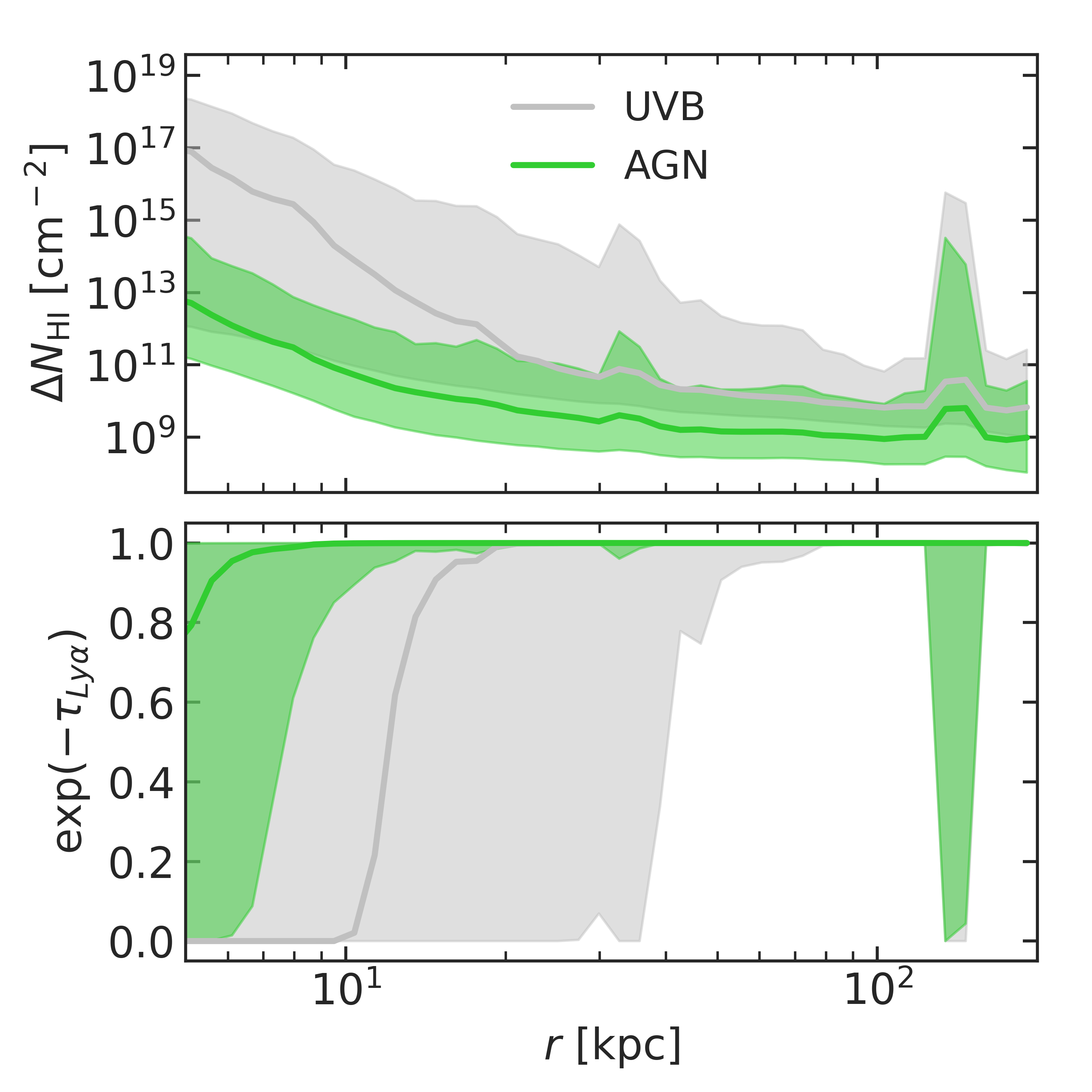}
     \caption{Local neutral hydrogen column density $\Delta N_{HI}$ as a function of distance to the center of the halo (top), and the corresponding absorption correction for the Ly$\alpha$ emission (bottom). Grey and green represent two different sources of photoionization as indicated in the legend and explained in the text.}
     \label{fig:NH_and_tau_lya}
\end{figure}

The local absorption approximation we use has a bigger effect on the gas in the case where the UVB is the only source of ionizing photons. The top panel of Figure~\ref{fig:NH_and_tau_lya} shows the median neutral hydrogen local column as a function of distance from the center of the halo, in two extreme cases: all gas particles are photoionized only by the UVB (grey), and all are photoionized only by a 10$^{\rm 9}$M$_{\rm\odot}$ SMBH accreting at 10\% of its Eddington rate (green). The bottom panel shows the corresponding distributions of local absorption corrections as functions of distance from the center. Shaded areas in both panels give the 16$^{\rm th}$ and 84$^{\rm th}$ percentiles. For the CGM region ($r>20$kpc), the local absorption is negligible for the AGN photoionization case for all gas particles, except the dense ones belonging to the infalling satellite at $r\sim r_{\rm 200}$. For the UVB photoionization case, the local absorption is very important, especially in the inner CGM region ($r\leq 60$kpc).   

\begin{figure}
 \centering
 \includegraphics[width=0.45\textwidth]{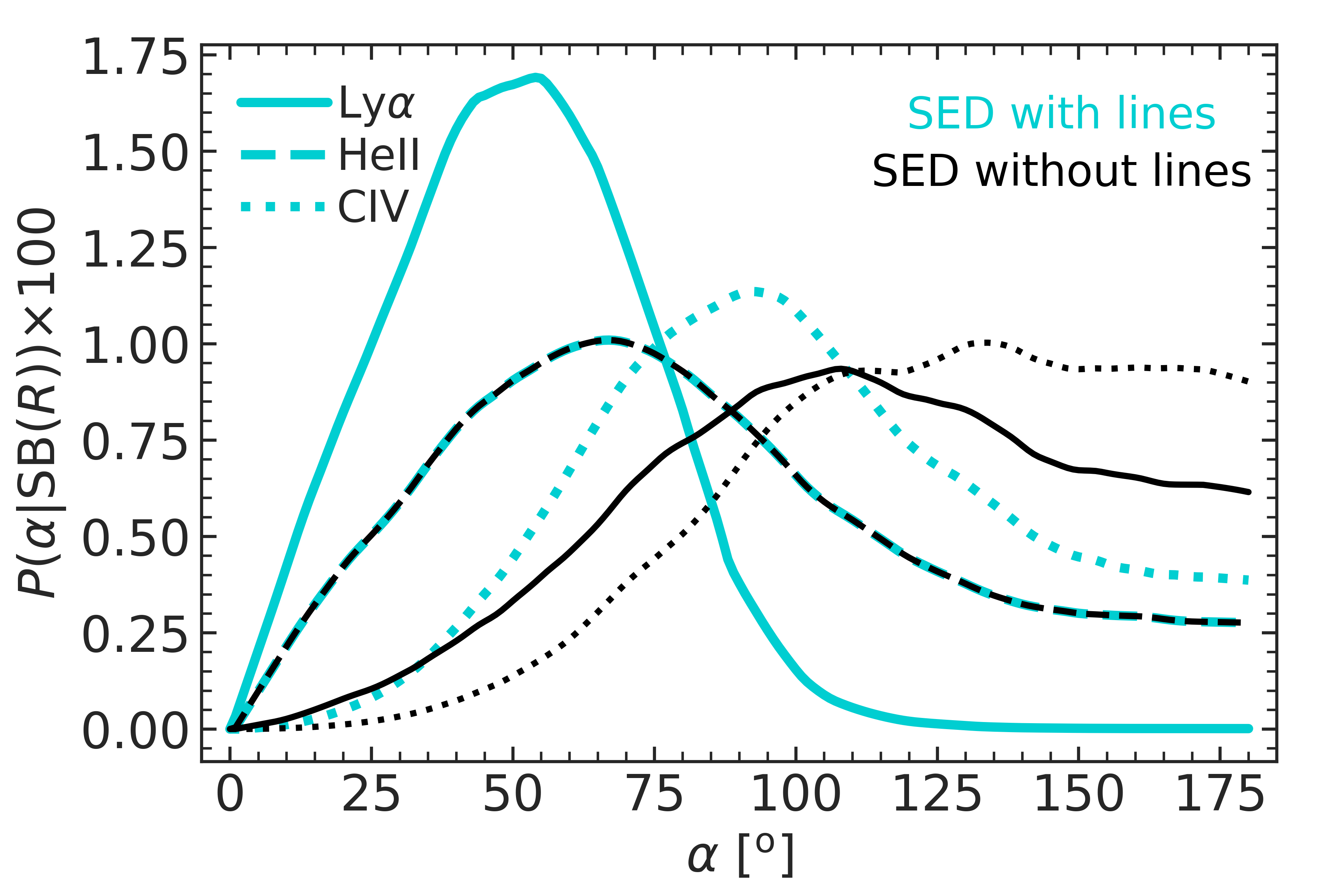}
 \caption{Comparison between the posterior probability distributions of $\alpha$ as constrained by Fossati's stacked emission line profiles above the noise level for the Ly$\alpha$ (solid), \ion{He}{II} (dashed) and \ion{C}{iv} (dotted), for our preferred AGN SED model ($M_{\rm\bullet}$=10$^{\rm 9}$M$_{\rm\odot}$ \& $\lambda=0.1$) with (turquoise) and without (black) the contribution of Lusso's stacked emission line AGN spectrum.}
\label{fig:nolines_Lya}
 \end{figure}

We tested how the UV line emission maps and derived surface brightness profiles change when we remove the line spectrum contribution \citep{Lusso:2015} from the AGN SED models. As expected, the \ion{He}{ii} emission is not impacted at all, while the SBs of Ly$\alpha$ and \ion{C}{iv} decrease without changing the overall shape of the profiles. To quantify the differences in SBs with and without the line spectrum contribution, we give in Figure~\ref{fig:nolines_Lya} the posterior probabilities of $\alpha$ for the three lines, for our preferred AGN model. If the input continuum does not include the line emission spectrum, higher ionization cone opening angles are needed to explain the SB profiles of Ly$\alpha$ and \ion{C}{iv}. E.g. the most likely $\alpha$ angle needed to explain the Ly$\alpha$ SB is $\sim$ twice as large as our default case. However, such large ionization cone opening angles are excluded by absorption studies as we discuss in Section~\ref{sec:absorbtion}.   

\begin{figure*}
    \centering
    \includegraphics[width=0.95\textwidth]{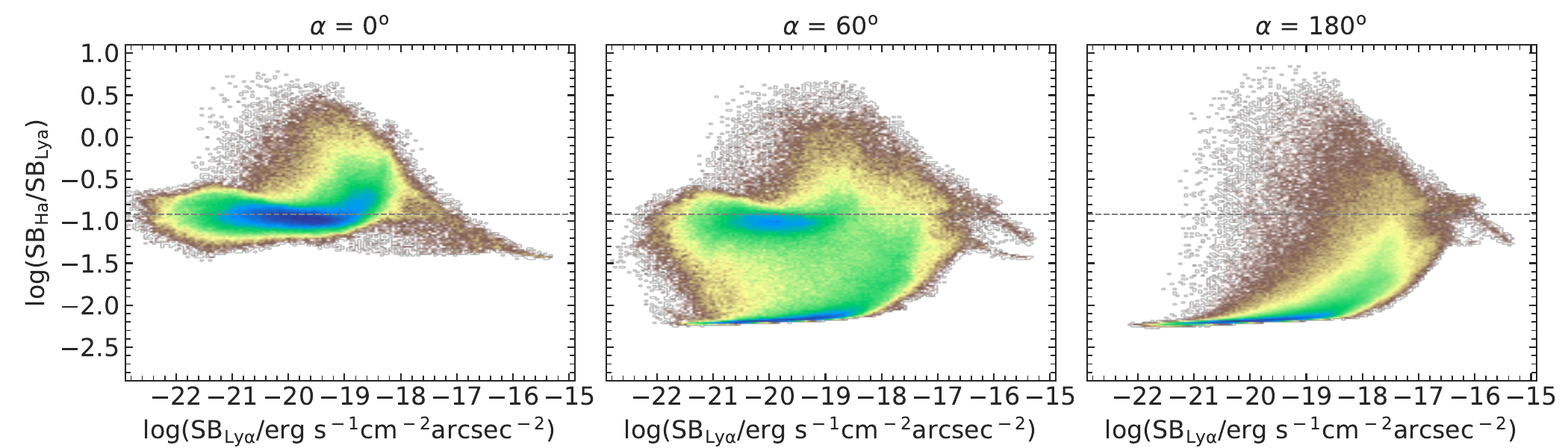}\\
    \includegraphics[width=0.95\textwidth]{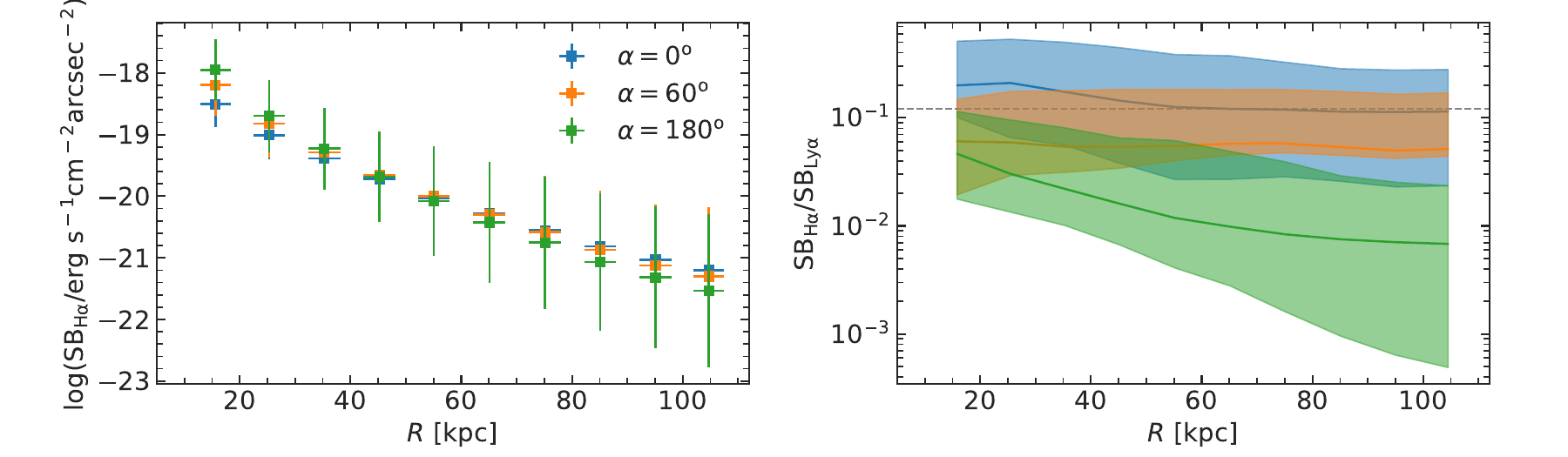}\\
    \caption{Same as Figure~\ref{fig:Ha_predictions} but using the AGN SED spectrum without emission lines.}    
    \label{fig:Ha_predictions_woLusso} 
\end{figure*}

\bsp	
\label{lastpage}
\end{document}